\documentclass{aastex631}
\usepackage{threeparttable}
\usepackage[nooneline]{subfigure}
\usepackage{amsmath}
\usepackage{comment}
\subfiguretopcaptrue
\usepackage{bm}       
\usepackage{braket}    
\usepackage{ulem}  
\DeclareRobustCommand{\erase}{\bgroup\markoverwith{\textcolor{red}{\rule[.5ex]{2pt}{0.4pt}}}\ULon} 

\usepackage{natbib}
\usepackage{etoolbox} 
\usepackage{hyperref}

\newtoggle{citedPaperI}
\togglefalse{citedPaperI}

\newcommand{\citepaperI}[1]{%
  \iftoggle{citedPaperI}{%
    \hyperref[cite.#1]{Paper I}%
  }{%
    \citet{#1}%
    \label{cite.#1}%
    \global\toggletrue{citedPaperI}%
  }%
}

\accepted{August 16, 2024}

\shorttitle{Filament-Filament Collision}
\shortauthors{Kashiwagi, Iwasaki, and Tomisaka}

\begin{document}

\title{Instability and Evolution of Shocked Clouds Formed by Orthogonal Collisions between Magnetized Filamentary Molecular Clouds}

\correspondingauthor{Raiga Kashiwagi}
\email{raiga.kashiwagi@gmail.com}

\author[0000-0002-1461-3866]{Raiga Kashiwagi}
\affiliation{Division of Science, National Astronomical Observatory of Japan, 2-21-1 Osawa, Mitaka, Tokyo 181-8588, Japan}

\author[0000-0002-2707-7548]{Kazunari Iwasaki}
\affiliation{Division of Science, National Astronomical Observatory of Japan, 2-21-1 Osawa, Mitaka, Tokyo 181-8588, Japan}
\affiliation{Center for Computational Astrophysics, National Astronomical Observatory of Japan, 2-21-1 Osawa, Mitaka, Tokyo 181-8588, Japan}
\affiliation{Astronomical Science Program, The Graduate University for Advanced Studies (SOKENDAI), 2-21-1 Osawa, Mitaka, Tokyo 181-8588,
Japan}

\author[0000-0003-2726-0892]{Kohji Tomisaka}
\affiliation{Division of Science, National Astronomical Observatory of Japan, 2-21-1 Osawa, Mitaka, Tokyo 181-8588, Japan}
\affiliation{Astronomical Science Program, The Graduate University for Advanced Studies (SOKENDAI), 2-21-1 Osawa, Mitaka, Tokyo 181-8588, Japan}



\begin{abstract}
Filamentary molecular clouds are recognized as primary sites for the formation of stars. 
Specifically, regions characterized by the overlapping point of multiple filaments, known as hub regions, 
often associated with active star formation.
However, the formation mechanism of this hub structure is not well understood.
Therefore, to understand the formation mechanism and star formation in hub structures, 
as a first step, we investigate the orthogonal collisions between two filaments using three-dimensional ideal magnetohydrodynamical simulations.
As a model of initial filaments, we use an infinitely long filament in magnetohydrostatic equilibrium under a global magnetic field running perpendicular to the filament axis. 
Two identical equilibrium filaments, sharing the same magnetic flux, are arranged with their long axes perpendicular to each other and given an initial velocity perpendicular to their long axes to replicate an orthogonal collision.
We find three types of evolution after 
the shocked cloud is formed: collapse, stable, and expansion modes. 
The energy balance just after the filaments completely collide explains the future evolution of the shocked cloud.
If the magnitude of gravitational energy is larger than the sum of the kinetic, thermal, and magnetic energies, the shocked cloud evolves in collapse mode. If the magnitude of gravitational energy is less than the sum of these energies, the cloud evolves in stable mode when the kinetic energy is relatively small and in expansion mode when the kinetic energy is sufficiently large.

\end{abstract}


\section{Introduction} \label{sec:intro}

The structure of high-density regions within molecular clouds is generally characterized as elongated filaments. 
Recent observations using the {\it {Herschel Space Telescope}} \citep{2010A&A...518L...1P} have unveiled the ubiquity of these filamentary structures within molecular clouds \citep{2010A&A...518L.102A}. 
Furthermore, it has become evident that the majority of protostars exist along these filaments \citep[see, e.g.,][]{2010A&A...524A..18B,2015A&A...584A..91K,2019A&A...621A..42A}.
These observational findings have brought attention to star formation throughout the filament, which is called the ``filament paradigm" \citep{2014prpl.conf...27A}, and establishing a novel star formation scenario involving the evolution from molecular clouds through filaments to molecular cloud cores has become a significant research subject \citep[see, e.g.,][]{2023ASPC..534..153H}.
Stars seem to form in filaments with a line mass exceeding $\lambda_\mathrm{crit}=2c_s^2/G\simeq 16M_\odot\mathrm{pc}^{-1}$, where $c_s\simeq 190~\mathrm{m\,s^{-1}}$ and $G$ represent the typical isothermal sound speed of the filaments and the gravitational constant, respectively \citep{1987PThPh..77..635N,1992ApJ...388..392I}.
This $\lambda_\mathrm{crit}$ is the critical line mass, obtained from the hydrostatic equilibrium of the filaments \citep{1963AcA....13...30S,1964ApJ...140.1056O}, which represents the maximum value of the line mass supported against gravity by thermal pressure and is believed to be a key parameter of initiating the star formation.

However, from the polarization observation, actual starforming filaments are penetrated by the interstellar magnetic fields \citep[e.g.,][]{2011ApJ...734...63S,2013A&A...550A..38P,2019PASJ...71S...7S}.
This fact is important for the balance of the filaments because magnetic fields can support them against the gravity, and magnetic fields are also believed to play significant roles in subsequent star formation \citep[see, e.g.,][]{2023ASPC..534..193P}.
\citet{2014ApJ...785...24T} studied the magnetized critical line mass ($\lambda_\mathrm{crit,B}$) for the isothermal filament with a lateral magnetic field and presented $\lambda_\mathrm{crit,B}$ as a function of the magnetic flux: 
\begin{equation}\label{eq:magnetized_critical_line_mass}
    \lambda_{\rm crit,B}\simeq 0.24 \frac{\Phi}{G^{1/2}}+1.66\frac{c_s^2}{G},
\end{equation}
where $\Phi$ denotes one-half of the magnetic flux threading a filament per unit length 
(see also \citet{2021ApJ...911..106K} for the equivalent empirical relation concerning a polytropic gas filament with a negative temperature gradient).
Equation (\ref{eq:magnetized_critical_line_mass}) indicates that the magnetized filament with a sufficiently strong magnetic field can support a larger line mass by the Lorentz force, in comparison to the nonmagnetized filaments.
If the line mass of a filament exceeds $\lambda_\mathrm{crit,B}$, no magnetohydrostatic equilibrium solution exists, and the filament will undergo radial collapse. 
Therefore, star formation inside filaments requires exceeding this value.

Recent observations have indicated that active star formation often occurs at the junction or overlap point of the filaments \citep[e.g.,][]{2014ApJ...791L..23N,2019ApJ...884...84D,2020A&A...642A..87K,2024MNRAS.527.5895D}. 
For instance, \citet{2014ApJ...791L..23N} proposed that the formation of a star cluster in the Aquila Serpens South region, which is one of the most active star-forming regions within nearby molecular clouds, was triggered by the collision between three filaments.  
\citet{2020A&A...642A..87K} reported that, in high-mass star-forming regions, all luminous clumps ($L\ge10^4L_\odot$) exist in the intersection of the filaments, known as hub-filamentary structures.
This property is also seen in the Large Magellanic Cloud \citep{2019ApJ...886...15T,2023ApJ...955...52T}.
Furthermore, \citet{2023ApJ...947L..29A} studied that the hub-filament structure is crucial for explaining the observed properties of the birth environment of the Sun.
Therefore, understanding the hub-filament structure is important for explaining various aspects of star formation.

One of the formation processes of these hub structures is thought to be through the coalescence of multiple filaments. 
Therefore, collisions between filaments may potentially trigger the formation of not only star clusters but also massive stars.
However, the formation scenario of the hub filament structure is still unresolved, 
and the physical characteristics of filament collision evolution are also not well understood.

Based on these observational findings, there are several theoretical studies of the induced star formation caused by the filament-filament collision \citep{2011A&A...528A..50D,2021MNRAS.507.3486H,2023ApJ...954..129K,2024MNRAS.532L..42H}.
In particular, \citepaperI{2023ApJ...954..129K} (hereafter \citepaperI{2023ApJ...954..129K})
studied the instability of shocked clouds resulting from filament collisions and identified conditions that lead to collapse.
\citepaperI{2023ApJ...954..129K} simulated a head-on (parallel) collision of isothermal filaments of identical mass, in a magnetohydrostatic equilibrium state, penetrated by a lateral magnetic field, with relative initial velocities ranging from $c_s$ to $10c_s$.
\citepaperI{2023ApJ...954..129K} indicated that collapse occurs, regardless of the initial velocity, if the initial total line mass exceeds the magnetized critical line mass ($\lambda_\mathrm{tot}>\lambda_\mathrm{crit,B}$); otherwise, the filaments maintain an oscillatory stable state. 
However, \citepaperI{2023ApJ...954..129K} was restricted to collisions such as the filament's long axes being parallel. 
This does not necessarily represent the full perspective of filament collision phenomena. 
Therefore, in this paper, we investigate collisions where the filament's long axes are orthogonal. 
This aims to generalize the conditions under which the shocked region becomes unstable in filament collisions.

The paper is organized as follows.
Section \ref{sec:method} explains a model 
for our numerical calculations.  
Section \ref{sec:result} introduces the simulation results obtained from the filament collision.
In section \ref{sec:discussion}, we discuss the condition of the instability of the shocked clouds.
Section \ref{sec:summary} summarizes the results and conclusions of this study.

\section{Methods}\label{sec:method}

This section explains the basic equations, initial, and boundary conditions, and parameters used to investigate orthogonal collisions of two identical filaments penetrated by a lateral magnetic field.
Although there are differences in the initial filament positions and boundary conditions, fundamentally, our study follows a similar approach as \citepaperI{2023ApJ...954..129K}. 
However, for the sake of clarity, we explain the methodology starting from the basic equations.

\subsection{Basic Equations}
To simulate orthogonal collisions, we solve the ideal magnetohydrodynamic equations in a conservative form as shown below:
\begin{equation}\label{eq:mass_conservation}
   \frac{\partial \rho}{\partial t}+\nabla\cdot(\rho \textit{\textbf{v}})=0,
\end{equation}

\begin{equation}\label{eq:equation_of_motion}
   \frac{\partial \rho\textit{\textbf{v}}}{\partial t}+\nabla\cdot\left(\rho \textit{\textbf{v}}\textit{\textbf{v}} -\frac{\textit{\textbf{B}}\textit{\textbf{B}}}{4\pi}+P_{\rm tot}\right)=-\rho\nabla \psi,
\end{equation}    
   
\begin{equation}\label{eq:induction_equation}
   \frac{\partial \textit{\textbf{B}}}{\partial t}-\nabla\times\left(\textit{\textbf{v}}\times \textit{\textbf{B}}\right)=0,
\end{equation}  
where $\rho$, $\bm{v}$, and $\bm{B}$ represent respectively the density, the velocity, and the magnetic field, and $P_{\rm tot}$ is the total pressure, including gas pressure $(p_\mathrm{gas})$ and magnetic pressure $(B^2/8\pi)$, expressed as
\begin{equation}\label{eq:total_pressure}
   P_{\rm tot}=p_{\rm gas}+\frac{B^2}{8\pi}.
\end{equation}
Here, $\psi$  of Equation (\ref{eq:equation_of_motion}) indicates the gravitational potential derived from solving the Poisson's equation, as shown below:
\begin{equation}\label{eq:poisson's_equation}
   \Delta \psi=4\pi G\rho.
\end{equation}  

We assume two isothermal filaments enclosed by an isothermal hot ambient medium, resulting in the coexistence of two gases at distinct temperatures.
We replicate this situation utilizing a scalar field, governed by the following equation of state:
\begin{equation}\label{eq:equation_of_state}
p_{\rm gas}=S \rho,
\end{equation}
where $S$ represents the scalar field proportional to the temperature.
We assume that $S$ evolves according to the advection of gas, as described below:
\begin{equation}
\frac{\partial S}{\partial t}+(\bm{v}\cdot \mathbf{\nabla})S=0.
\label{advection_scalar}
\end{equation}
Further details regarding the scalar field are provided in the following section.


\subsection{Initial Condition}\label{sec:initial_condition}

\begin{figure*}
    \centering

     \begin{tabular}{ccc}
         \subfigure[]{
         \includegraphics[keepaspectratio,scale=0.25]{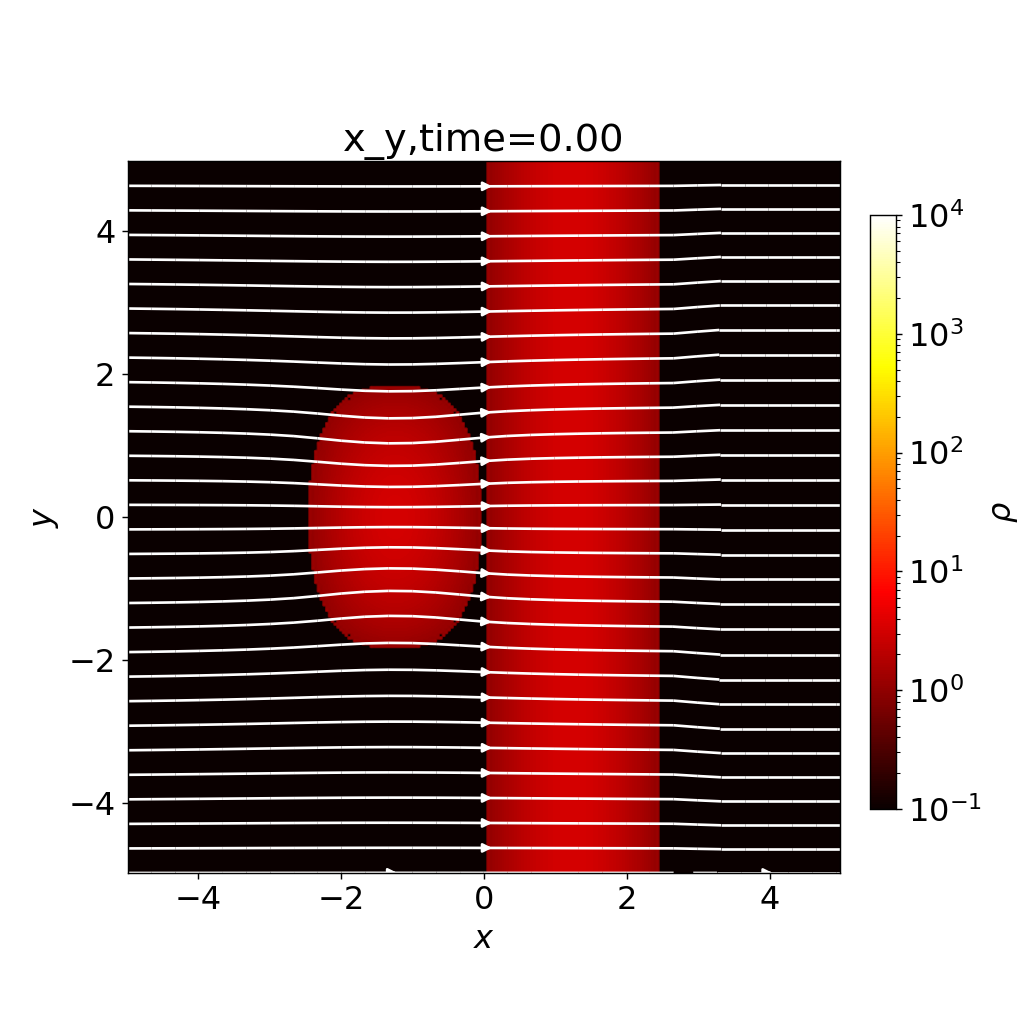}  
         }&
         \subfigure[]{
         \includegraphics[keepaspectratio,scale=0.25]{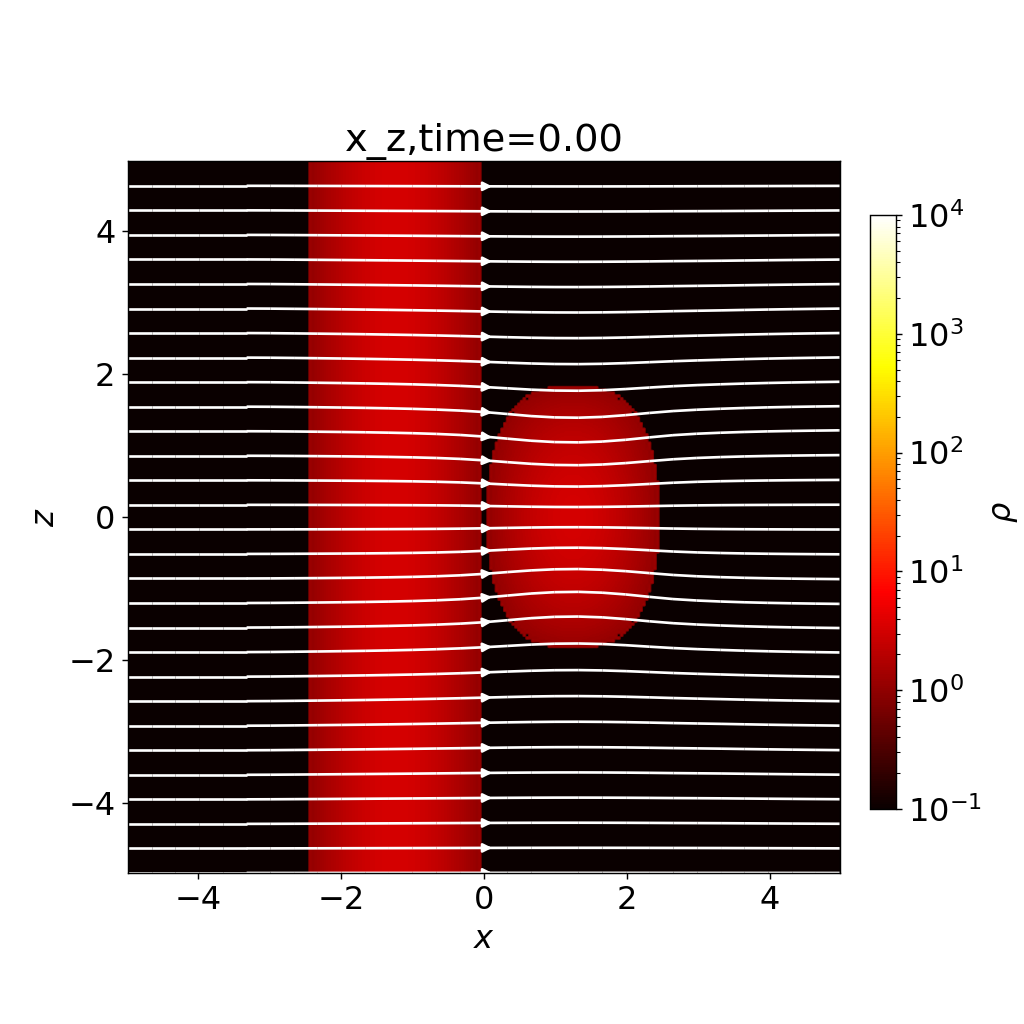}
         }&
         \subfigure[]{
         \includegraphics[keepaspectratio,scale=0.25]{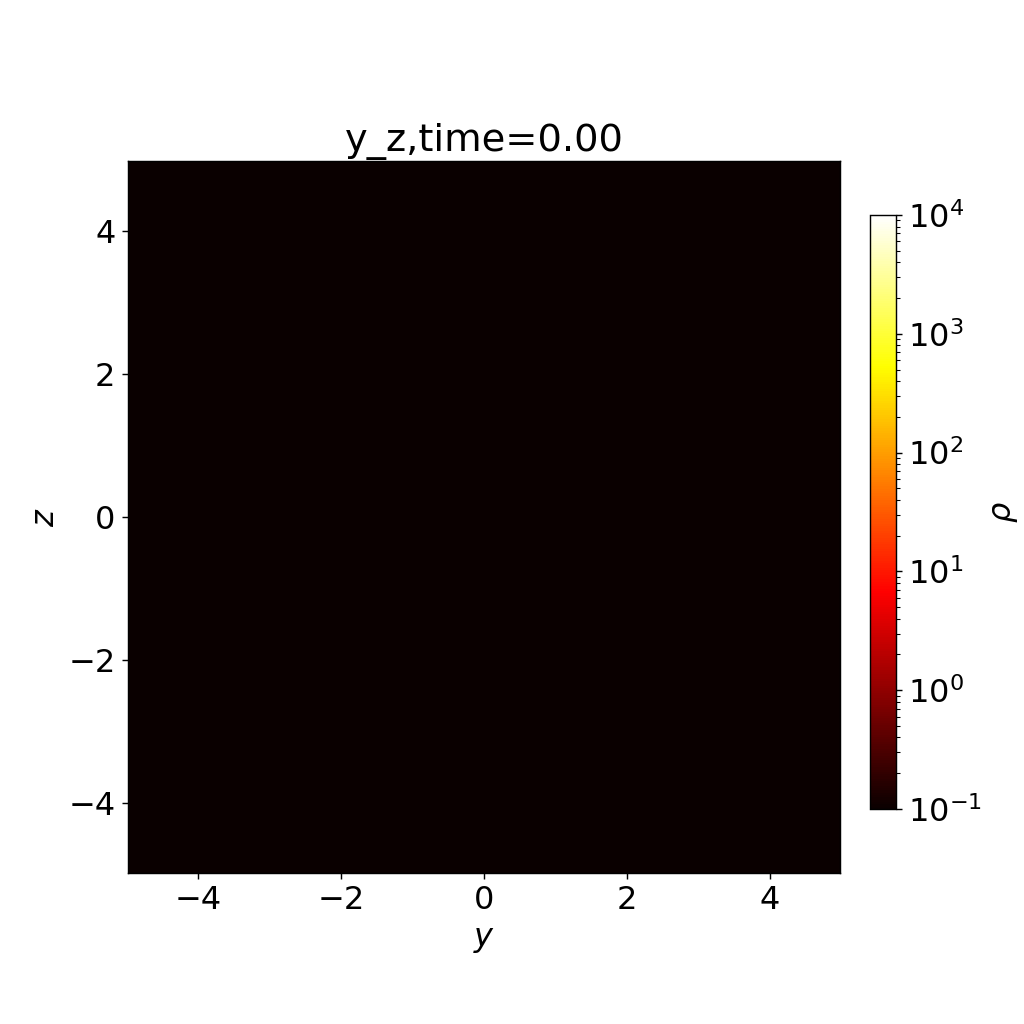} 
         }\\                  
     \end{tabular}
      \caption{An example of the initial conditions for orthogonal collisions.
      Two-dimensional slice of the model B1L05V1 (see Table \ref{tab:model_parameters}). 
      The panels (a), (b), and (c) correspond to the slices at $z=0$, $y=0$, and $x=0$ planes, respectively.
      The color scale represents the density, and the white lines correspond to the magnetic field lines.
      }
     \label{fig:initial_condition}
\end{figure*}

Figure \ref{fig:initial_condition} depicts one of the initial configurations for the orthogonal collisions, represented in the Cartesian coordinate system $(x,y,z)$. 
Initially, each filament is assumed to be in magnetohydrostatic equilibrium, with a global magnetic field oriented perpendicular to its elongated axis, and this equilibrium state is obtained through numerical calculations \citep{2014ApJ...785...24T,2021ApJ...911..106K}.
The magnetic field outside the filament is assumed to be force-free and converged to the uniform magnetic field $\bm{B}_0$ with increasing distance from the filament, following \citet{2014ApJ...785...24T}.
This equilibrium filament is specified with three parameters: the uniform magnetic field strength $B_0$, a radius of a parental (cylindrical) filament $R_0$, and a line mass (details are shown in Section \ref{sec:parameters}).
Then the two identical filaments are positioned adjacent to each other and collide in the $\pm x$-direction.
As shown in Figure \ref{fig:initial_condition}, since the two filaments are identical, slices at $z=0$ (a) and at $y=0$ (b) are mirror symmetric to each other. 
Thus, hereafter, we display only the $z=0$ and $x=0$ two-dimensional slices to show the evolution.

In the context of orthogonal collisions, it is necessary to provide a density structure and a magnetic field distribution where the angle formed by the axes of the two filaments is 90 degrees.
Therefore, the filament in $x<0$ aligns its long axis with the $z$-axis, while the other filament in $x>0$ aligns its long axis with the $y$-axis.

To set the initial magnetic field so that 
$\bm{\nabla}\cdot {\bm B}$ is zero in 
round-off errors, the initial magnetic field $\bm{B}$ is obtained by numerically differentiating the vector potential $\bm{A}^\mathrm{3D}(x,y,z)=(0,A_y^\mathrm{3D}(x,z),A_z^\mathrm{3D}(x,y))$ which is derived using the magnetohydrostatic solution (i.e., $\bm{B}=\nabla\times \bm{A}$).
The three-dimensional vector potential is given by extending the two-dimensional ones of the magnetohydrostatic equilibrium solutions along the long axis of the filaments.

We divide the vector potential (and consequently the magnetic field) into two components: 
the uniform magnetic field $\bm{B}_0$ aligned with the $x$-axis, and the residual component $\bm{B}_\mathrm{curv}$.
The vector potential used to derive $\bm{B}_\mathrm{curv}$ is obtained by superposing $\bm{A}(x,y,z) = (0,0,A_z^\mathrm{2D,curve}(x,y))$ which is made by the current $j_z$ flowing along the filament located in $x<0$ and $\bm{A}(x,y,z) = (0,A_y^\mathrm{2D,curve}(x,z),0)$ which is made by the current $j_y$ flowing along the filament located in $x>0$.
Here, $A_z^\mathrm{2D,curv}$ represents the bending component of the magnetic field in the magnetohydrostatic solution, given by $A_z^\mathrm{2D,curv}(x,y) = A_z^\mathrm{2D}(x,y) - B_0y$, while, $A_y^\mathrm{2D,curve}(x,z)$ is obtained by rotating $A_z^\mathrm{2D,curve}(x,y)$ by 90 degrees around the $x$-axis.
After deriving $\bm{B}_\mathrm{curv}$ from the vector potential, the initial magnetic field $\bm{B}$ is determined by adding $\bm{B}_\mathrm{curv}$ to the uniform magnetic field $\bm{B}_0=(B_0,0,0)$.
This approach allows us to derive a magnetic field configuration devoid of spurious current sheets at $x=0$.

In our simulation, we assumed a constant gas temperature within the molecular cloud, approximately $T\simeq 10\,\mathrm{K}$, therefore using the isothermal equation of state.
While in reality, filaments exist within molecular clouds, our calculation assumes the filament is surrounded by hot, low-density gas. 
We reproduce this situation using a scalar field $S$ in the equation of state. 
This assumption attempts to replicate the magnetohydrostatic equilibrium state described by \citet{2014ApJ...785...24T}, where the filament is immersed in ambient pressure $p_\mathrm{ext}$.
Initially, we assume pressure equilibrium between the filament and the external medium at their interface, described as follows:
 \begin{equation}
     p_s= S_\mathrm{fil}\rho_s= S_\mathrm{ext}\rho_\mathrm{ext},
 \end{equation}
 where $p_s$ denotes the pressure at the filament surface,
 $\rho_s$ is the density at the filament surface, $\rho_\mathrm{ext}$ is the density of the external medium, and
 $S_\mathrm{fil}$ and $S_\mathrm{ext}$ indicate the scalar fields inside and outside of the filament, respectively.
 In this study, the density contrast between the external medium and the filament surface is set as $\rho_\mathrm{ext}/\rho_s= 0.01$. 
 Consequently, the scalar $S$ for the external medium is $S_\mathrm{ext}=100c_s^2$, and for the interior of the filament, it is $S_\mathrm{fil}=c_s^2$.
 In other words, in our simulation, the value of the sound speed of the external medium is maintained at $S_\mathrm{ext}^{1/2}=10c_s$ by solving the advection equation for $S$ (Equation (\ref{advection_scalar})).

\subsection{Normalization}\label{sec:normalization}

\begin{table*}
    \caption{Units used in normalization}          
    \label{tab:normalization}      
    \centering          
    \begin{tabular}{ll}
    \hline 
        Unit of pressure\dotfill &  External pressure, $p_{\rm ext}$ \\
        Unit of density\dotfill &  Density at the surface, $\rho_s=p_\mathrm{ext}/c_s^2$\\
        Unit of time\dotfill &  Free-fall time, $t_{\rm ff}=(4\pi G\rho_s)^{-1/2}$ \\ 
        Unit of speed\dotfill & Isothermal sound speed in the filament, $c_s$ \\
        Unit of magnetic field strength \dotfill & $B_u=(4\pi p_{\rm ext})^{1/2}$\\
        Unit of length\dotfill & $L=c_{s}t_{\rm ff}=c_{s}/(4\pi G\rho_s)^{1/2}$\\
    \hline                  
    \end{tabular}
    \end{table*}

In this study, physical variables are normalized by using the following quantities: the external pressure ($p_\mathrm{ext}=p_s$), the filament surface density ($\rho_s$), and the sound speed of the filament ($c_s$).
The scale length $L$ is defined by the free-fall time $t_\mathrm{ff}$ at the filament surface density $\rho_s$ and $c_s$ as $L=c_s t_\mathrm{ff}=c_s/(4\pi G\rho_s)^{1/2}$.
Table \ref{tab:normalization} provides the physical scales characterizing the system.
Normalized variables are defined as follows: $\bm{v}'\equiv \bm{v}/c_s$, $p'\equiv p_\mathrm{gas}/p_\mathrm{ext}$, $\rho' \equiv \rho/\rho_s$, $\Phi'\equiv \Phi/(B_uL)$, $\bm{B}'\equiv \bm{B}/B_u$, $\lambda'\equiv \lambda/(\rho_sL^2)$, $S'\equiv S/c_s^2$, $x' \equiv x/L$, $y' \equiv y/L$, and $z' \equiv z/L$, where the prime denotes the normalized variables.
It is noted that in this study, the magnetic flux $(\Phi)$ is given by the dimension $[\Phi]=[B][L]$ and is treated as the magnetic flux per unit length. 

\subsection{Parameters}\label{sec:parameters}

In this study, we consider three parameters: the magnetic field strength $B_0$, the initial line mass $\lambda_0$, and the initial velocity $V_\mathrm{int}$. The first two parameters define the magnetohydrostatic state of each filament, while $V_\mathrm{int}$ determines the collision condition.

The dimensionless representation of the magnetic field strength $B_0$ is given by $B'_0=B_0/(4\pi p_\mathrm{ext})^{1/2}=\sqrt{2}\beta_0^{-1/2}$, where $\beta_0 = 8\pi p_\mathrm{ext}/B_0^2$ represents the plasma beta of the external gas, located far from the filaments. In this study, we employ $\beta_0$ instead of $B_0$ to specify the field strength. We investigate three plasma beta values: $\beta_0=1$ (weak), $\beta_0=0.1$ (normal), and $\beta_0=0.01$ (strong).

The line mass of single filaments $(\lambda'_0)$ is 
chosen smaller than the critical line mass $(\lambda_\mathrm{crit,B})$, where the dimensionless form of Equation (\ref{eq:magnetized_critical_line_mass}) is represented by
\begin{equation}\label{eq:dimensionless_magnetized_critical_line_mass}
    \lambda'_{\rm crit,B}\simeq 3.04\Phi'+20.8,
\end{equation}
where the non-dimensional magnetic flux 
$\Phi' = R_0'\sqrt{2/\beta_0}$ 
is a function of a radius of the parental filament $R_0'$ and $\beta_0$, that is, $\lambda'_0 \le \lambda_\mathrm{crit,B}'$.
To focus on the role of the magnetic field strength in the magnetized critical line mass, we fix $R_0' = 2$ to minimize the contribution of the radius to the magnetic flux and similar to the fiducial models of \citepaperI{2023ApJ...954..129K}.
This $R_0$ determines the cross-section of the intersection and is therefore important in understanding the filament collision results described later.
The magnetized critical line masses for the three
$\beta_0$ are represented as $\lambda'_{\rm crit,B}(\beta_0=1)=
1.17\lambda'_{\rm crit}$,
$\lambda'_{\rm crit,B}(\beta_0=0.1)=
1.91\lambda'_{\rm crit}$, and 
$\lambda'_{\rm crit,B}(\beta_0=0.01)=
4.25\lambda'_{\rm crit}$,
where 
$\lambda'_\mathrm{crit}$ is the 
non-dimensional form of the 
nonmagnetized critical line mass as $\lambda'_\mathrm{crit}=\lambda_\mathrm{crit}/(\rho_sL^2)=8\pi$.

To summarize, in this study, we consider three plasma beta values: $\beta_0=1$, 0.1, and 0.01. We vary the initial line mass of the single filament $\lambda_0$ within a range spanning from 0.4 to 0.9 times the magnetic critical line mass ($\lambda_\mathrm{crit,B}$).
In addition to this, in $\beta_0=1$, we calculate $\lambda_0=0.34\lambda_\mathrm{crit,B}$ to investigate in detail.  
The initial velocity is taken $V_\mathrm{int}=\pm0.5$, $\pm2.5$,  and $\pm5$ times the speed of sound ($c_s$).
The number of our models is 48 in total as shown in Table \ref{tab:model_parameters}.

\begin{longtable*}[c]{lccccccc} 
\caption{Model Parameters for the Orthogonal Collisions}
\label{tab:model_parameters}  
\\
\hline $\mathrm{Model}^{(1)}$& ${\beta_0}^{(2)}$ & ${\lambda_0/\lambda_{\rm crit}}^{(3)}$ & ${V_{\rm int}/c_s}^{(4)}$  & ${\rho_{c}/\rho_s}^{(5)}$& ${\rho_{\rm lim}/\rho_s}^{(6)}$ &${N^3_{\rm mesh}}^{(7)}$ & $\mathrm{Result}^{(8)}$\\ \hline
\endfirsthead
\hline $\mathrm{Model}^{(1)}$& ${\beta_0}^{(2)}$ & ${\lambda_0/\lambda_{\rm crit}}^{(3)}$ & ${V_{\rm int}/c_s}^{(4)}$  & ${\rho_{c}/\rho_s}^{(5)}$& ${\rho_{\rm lim}/\rho_s}^{(6)}$ &${N^3_\mathrm{mesh}}^{(7)}$ & $\mathrm{Result}^{(8)}$\\ \hline
\endhead
    B1L034V1  & 1& 0.41  & $\pm0.5$ & 2.00 & $6.48\times10^3$& $512^3$ & S\\
    B1L034V5 & 1& 0.41 & $\pm2.5$ & 2.00 & $6.48\times10^3$ & $512^3$ & E\\
    B1L034V10 & 1& 0.41  & $\pm5.0$ & 2.00 & $6.48\times10^3$ & $512^3$ &  E\\
    B1L04V1  & 1& 0.47  & $\pm0.5$ & 2.40 & $6.48\times10^3$& $512^3$ & MC\\
    B1L04V5 & 1& 0.47  & $\pm2.5$ & 2.40 & $6.48\times10^3$ & $512^3$ & E\\
    B1L04V10 & 1& 0.47  & $\pm5.0$ & 2.40 & $6.48\times10^3$ & $512^3$ & E\\
    B01L04V1 & 0.1&  0.76   & $\pm0.5$ & 4.30 & $6.48\times10^3$ & $512^3$ & S\\
    B01L04V5  & 0.1& 0.76  & $\pm2.5$ & 4.30 & $6.48\times10^3$ & $512^3$ & S\\
    B01L04V10  & 0.1& 0.76  & $\pm5.0$ & 4.30 & $6.48\times10^3$ & $512^3$ & S\\
    B001L04V1 & 0.01 & 1.70 & $\pm0.5$ & 21.00 & $6.48\times10^3$ & $512^3$ & S\\
    B001L04V5 & 0.01 & 1.70 & $\pm2.5$ & 21.00 & $6.48\times10^3$ & $512^3$ & S\\
    B001L04V10 & 0.01& 1.70  & $\pm5.0$ & 21.00 & $6.48\times10^3$ & $512^3$ & S\\
    B1L045V1 & 1& 0.53  & $\pm0.5$ & 2.80 & $6.48\times10^3$ & $512^3$ & MC\\
    B1L045V5 & 1& 0.53  & $\pm2.5$ & 2.80 & $6.48\times10^3$ & $512^3$ & E\\
    B1L045V10 & 1& 0.53  & $\pm5.0$ & 2.80 & $6.48\times10^3$ & $512^3$ & E\\
    B01L045V1 & 0.1& 0.86  & $\pm0.5$ & 5.40 & $6.48\times10^3$ & $512^3$ & MC\\
    B01L045V5 & 0.1& 0.86  & $\pm2.5$ & 5.40 & $6.48\times10^3$ & $512^3$ & MC\\
    B01L045V10 & 0.1& 0.86  & $\pm5.0$ & 5.40 & $6.48\times10^3$ & $512^3$ & E\\
    B001L045V1 & 0.01& 1.92  & $\pm0.5$ & 28.10& $6.48\times10^3$ & $512^3$ & MC\\
    B001L045V5 & 0.01& 1.92  & $\pm2.5$ & 28.10 & $6.48\times10^3$ & $512^3$ & MC\\
    B001L045V10 & 0.01& 1.92  & $\pm5.0$ & 28.10 & $6.48\times10^3$ & $512^3$ & MC\\
    B1L05V1 & 1& 0.59  & $\pm0.5$ & 3.30 & $6.48\times10^3$ & $512^3$ & MC\\
    B1L05V5 & 1& 0.59  & $\pm2.5$ & 3.30 & $6.48\times10^3$ & $512^3$ & MC\\
    B1L05V10 & 1& 0.59  & $\pm5.0$ & 3.30 & $6.48\times10^3$ & $512^3$ & E\\
    B01L05V1 & 0.1& 0.96  & $\pm0.5$ & 6.80 & $6.48\times10^3$ & $512^3$ & RC\\
    B01L05V5 & 0.1& 0.96  & $\pm2.5$ & 6.80 & $6.48\times10^3$ & $512^3$ & MC\\
    B01L05V10 & 0.1& 0.96  & $\pm5.0$ & 6.80 & $6.48\times10^3$ & $512^3$ & E\\
    B001L05V1 & 0.01& 2.13  & $\pm0.5$ & 37.00 & $6.48\times10^3$ & $512^3$ & MC\\
    B001L05V5 & 0.01& 2.13  & $\pm2.5$ & 37.00 & $6.48\times10^3$ & $512^3$ & MC\\
    B001L05V10 & 0.01& 2.13  & $\pm5.0$ & 37.00 & $6.48\times10^3$ & $512^3$ & MC\\
    B1L075V1 & 1& 0.88  & $\pm0.5$ & 11.01 & $6.48\times10^3$ & $512^3$ & RC\\
    B1L075V5 & 1& 0.88  & $\pm2.5$ & 11.01 & $6.48\times10^3$ & $512^3$ & MC\\
    B1L075V10 & 1& 0.88  & $\pm5.0$ & 11.01 & $6.48\times10^3$ & $512^3$ & MC\\
    B01L075V1 & 0.1& 1.44  & $\pm0.5$ & 24.50 & $6.48\times10^3$ & $512^3$ & RC\\
    B01L075V5 & 0.1& 1.44  & $\pm2.5$ & 24.50 & $6.48\times10^3$ & $512^3$ & MC\\
    B01L075V10 & 0.1& 1.44  & $\pm5.0$ & 24.50 & $6.48\times10^3$ & $512^3$ & MC\\
    B001L075V1 & 0.01& 3.19  & $\pm0.5$ & 139.91 & $1.65\times10^6$ & $512^3$(AMR) & MC\\
    B001L075V5 & 0.01& 3.19  & $\pm2.5$ & 139.91 & $1.65\times10^6$ & $512^3$(AMR) & MC\\
    B001L075V10 & 0.01& 3.19  & $\pm5.0$ & 139.91 & $1.65\times10^6$ & $512^3$(AMR) & MC\\
    B1L09V1 & 1& 1.06  & $\pm0.5$ & 46.95 & $6.48\times10^3$ & $512^3$ & RC\\
    B1L09V5 & 1& 1.06  & $\pm2.5$ & 46.95 & $6.48\times10^3$ & $512^3$ & RC\\
    B1L09V10 & 1& 1.06  & $\pm5.0$ & 46.95 & $6.48\times10^3$ & $512^3$ & MC\\
    B01L09V1 & 0.1& 1.72  & $\pm0.5$ & 81.01 & $6.48\times10^3$ & $512^3$ & RC\\
    B01L09V5 & 0.1& 1.72  & $\pm2.5$ & 81.01 & $6.48\times10^3$ & $512^3$ & MC\\
    B01L09V10 & 0.1& 1.72  & $\pm5.0$ & 81.01 & $6.48\times10^3$ & $512^3$ & MC\\
    B001L09V1 & 0.01& 3.83  & $\pm0.5$ & 406.24 & $1.65\times10^6$ & $512^3$(AMR) & MC\\
    B001L09V5 & 0.01& 3.83  & $\pm2.5$ & 406.24 & $1.65\times10^6$ & $512^3$(AMR) & MC\\
    B001L09V10 & 0.01& 3.83  & $\pm5.0$ & 406.24 & $1.65\times10^6$ & $512^3$(AMR) & MC\\
    \hline
\end{longtable*}
\begin{threeparttable}
\begin{tablenotes}
\item {\bf Notes.} 
The columns in the table are as follows: (1) Name of the model, (2) plasma beta value,  (3) the line mass of single filament normalized by $\lambda_\mathrm{crit}$, (4) initial velocity, (5) the initial central density of the filament, (6) maximum density due to numerical restriction, (7) the total number of grid points, and (8) dynamical state of the merged filament: S= stable, E= expansion, MC= moderate collapse, and RC= rapid collapse.
In all models, the size of the numerical box is set to $L'_\mathrm{box}=10$.
Some models with $N^3_\mathrm{mesh}$ written as $512^3$(AMR) used Adaptive Mesh Refinement, to ensure sufficient density range. 
\end{tablenotes}
\end{threeparttable}

\subsection{Numerical Methods}

To solve the basic equations including the initial condition under given parameters, we 
employ Athena++ \citep{2020ApJS..249....4S} with the following setup: we utilize a 
third-order Runge--Kutta time integration scheme \citep{2009JSCom.38..251G} 
for solving the magnetohydrodynamic equations.
The numerical fluxes are calculated
using the Harten-Lax-van Leer discontinuities (HLLD) method \citep{2005JCoPh.208..315M}. 
Spatial construction involves employing the piecewise linear method for the characteristic variables. 
To maintain the $\nabla \cdot \bm{B}=0$ condition, we use the constrained transport method \citep{1988ApJ...332..659E,2008JCoPh.227.4123G}. 
The multi-grid algorithm is utilized for solving Poisson's equation (Equation (\ref{eq:poisson's_equation})), implemented in Athena++ by \citet{2023ApJS..266....7T}. 

For boundary conditions, the magnetohydrodynamic variables are set with outflow boundary conditions in all directions, while the gravitational potential is assumed to have periodic boundary conditions.

The computational domain is taken as $-5 \le x', y', z' \le 5$, 
i.e., the box size is $L'_\mathrm{box}=10$.
To minimize the influence of the boundary conditions, 
$L'_\mathrm{box}$ are set sufficiently larger than the filament width $\simeq2R_0'$. 
As is mentioned in Equation (\ref{eq:dimensionless_magnetized_critical_line_mass}), 
we assume here $R'_0=2$ (see Section \ref{sec:parameters}). 
Since the half-width of the magnetohydrostatic filaments is approximately equal to $R'_0$, the
initial filaments are resolved with a sufficient spatial resolution to maintain the mechanical balance, using a density and vector potential distribution with $\Delta x'=L'_\mathrm{box}/N_\mathrm{mesh}\simeq 1.95\times 10^{-2}(L'_\mathrm{box}/10)(N_\mathrm{mesh}/512)^{-1}$. 
Then, these filaments are arranged for orthogonal collision as described in Section \ref{sec:initial_condition}.

Additionally, the Jeans criterion \citep{1997ApJ...489L.179T} is employed to prevent unexpected numerical fragmentation. 
The calculations are terminated when the relationship between the Jeans length $L_{\rm J}=\sqrt{\pi c_s^2/(G\rho)}$ and the grid spacing $\Delta x$ breaks
the condition $4\Delta x\leq L_{\rm J}$.
Therefore, the maximum density due to numerical restriction is determined by the Jeans criterion as $\rho'_\mathrm{lim}\simeq 6.48\times 10^3(L'_\mathrm{box}/10)^{-2}(N_\mathrm{mesh}/512)^2$.
In some models, we employ the adaptive mesh refinement (AMR) technique \citep{1989JCoPh..82...64B
}, which is implemented 
in Athena++ by \citet{2023ApJS..266....7T}, to address issues arising from insufficient resolution during the evolution and proceeding with the collapse.
The refinement condition is defined such that a finer level is created when the grid size is $16\Delta x > L_{\rm J}$, with up to 5 levels (including the root level and 4 finer levels).
Therefore, calculations are conducted to the grid spacing equivalent to that of $N_\mathrm{mesh}=8192$ ($\Delta x'=L'_\mathrm{box}/N_\mathrm{mesh}=10/8192\simeq 0.001$), and $\rho'_\mathrm{lim}\simeq 1.65\times 10^6$.
In the AMR models, the initial uniform grid structure evolves and finer grids are generated to cover the dense region of the intersection.  
The grid structure is similar to that expected for the nested grid (or a static version of the AMR) in which a connected finer-grid domain is surrounded by a connected coarser-grid domain \citep{1998ApJ...502L.163T,2004MNRAS.348L...1M}.
The calculation settings for each model are listed in Table  \ref{tab:model_parameters}.

Hereafter, we omit $'$ in normalization variables unless the quantity is misinterpreted as a dimensional variable.

\section{Results}\label{sec:result}
Evolution of the orthogonal collisions is classified into three distinct modes: a collapse mode characterized by a monotonic increase in the intersection's density, a stable mode indicating oscillations and stability without collapse, and an expansion mode where the intersection begins to expand.
We begin with the typical evolution of the collapse mode.

\begin{figure*}
    \centering
     \begin{tabular}{ccc}
         \subfigure[]{
         \includegraphics[keepaspectratio,scale=0.25]{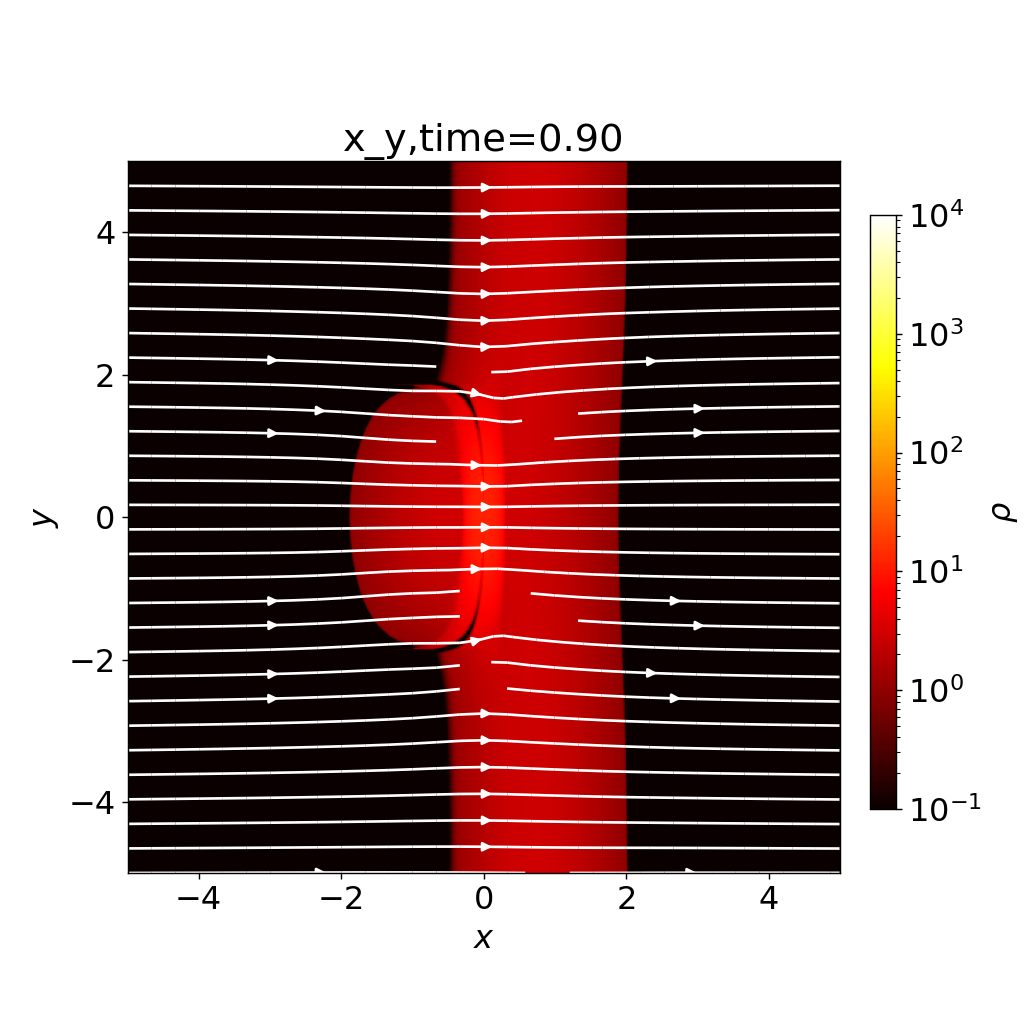}  
         }&
         \subfigure[]{
         \includegraphics[keepaspectratio,scale=0.25]{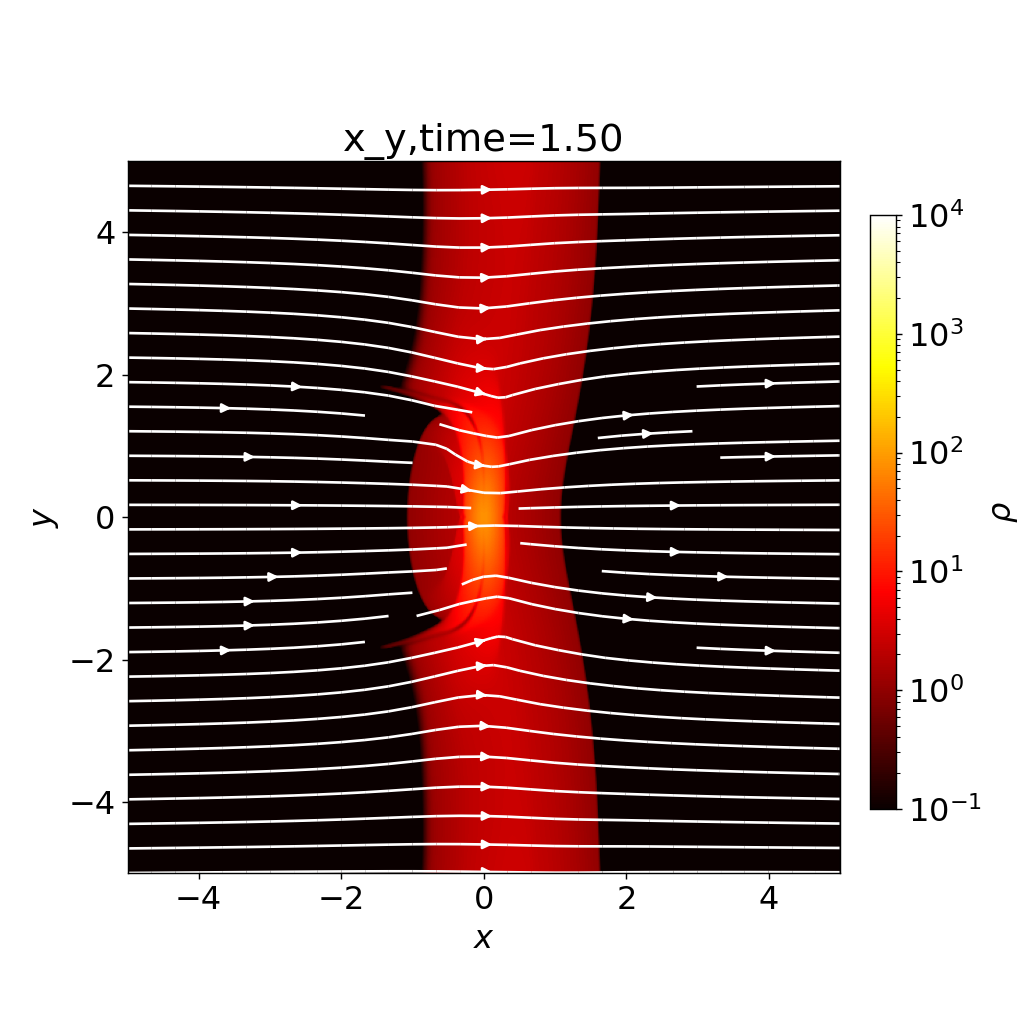}
         }&
         \subfigure[]{
         \includegraphics[keepaspectratio,scale=0.25]{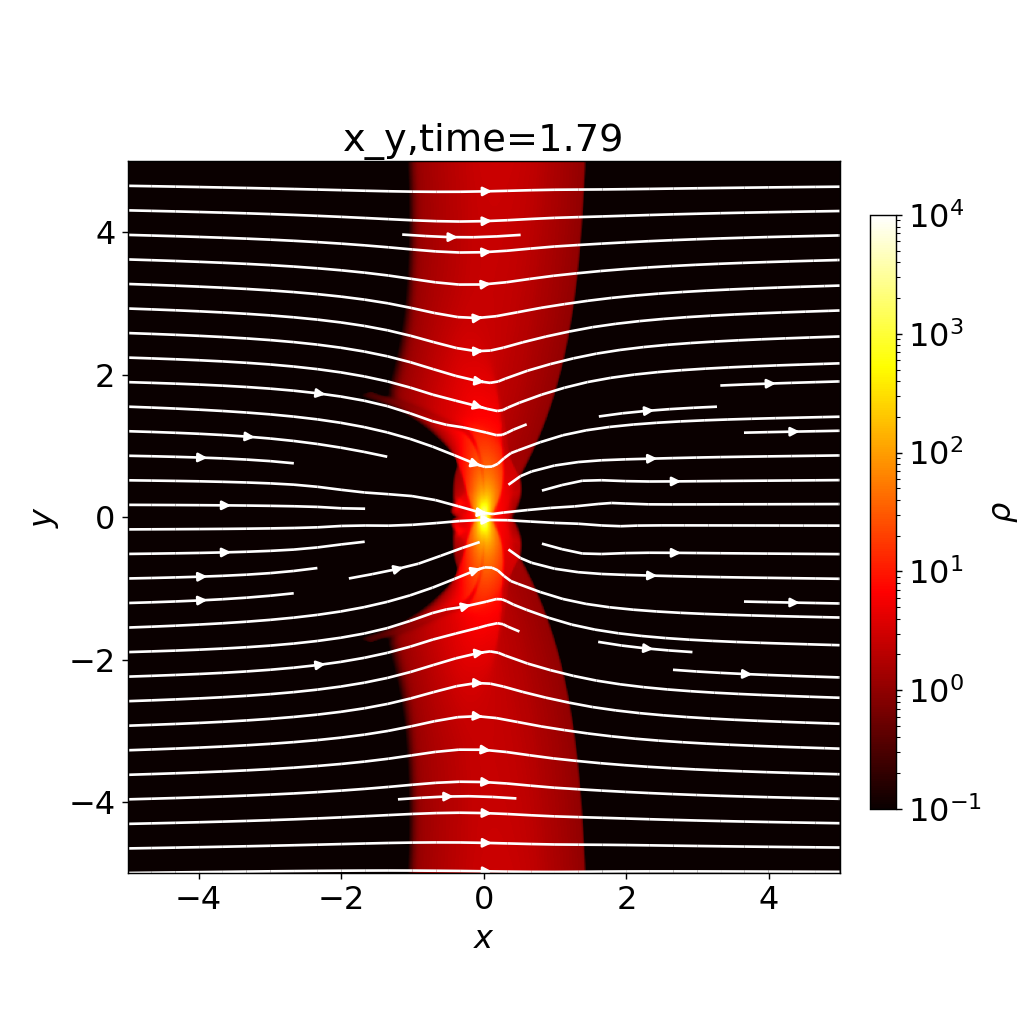} 
         }\\
         \subfigure[]{
         \includegraphics[keepaspectratio,scale=0.25]{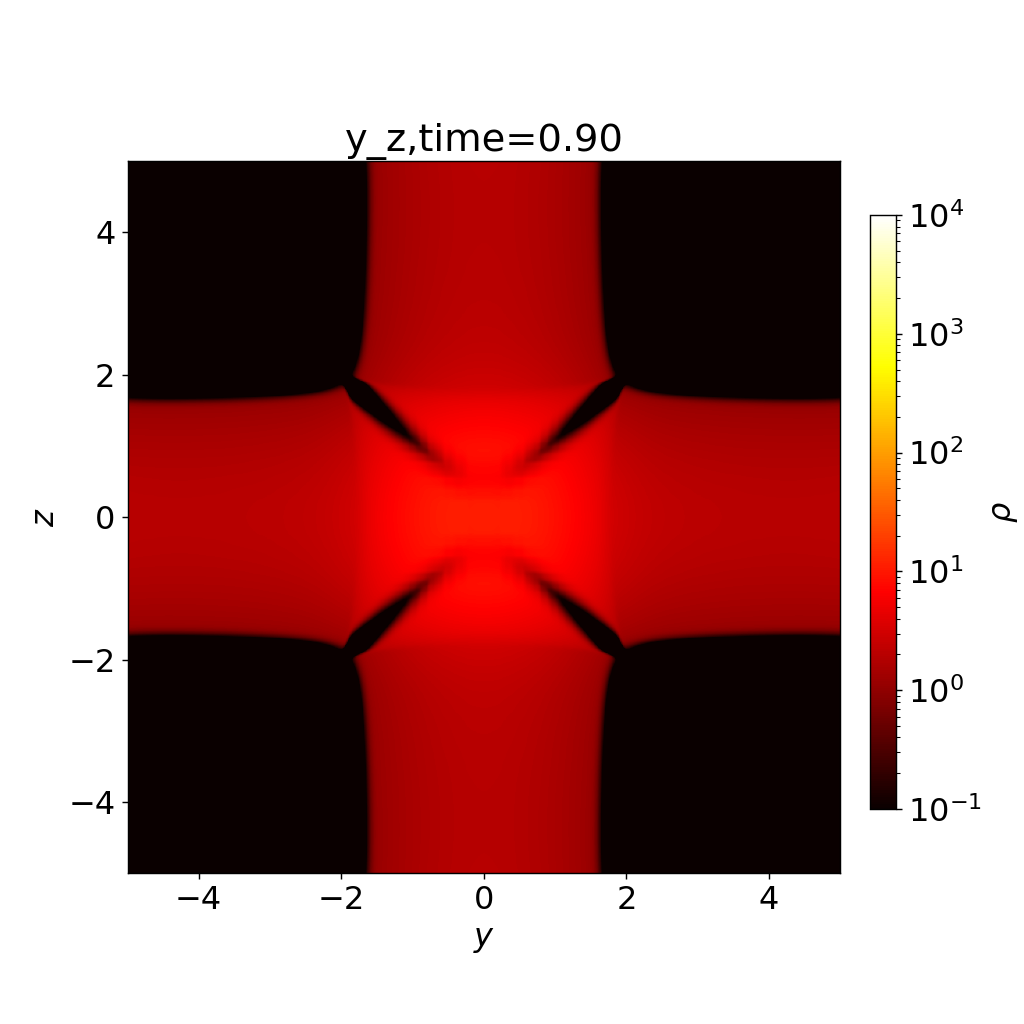}
         }&
         \subfigure[]{
         \includegraphics[keepaspectratio,scale=0.25]{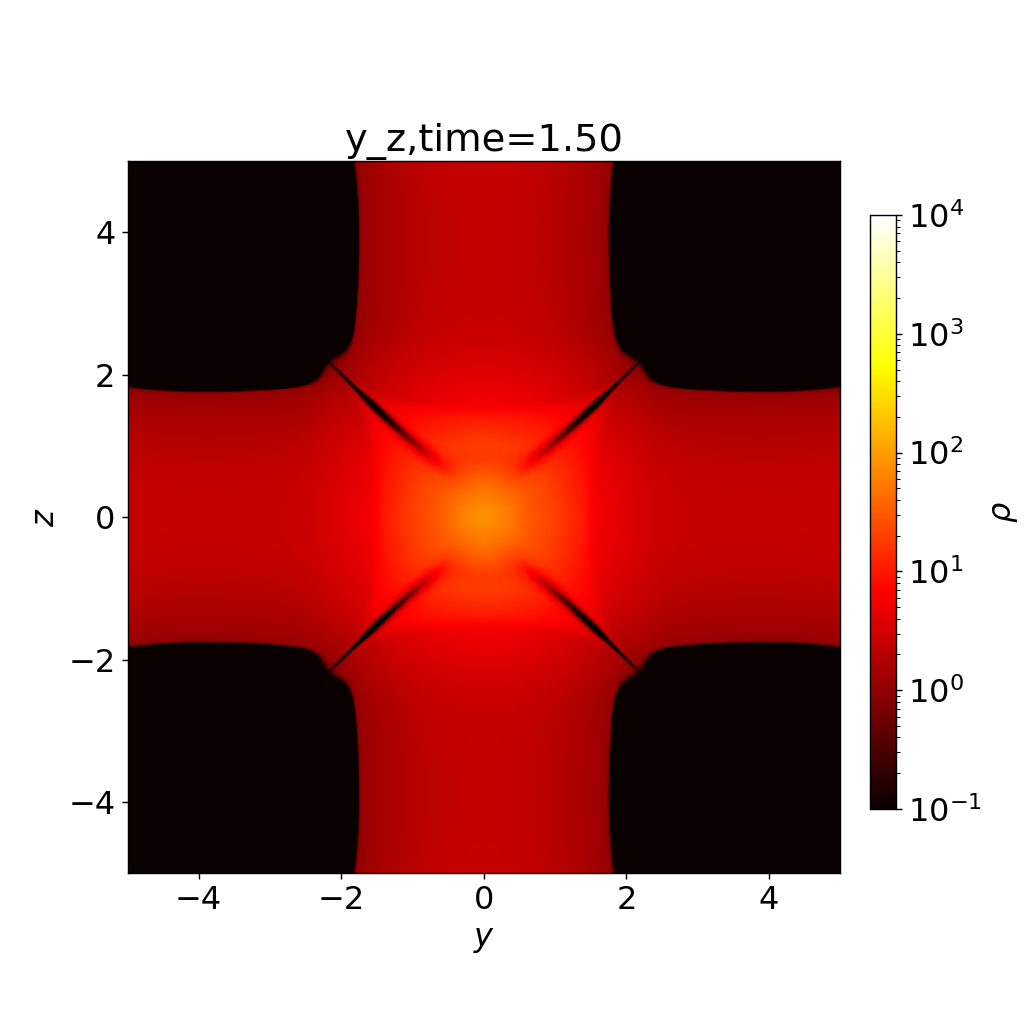}
         }&
         \subfigure[]{
         \includegraphics[keepaspectratio,scale=0.25]{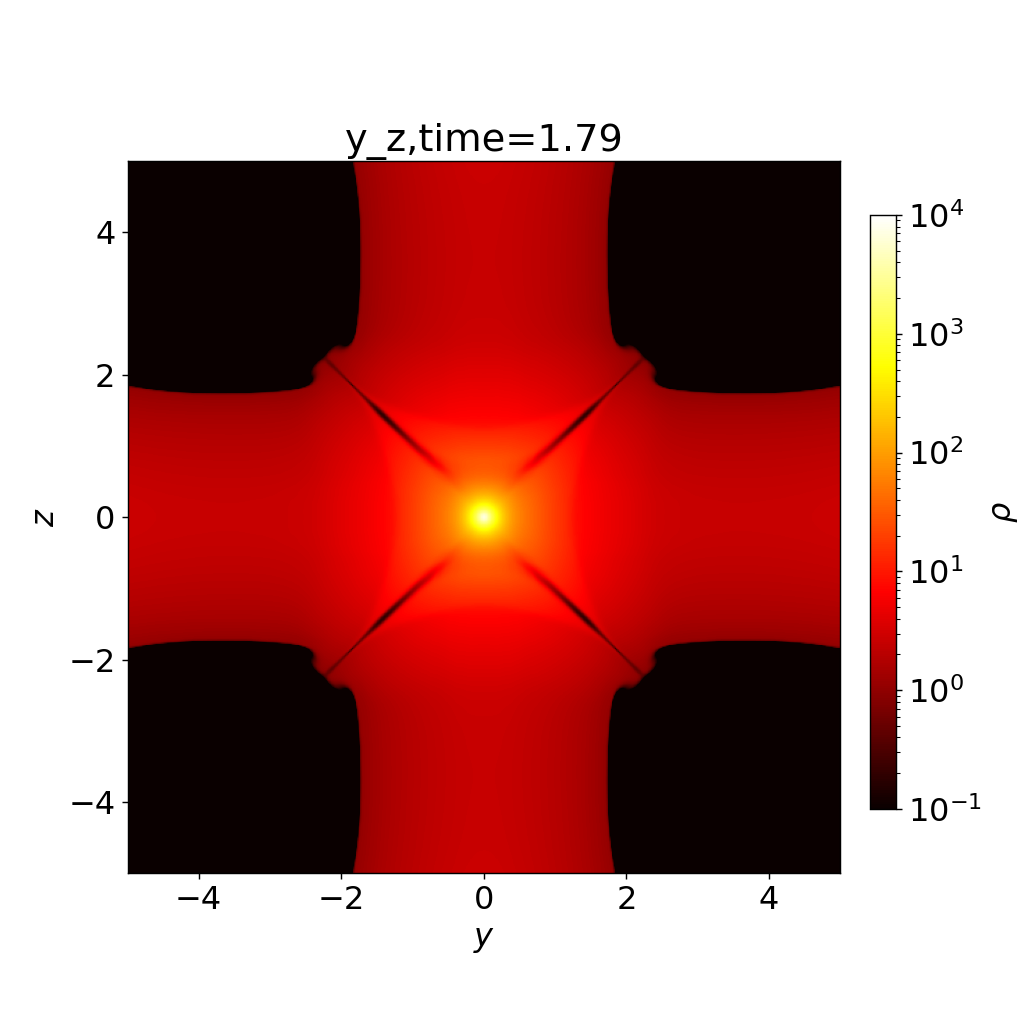}
         }

     \end{tabular}
      \caption{
      Two-dimensional slice of the collapse model (B1L05V1).  
      The rows correspond to the slices at $z=0$ and $x=0$ planes from top to bottom.
      The columns correspond to the three representative epochs $t=0.90$, $1.50$, and $1.79$ from left to right.
      The color scale corresponds to the density, and the white lines depict magnetic field lines.
      As we mentioned in Section \ref{sec:initial_condition}, due to symmetry, we omit the two-dimensional slice of $x-z$ plane.
      }
     \label{fig:3d_L05B1_collapse}
\end{figure*} 

\begin{figure*}
    \centering
     \begin{tabular}{cc}
         \subfigure[]{
         \includegraphics[keepaspectratio,scale=0.4]{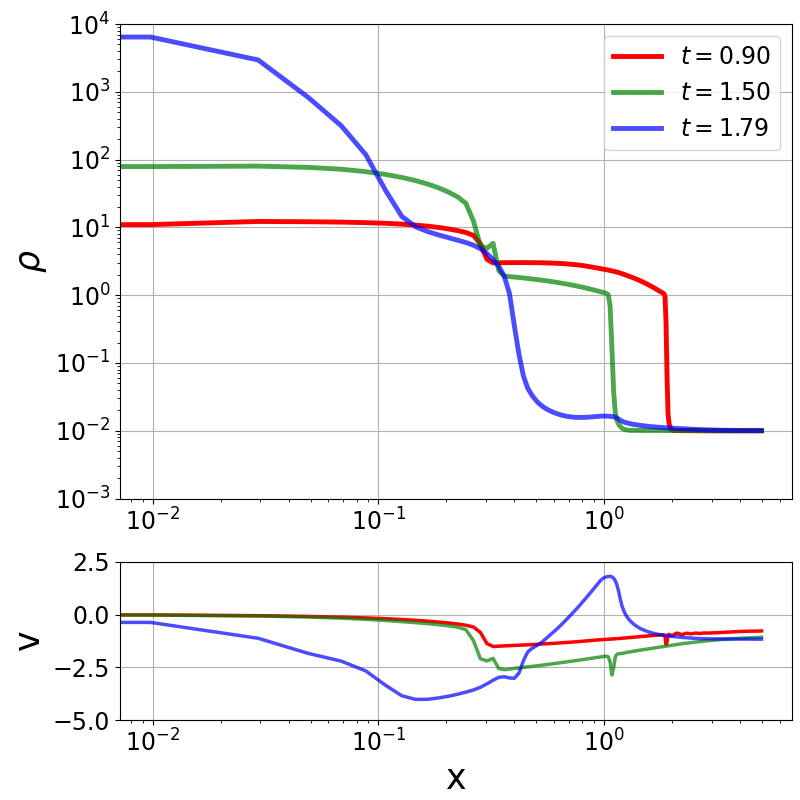}  
         }&
         \subfigure[]{
         \includegraphics[keepaspectratio,scale=0.4]{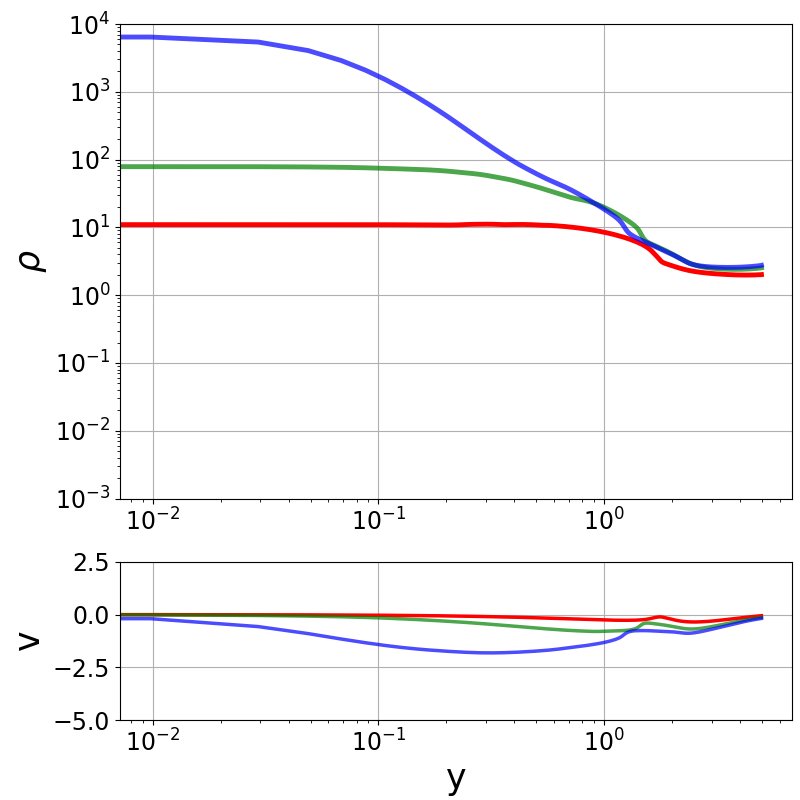}
         }
     \end{tabular}
      \caption{
      The one-dimensional profiles on the  $x$-(left) and  $y$-(right) axes for the collapse model (B1L05V1) at three different epochs: $t=0.90$ (red), $1.50$ (green), and $1.79$ (blue).
      The density and velocity profiles on the $x$- and $y$-axes are shown in (a) and (b), respectively,
      in which the lines in the upper panels represent 
      the density distributions ($\rho(x)$ in (a) and $\rho(y)$ in (b))
      whereas the lines in the lower panels represent the velocity profiles ($v_x$ in (a) and $v_y$ in (b)).
      }
     \label{fig:1d_L05B1_collapse}
\end{figure*}

\subsection{Collapse Mode}\label{sec:collapse_mode}
The model with $\beta_0=1,\lambda_0=0.59 \lambda_\mathrm{crit}$, and $V_\mathrm{int}=\pm0.5$ is selected as the fiducial model for the collapse mode and corresponds to B1L05V1 in Table \ref{tab:model_parameters}.

Figure \ref{fig:3d_L05B1_collapse} shows the density color maps on the two orthogonal planes passing through the origin at three different epochs.
Figures \ref{fig:1d_L05B1_collapse}(a) and (b) 
show the density and the velocity profiles on the $x$- and $y$-axes, respectively. 
Since due to the symmetry, these one-dimensional distributions on the $y$- and $z$-axes are the same, we omit the distribution on the $z$-axis.

Figures \ref{fig:3d_L05B1_collapse}(a) and (d) represent the density distributions at $t=0.90$ that correspond to the early stage of the collision.  
A shocked cloud is seen at the intersection of the two filaments near the center of each panel (the shocked cloud with a thickness of $\simeq 0.6$ and a width of $\simeq 4$).
The structure is very narrow in the $x$-direction, 
whereas it extends as wide as the width of the filaments in the $y$- and $z$-directions.
This structure can also be observed in Figure \ref{fig:1d_L05B1_collapse}. 
In the upper panel of Figure \ref{fig:1d_L05B1_collapse}(a), a shock front is identified as the density jump at $x\sim 0.3$.  
In the upper panel of Figure \ref{fig:1d_L05B1_collapse}(b), 
there are no such shock-like structures. 
The lower panel of (a) shows that the velocity along the $x$-axis suddenly decelerates around $x\sim 0.3$ where the density jump exists, while there are no prominent shock-like structures along the $y$-axis at $t=0.90$.
Hence, the structure of the shocked region at the intersection resembles a sheet-like structure that is flattened along the $x$-axis.

In Figures \ref{fig:3d_L05B1_collapse}(b) and (e), the density distributions at $t=1.50$ indicate that 
the shocked region becomes denser, and the magnetic field lines are dragged towards the center.
In the lower panel of Figure \ref{fig:1d_L05B1_collapse}(a), the velocity distribution along the $x$-axis shows that the infall speed increases to $|v_x|\simeq2.5$, suggesting that it is accelerated by self-gravity compared to the initial velocity.
The velocity distributions along the $y$-axis shown in the lower panel of Figure \ref{fig:1d_L05B1_collapse}(b) exhibit velocity components directed towards the center, originating from the boundary of the filament at $y=5$, indicating the presence of infalling flows along the axis of the filament.

In Figures \ref{fig:3d_L05B1_collapse}(c) and (f), at $t=1.79$, the shocked region collapses gravitationally to form a centrally concentrated density structure. 
Figure  \ref{fig:3d_L05B1_collapse}(c) shows that 
the two filaments have completed collision and the magnetic field lines are dragged towards the center made by gravitational collapse.
In Figure \ref{fig:3d_L05B1_collapse}(f),  
the structure of the shocked region in the $y$-$z$ plane appears to be circular.
The density profiles at $t=1.79$ are also plotted in Figures \ref{fig:1d_L05B1_collapse}(a) and (b). 
Especially, the density structure along the $x$-axis, appears to be narrower than those observed along the $y$-axis, indicating that the collapsing shocked cloud has an oblate shape.
In the lower panel of Figure \ref{fig:1d_L05B1_collapse}(a), the velocity structure along the $x$-axis shows a clear infall motion, and the infall speed has a maximum value of $|v_x|\simeq 4.5$ around $x\sim 0.15$.
Then, the outward shock wave ($x\simeq 1$) has passed through the filament and is propagating into the ambient medium, indicating that the collision of filaments is completed.
Therefore, this model corresponds precisely to a moderate collapse (details are shown in Section \ref{sec:rapid_collapse}).
In addition to this, in the lower panel of Figure \ref{fig:1d_L05B1_collapse}(b), a global gravitational 
collapse is also seen along the $y$-axis towards the center at a maximum infall speed of $|v_y|\simeq 2$.
The flows along the $y$-axis are commonly referred to as the flow along the filament.

\begin{figure}
    \centering
    \includegraphics[keepaspectratio,scale=0.4]{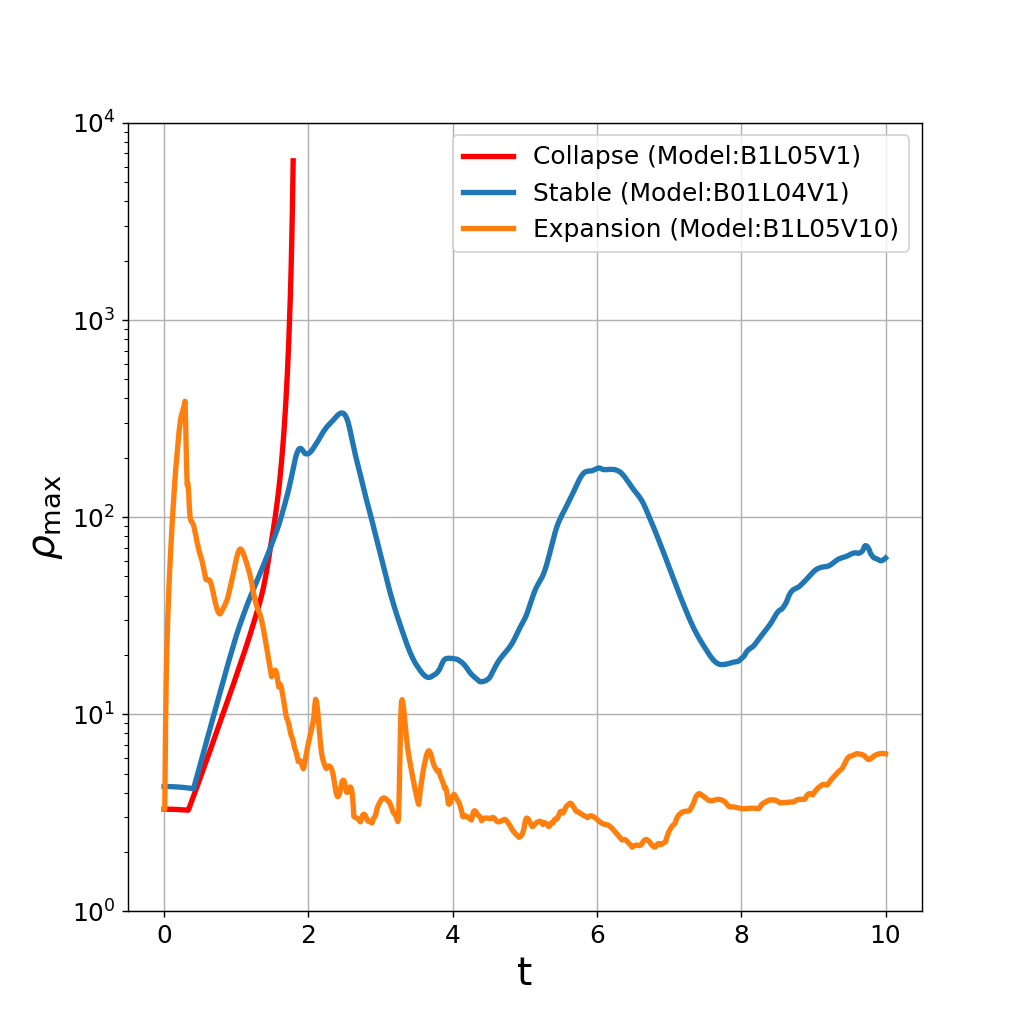}
      \caption{
      The maximum density($\rho_\mathrm{max}$) evolution as a function of the elapsed time. 
      The red, blue, and orange lines represent the collapse (B1L05V1), stable (B01L04V1), and expansion (B1L05V10) modes, respectively.
      }
    \label{fig:maxden_for_3d}
\end{figure}

Figure \ref{fig:maxden_for_3d} represents the time evolutions of the maximum density.
In Figure \ref{fig:maxden_for_3d}, the maximum density of the collapse mode exhibits a continuous increase over time, and the contraction never stops.

Consequently, in the collapse mode, the structure of the shocked cloud evolves from a sheet-like structure to an oblate shape that is flattened along the global magnetic field lines, resulting in a state of runaway collapse.

\subsection{Stable Mode}\label{sec:stable_mode}

\begin{figure*}
    \centering
     \begin{tabular}{ccc}
         \subfigure[]{
         \includegraphics[keepaspectratio,scale=0.25]{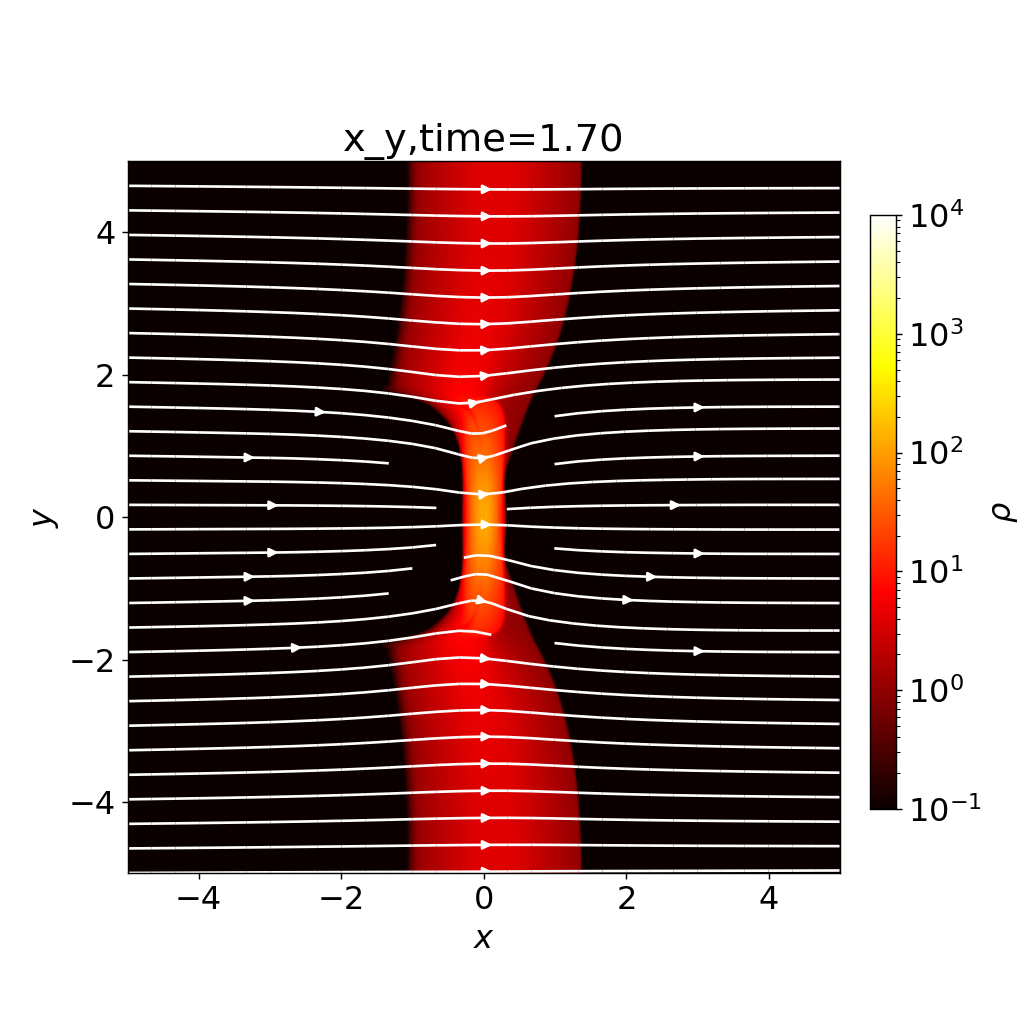} }& 
         \subfigure[]{
         \includegraphics[keepaspectratio,scale=0.25]{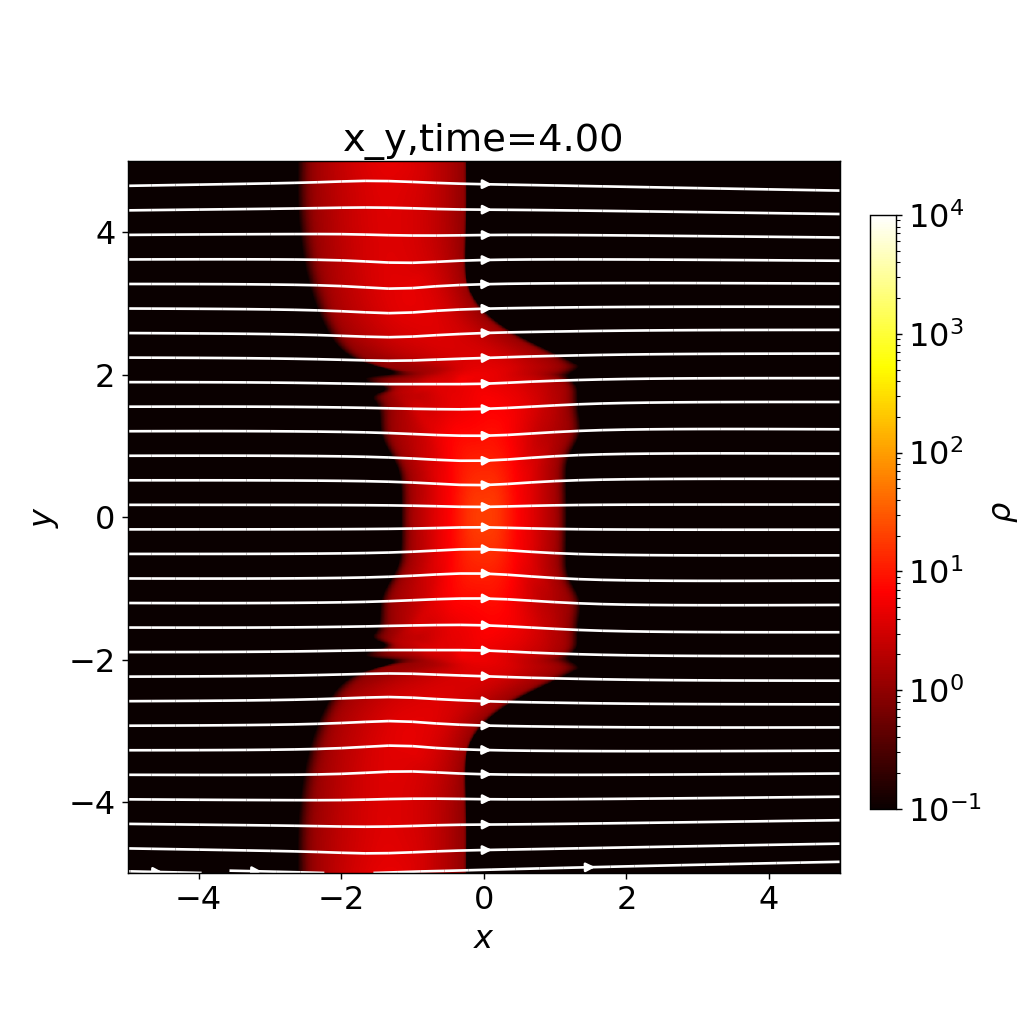}}&
         \subfigure[]{
         \includegraphics[keepaspectratio,scale=0.25]{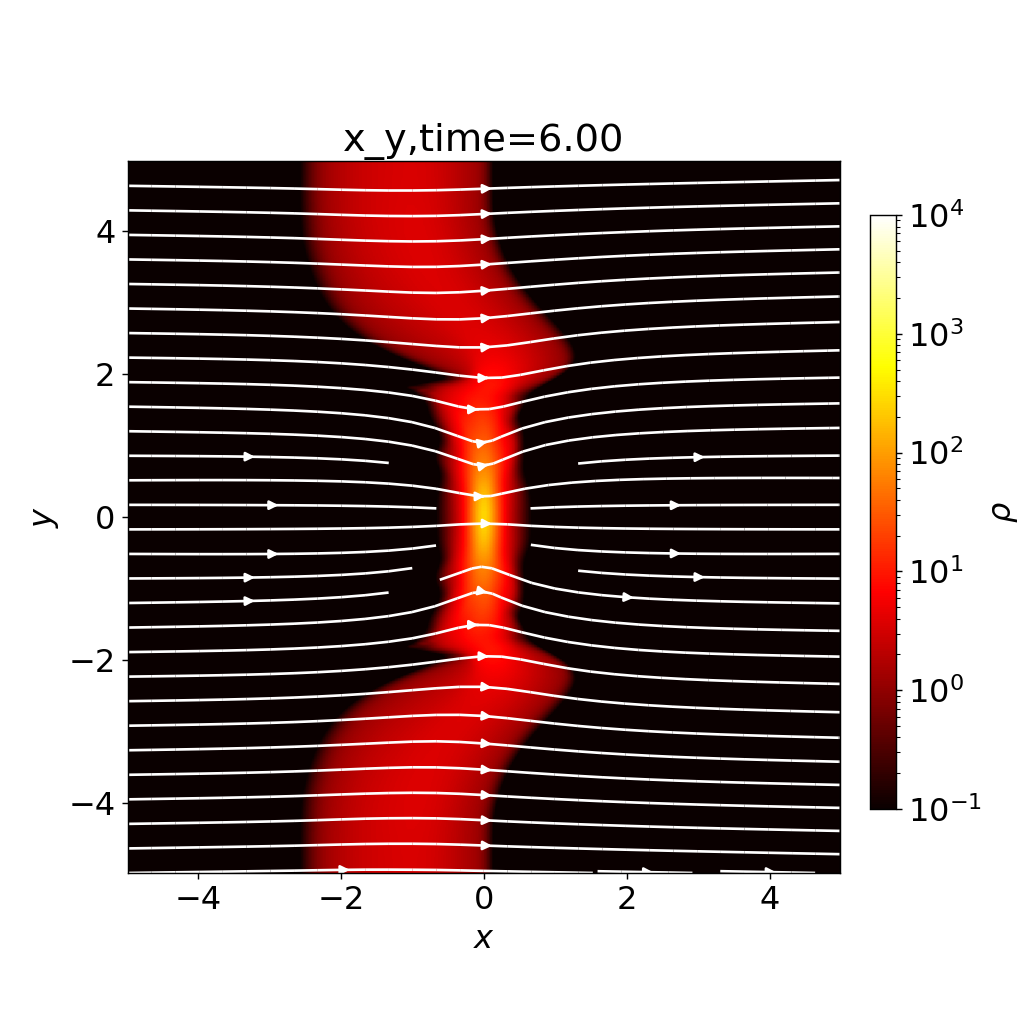}}\\ 
         \subfigure[]{
         \includegraphics[keepaspectratio,scale=0.25]{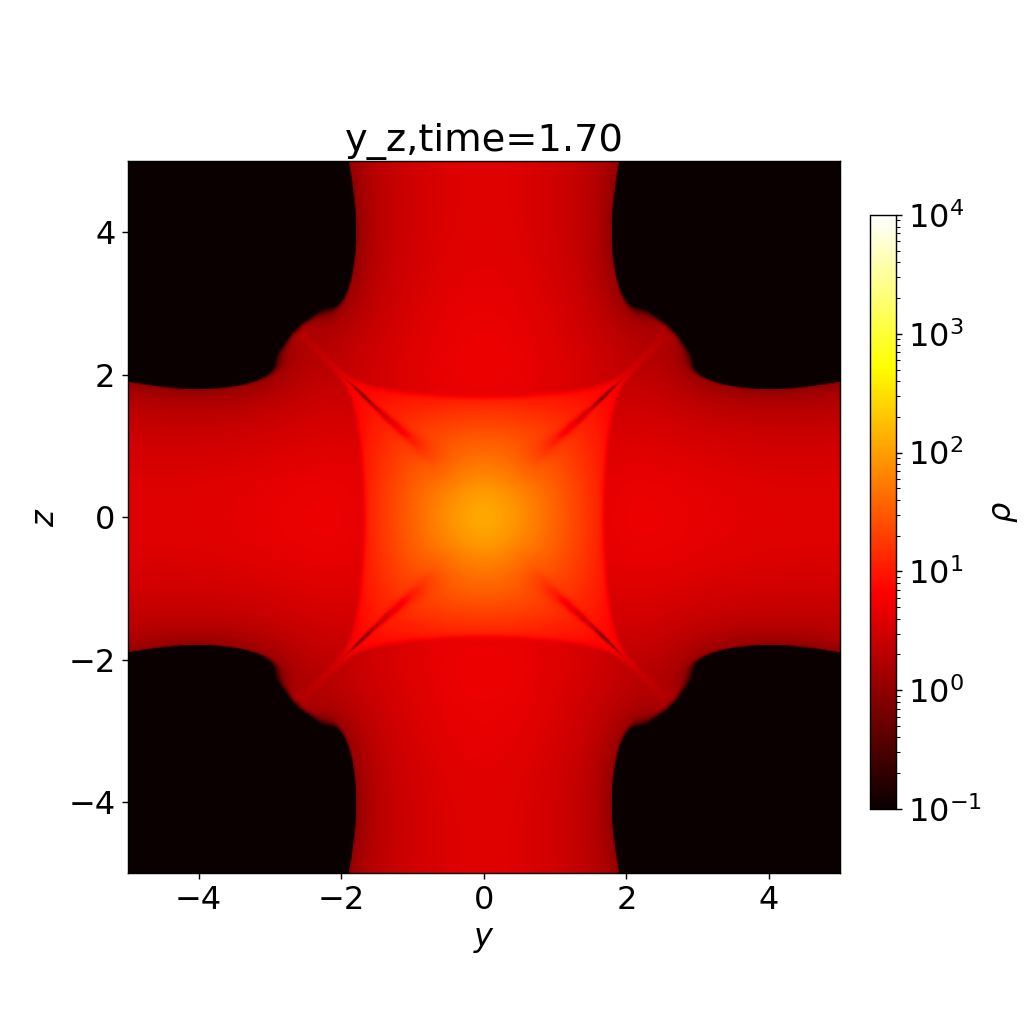} }& 
         \subfigure[]{
         \includegraphics[keepaspectratio,scale=0.25]{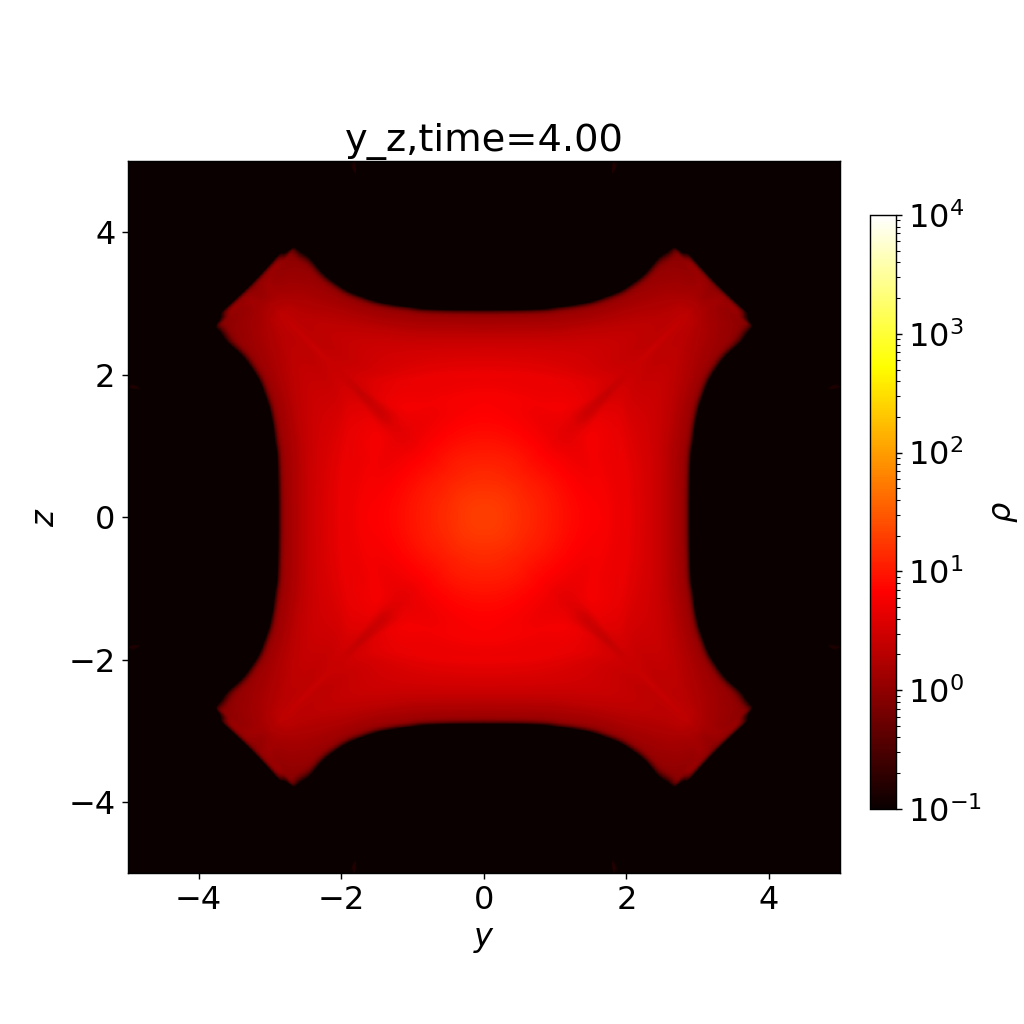}}&
         \subfigure[]{
         \includegraphics[keepaspectratio,scale=0.25]{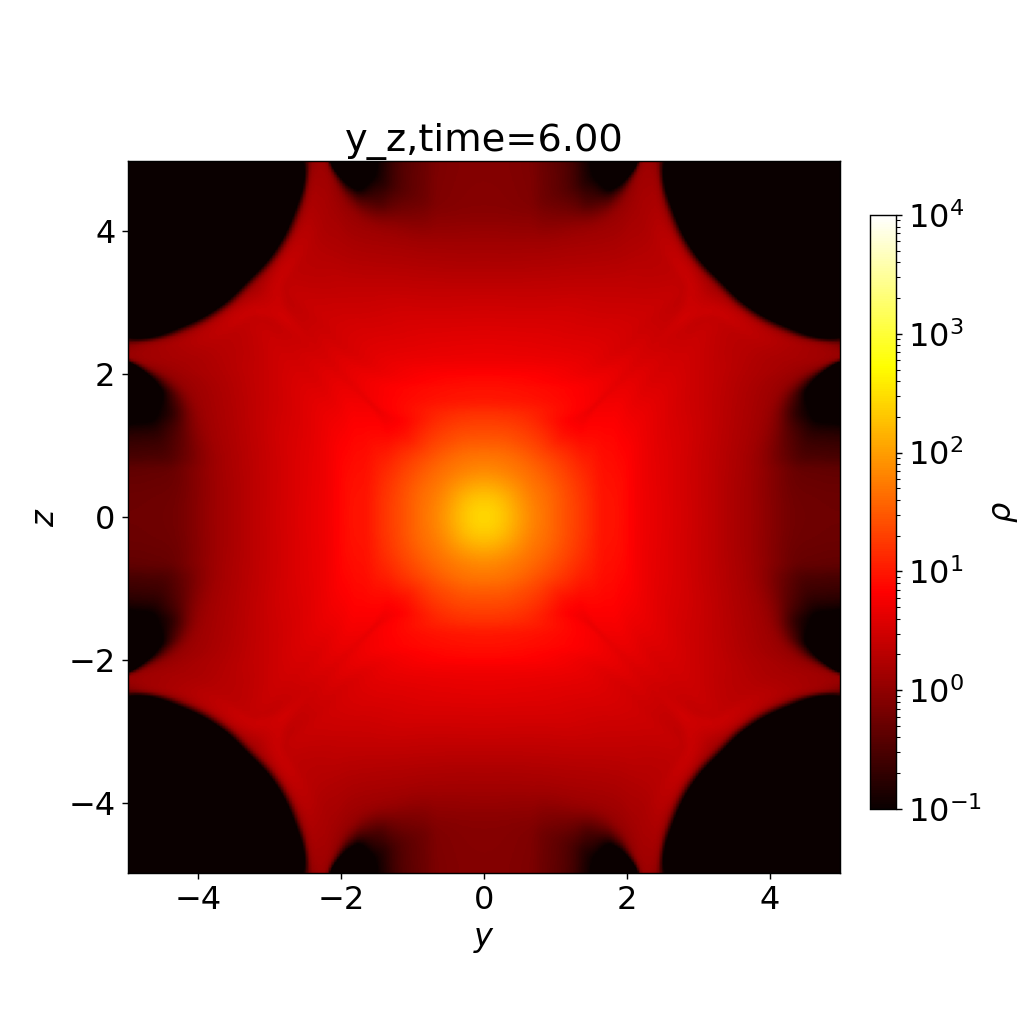}}          
     \end{tabular}
      \caption{
      Same as Figure \ref{fig:3d_L05B1_collapse}, but for the stable model (B01L04V1). 
      The columns correspond to respective times $t=1.70$, 4.00, and 6.00 from 
      left to right.
      }
     \label{fig:3d_L04B01_stable}
     \end{figure*}

In this section, we describe the typical evolution of the stable mode with $\beta_0=0.1,\lambda_0=0.76 \lambda_\mathrm{crit}$ and $V_\mathrm{int}=\pm 0.5$, corresponding to B01L04V1 in Table \ref{tab:model_parameters}.

The evolution of maximum density $\rho_\mathrm{max}(t)$ of this stable model is also shown in Figure \ref{fig:maxden_for_3d}.
The central density of the stable mode clearly shows an oscillation with a finite amplitude.
Figure \ref{fig:3d_L04B01_stable} shows the evolution of the density and magnetic field distributions for the stable model B01L04V1.

In Figures \ref{fig:3d_L04B01_stable}(a) and (d), 
at time $t=1.70$ the sheet-like shocked cloud formed at the intersection, as seen in the collapse mode (see Section \ref{sec:collapse_mode}).

The epoch of $t=4.00$ corresponds to one of the most expanded phases shown in Figure \ref{fig:maxden_for_3d}.
Comparing the left and middle panels of Figure \ref{fig:3d_L04B01_stable} shows that the shocked region expands in the evolution from $t=1.70$ to $t=4.00$, as evidenced by its increased width in the $x$-$y$ plane.

The epoch of $t=6.00$ corresponds to one of the most contracted phases shown in Figure \ref{fig:maxden_for_3d}.
In Figures \ref{fig:3d_L04B01_stable}(c) and (f), at $t=6.00$, the shocked cloud undergoes a contraction phase once more, showing the reformation of a dense region at the intersection, similar to what was observed at $t=1.70$. 
Especially, in Figures \ref{fig:3d_L04B01_stable}(f), the structure of the shocked region in the $y$-$z$ plane appears to more clearly resemble a circular shape compared to its structure at $t=1.70$.
Thus the shocked region evolves into the oblate shape. 

As a result, the structure of the shocked region in the stable mode does not collapse but oscillates with time instead.

\begin{figure*}
    \centering
     \begin{tabular}{ccc}
         \subfigure[]{
         \includegraphics[keepaspectratio,scale=0.25]{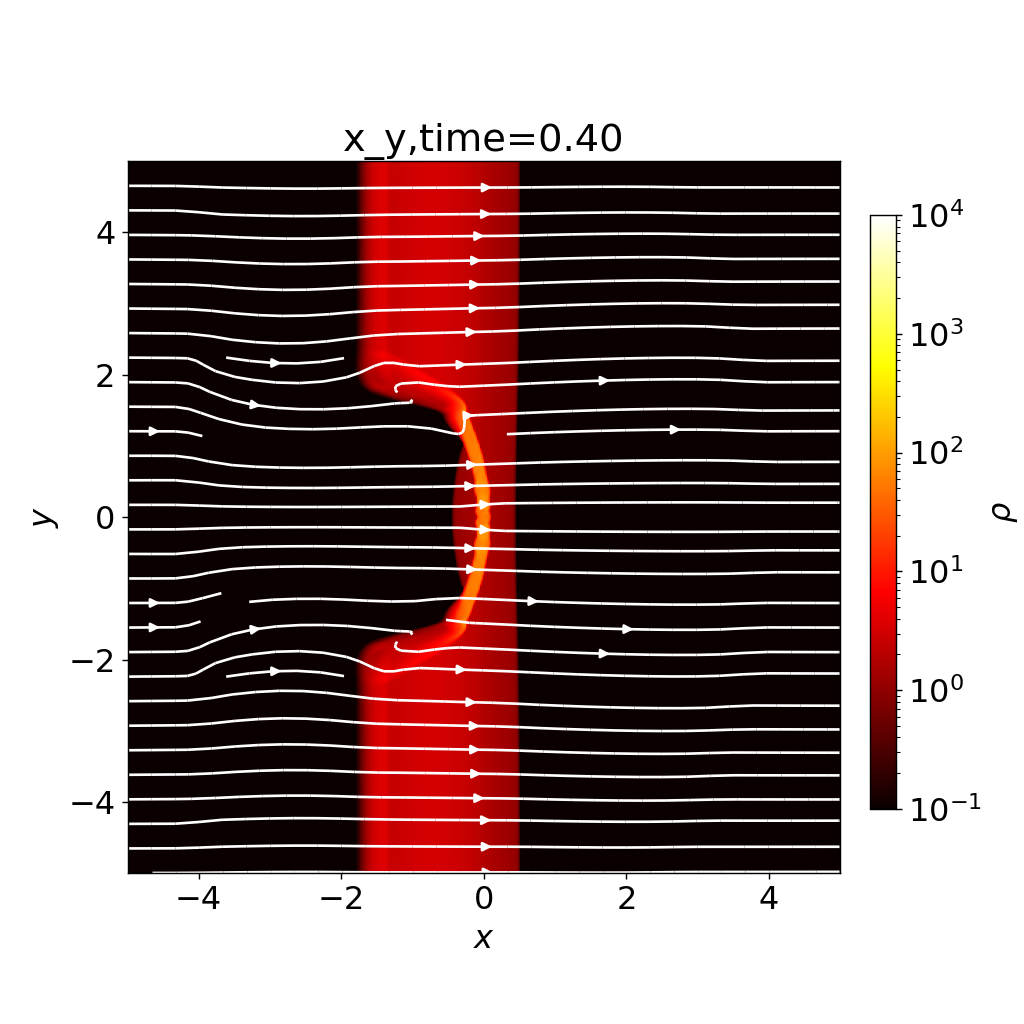} }& 
         \subfigure[]{
         \includegraphics[keepaspectratio,scale=0.25]{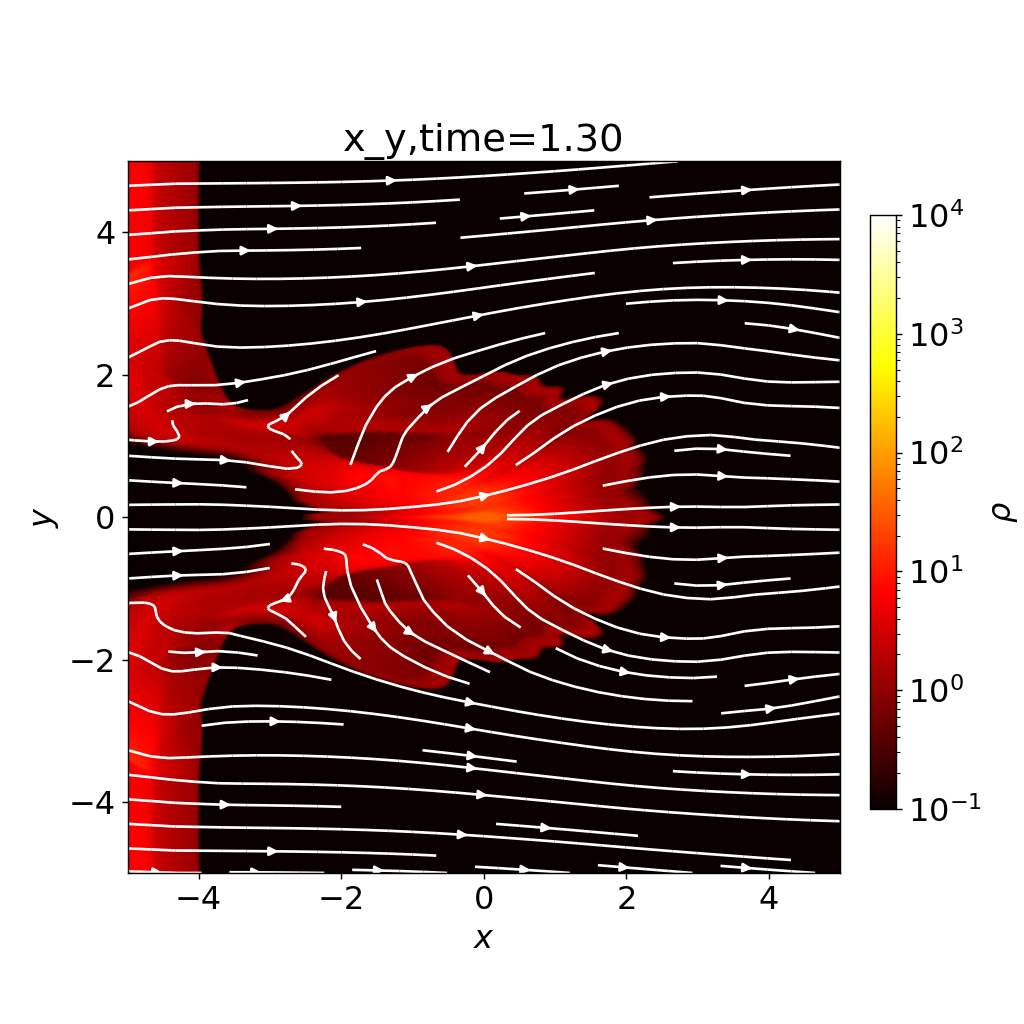}}&
         \subfigure[]{
         \includegraphics[keepaspectratio,scale=0.25]{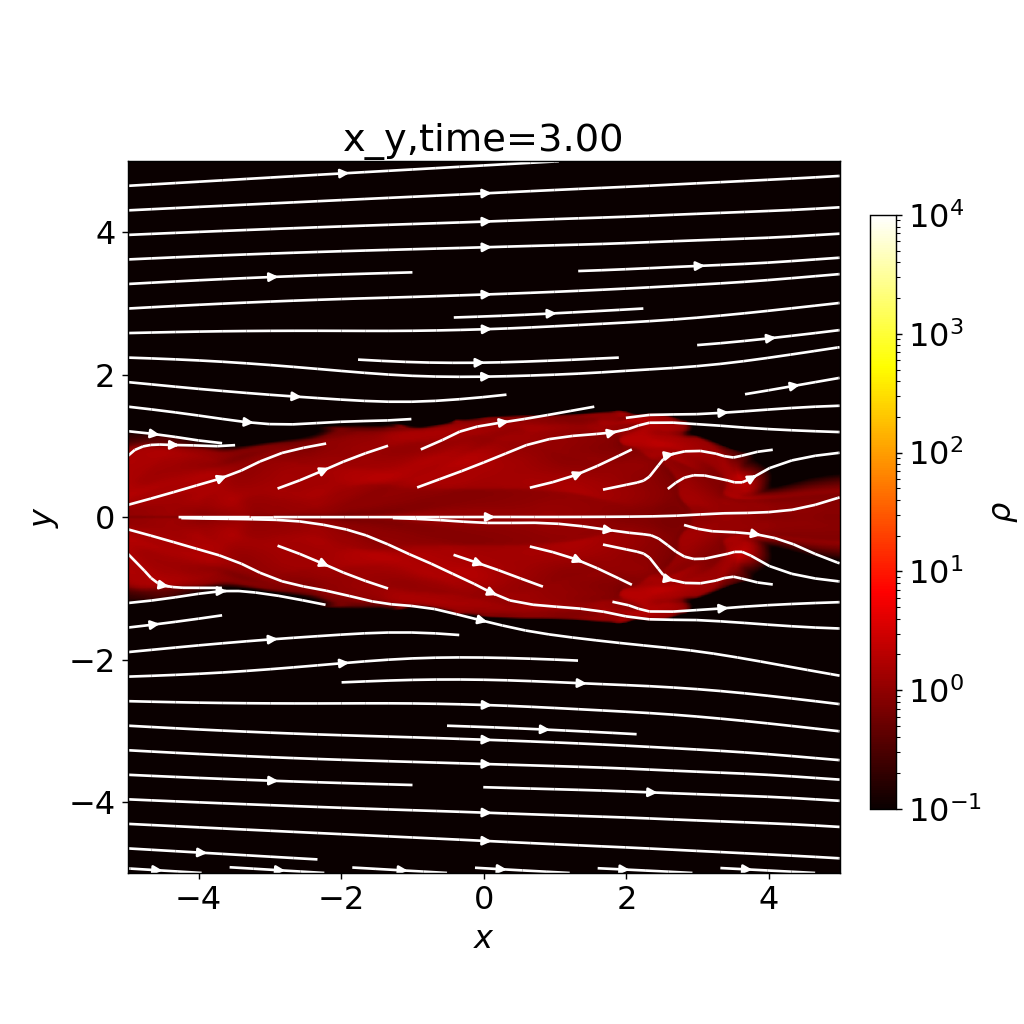}}\\ 
         \subfigure[]{
         \includegraphics[keepaspectratio,scale=0.25]{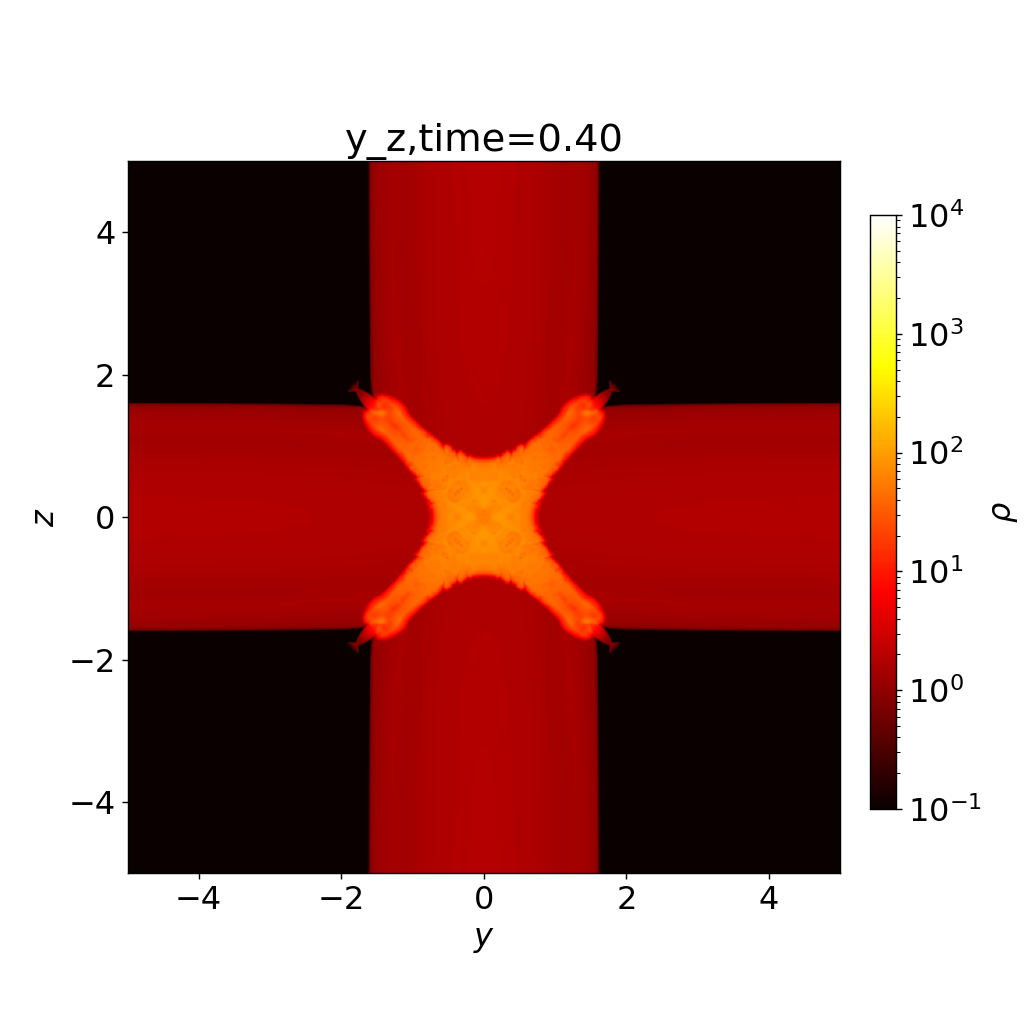} }&
         \subfigure[]{
         \includegraphics[keepaspectratio,scale=0.25]{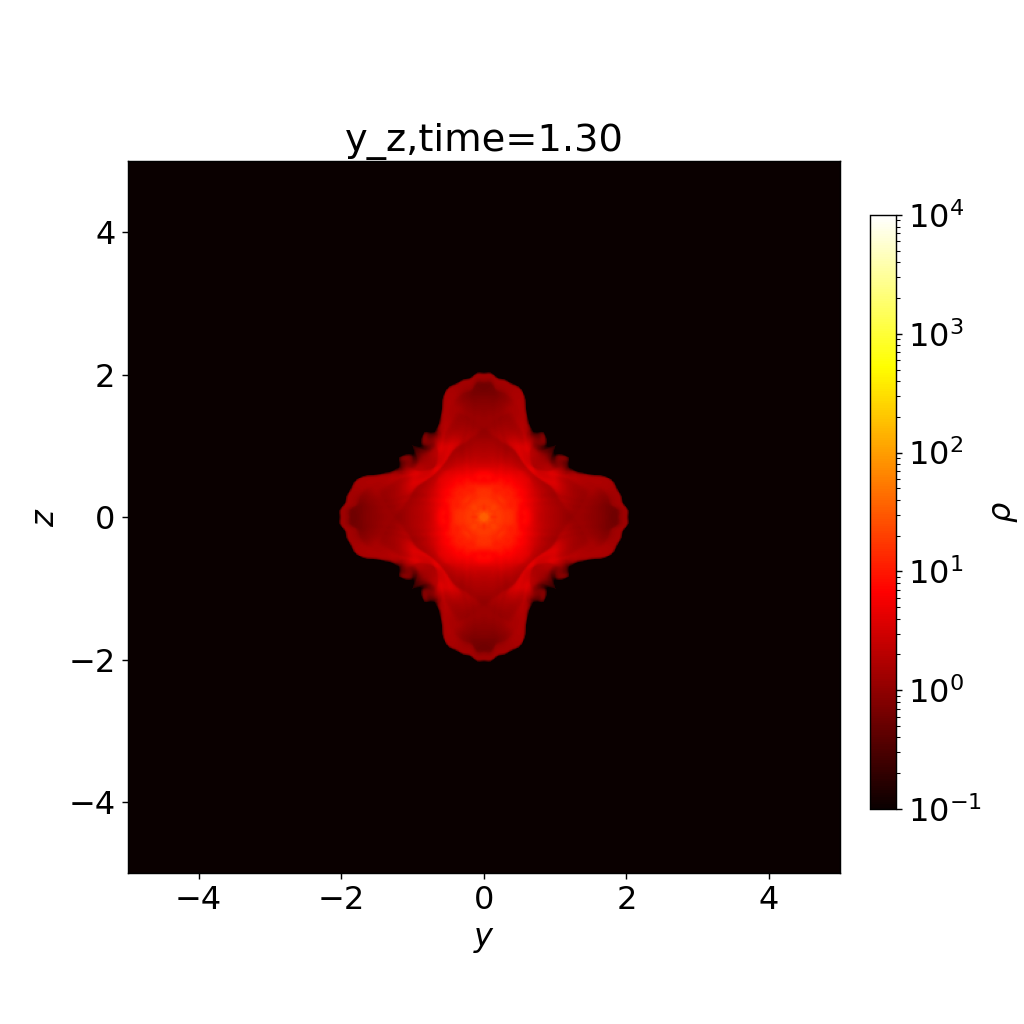}}&
         \subfigure[]{
         \includegraphics[keepaspectratio,scale=0.25]{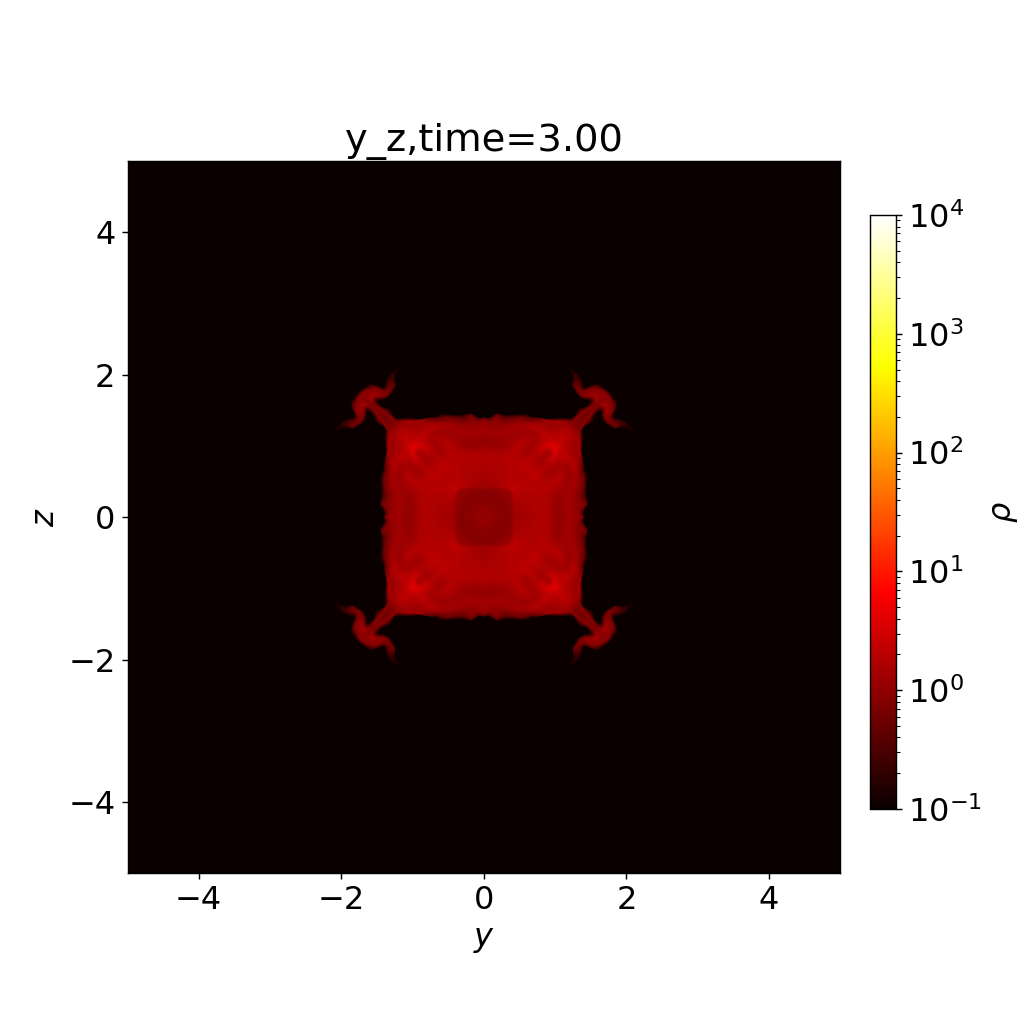}}           
     \end{tabular}
      \caption{
      Same as Figure \ref{fig:3d_L05B1_collapse}, but the expanding model (B1L05V10). 
      The columns correspond to respective times $t=0.40$, 1.30, and 3.00 from left to right.
       }
     \label{fig:3d_L05B1_expand}
\end{figure*} 

\subsection{Expansion Mode}\label{sec:expansion_mode}

Finally, we describe the typical evolution of the expansion mode.
This expansion mode is observed only in the orthogonal collisions, not in head-on (parallel) collisions (\citepaperI{2023ApJ...954..129K}).
The model with $\beta_0=1,\lambda_0=0.59 \lambda_\mathrm{crit}$, and $V_\mathrm{int}=\pm5$ is selected as a fiducial model corresponding to B1L05V10 in Table \ref{tab:model_parameters}.

From the maximum density evolution in Figure \ref{fig:maxden_for_3d}, it is evident that the expansion mode experiences a rapid increase in the density at first due to its fast initial velocity, then transitions into an expansion, and finally $\rho_\mathrm{max}$ maintains a relatively constant level around the initial central density of the filaments.

Figures \ref{fig:3d_L05B1_expand}(a) and (d) 
indicate that at $t=0.4$ the filaments have almost completely collided and a thin and dense structure is seen in the $x$-$y$ plane.

In Figures \ref{fig:3d_L05B1_expand}(b) and (e)
it becomes evident that at $t=1.3$, a shocked region has expanded in the $x$-direction, 
but in the $y$-$z$ plane, the density around the center is decreased and the size of the structure is not changed significantly compared with the previous epoch $t=0.4$.
Furthermore, in panel (b), the region of the filaments that did not collide ($|y|\gtrsim 2$) maintains its velocity and moves away from the shocked region around the center.

In Figures \ref{fig:3d_L05B1_expand}(c) and (f), 
at $t=3.00$, the shocked region evolves into a more expanded structure compared with the previous epoch.
Figure \ref{fig:3d_L05B1_expand}(c) reveals that this structure is elongated and parallel to the global magnetic field lines.
In Figure \ref{fig:3d_L05B1_expand}(f), the shocked region seen in the $y-z$ plane looks circular, if we ignore the structure of the outer edge.

Therefore, in the expanding mode, a sheet-like structure is initially formed, which then evolves into a low-density structure extending in the direction parallel to the global magnetic field lines and the collision direction.

\begin{figure*}
    \centering
     \begin{tabular}{cc}
         \subfigure[]{
         \includegraphics[keepaspectratio,scale=0.4]{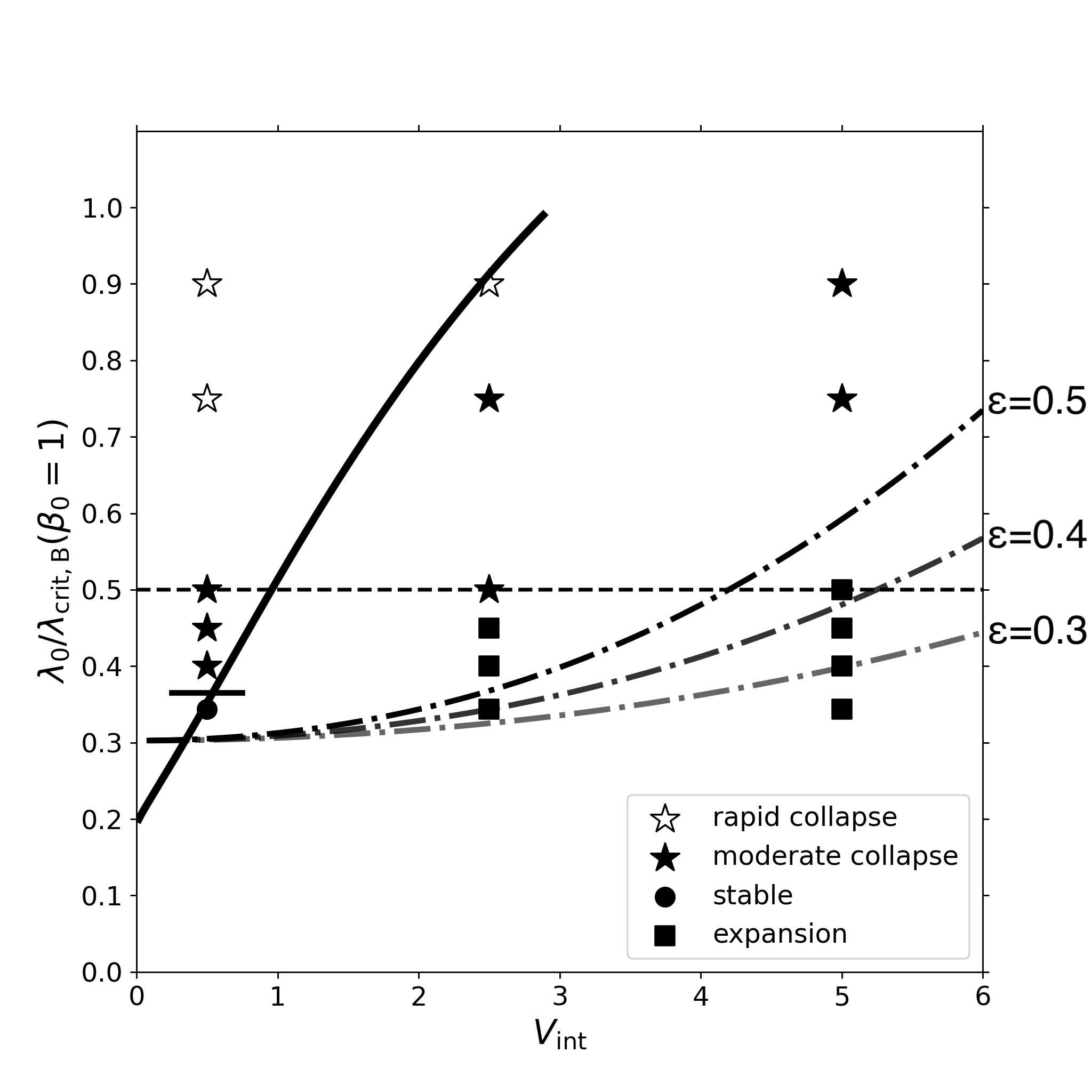}}&
         \subfigure[]{
         \includegraphics[keepaspectratio,scale=0.4]{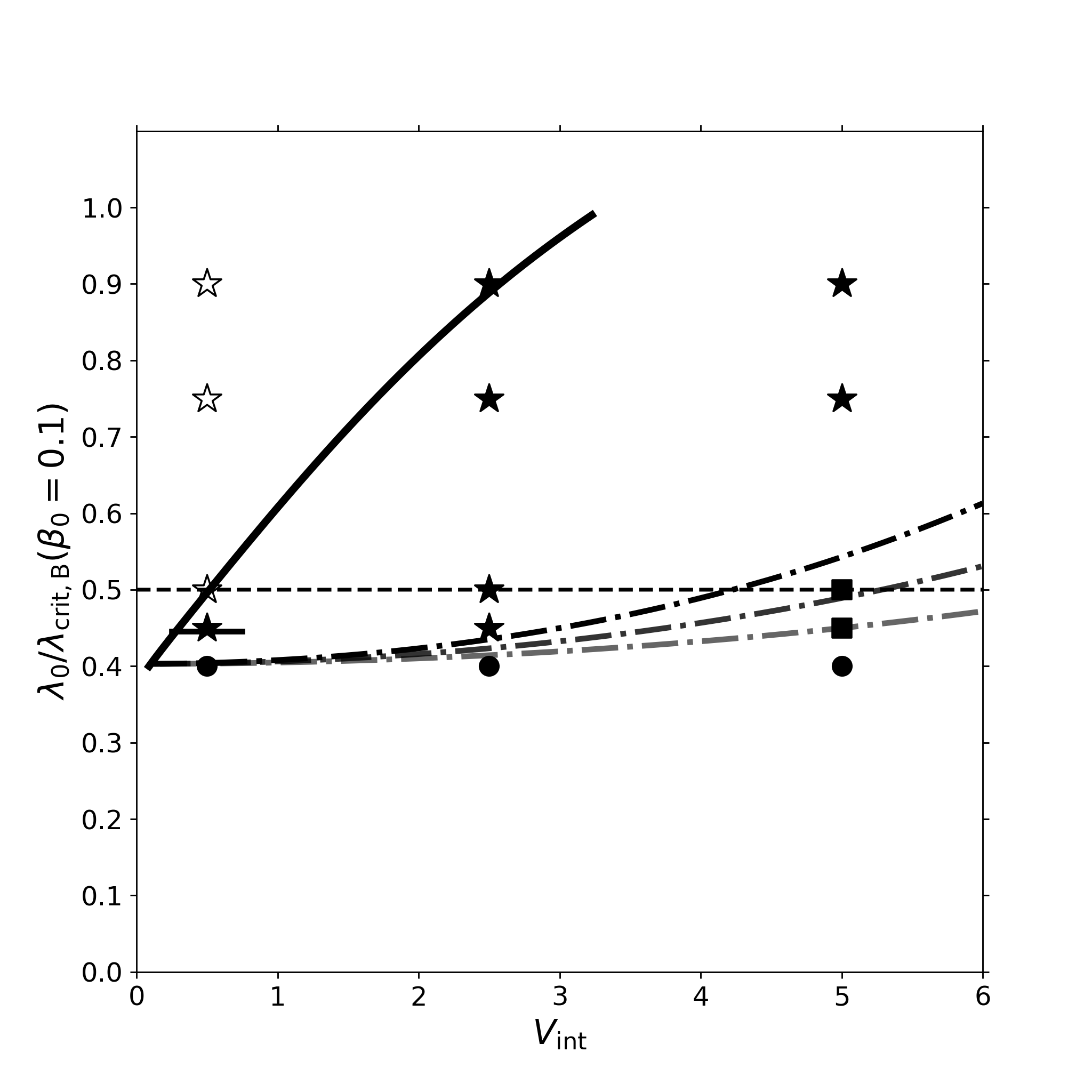}}\\
         &\\
         \multicolumn{2}{c}{\subfigure[]{
         \includegraphics[keepaspectratio,scale=0.4]{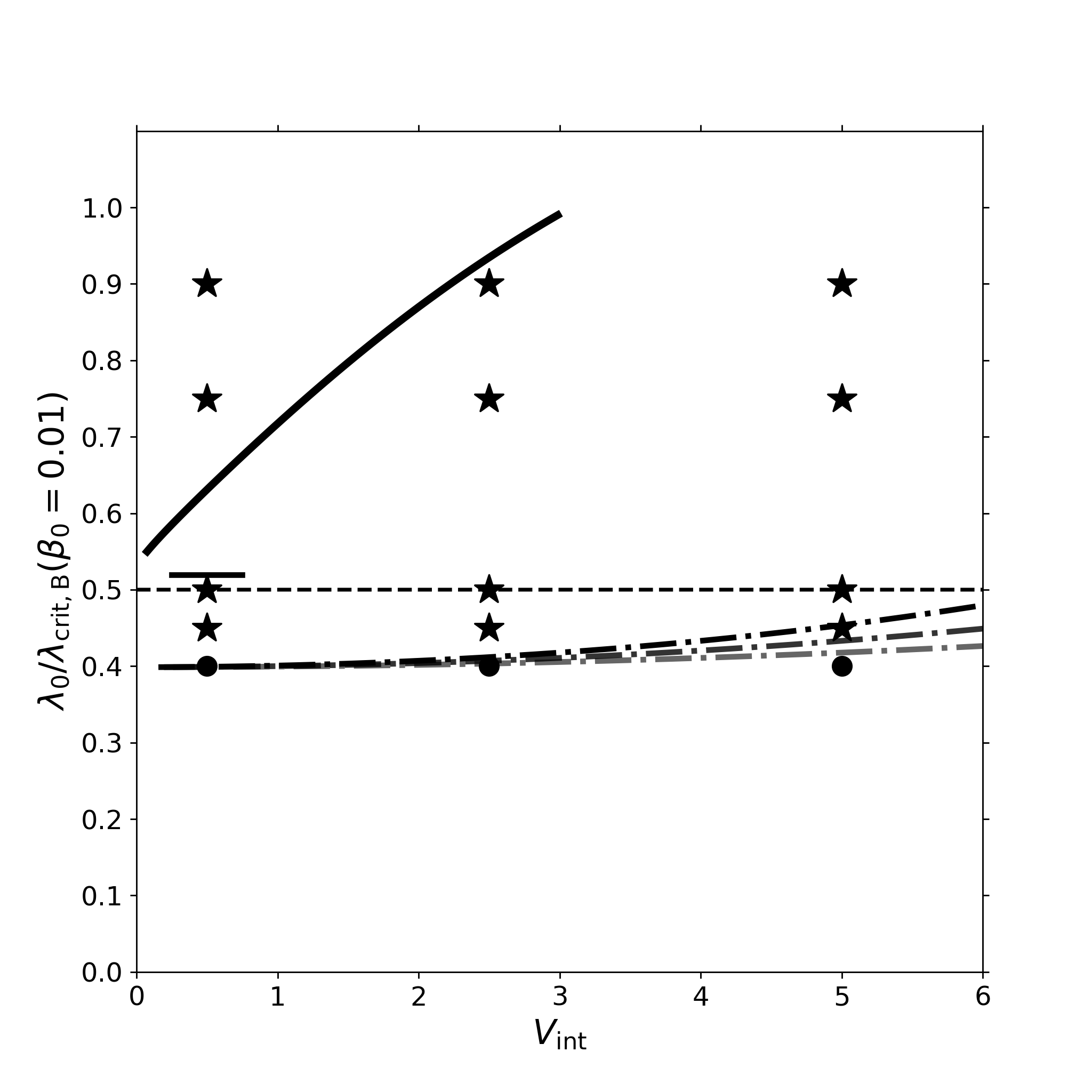}}}
     \end{tabular}
      \caption{Final outcome of filament collisions.
      The evolution modes are
      displayed on the $V_\mathrm{int}$-$\lambda_\mathrm{0}/\lambda_\mathrm{crit,B}$ plane.
      Panels (a), (b), and (c) correspond to the models with $\beta_0=1$, $0.1$, and $0.01$, respectively.
      The open star, filled star, circle, and square symbols represent the rapid collapse, moderate collapse, stable, and expansion modes, respectively.
      The solid line near $V_\mathrm{int}=0.5$ corresponds to the line mass in which the mass of the intersection is equal to the magnetized critical mass $M_\mathrm{cross}=M_\mathrm{crit,B}$ \citep{1988ApJ...335..239T}, and details are seen in Appendix \ref{magnetized_mass}.
      The dashed horizontal line corresponds to the magnetized critical line mass $2\lambda_0=\lambda_\mathrm{crit,B}$ \citep{2014ApJ...785...24T}.
      The dash-dotted lines represent $\lambda_0/\lambda_\mathrm{crit,B}$ where the total energy equals to zero in Equation (\ref{eq:KE_after_PE_ME_GE}) for respective $V_\mathrm{int}$, in which three models of the coefficient of restitution as $\epsilon=0.5$ (upper), 0.4 (middle), and 0.3 (lower) are included.
      The solid line represents the relation of $t_\mathrm{comp}=t_m$ in Equation (\ref{eq:t_comp_vs_t_m_nondim}).
      Here, to derive the dash-dotted and solid lines, the filament width is fixed as $R_0=2$. Additionally, for the dash-dotted lines, the geometric factor of the initial filament is set to the arithmetic average for each beta: $\theta=0.65\,(\beta_0=1)$, $\theta=0.52\,(\beta_0=0.1)$, and $\theta=0.30\,(\beta_0=0.01)$.
      }
     \label{fig:stability_for_orthogonal}
\end{figure*}

\subsection{Parameter Dependence of the Evolutionary Modes}\label{sec:effect_of_parameters}
In the preceding sections, we have shown the typical evolutional modes derived by the orthogonal collision of filaments.
Figure \ref{fig:stability_for_orthogonal} summarizes the evolutionary modes on the plane $(V_\mathrm{int},\lambda_0/\lambda_\mathrm{crit,B})$ for the three plasma beta values $\beta_0=1$, 0.1, and 0.01.
In this section, we examine the impact of the respective parameters (e.g., initial velocity $V_\mathrm{int}$, line mass $\lambda_0$, and plasma beta $\beta_0$) on the evolutionary process of the shocked cloud.

\subsubsection{Transition from Moderate Collapse Mode to Rapid Collapse Mode}\label{sec:rapid_collapse}

In Figure \ref{fig:stability_for_orthogonal}, the collapse mode can be further divided into two modes, depending on the ratio of the time required for gravitational 
collapse to the time required for the collision to complete.
The collision is completed when the two shock waves generated by the collision propagate throughout the two filaments.
The collapse mode where the shocked region collapses gravitationally before (after) the collision is completed is called 
the rapid (moderate) collapse mode.

\begin{figure*}
    \centering
     \begin{tabular}{cc}
         \subfigure[]{
         \includegraphics[keepaspectratio,scale=0.3]{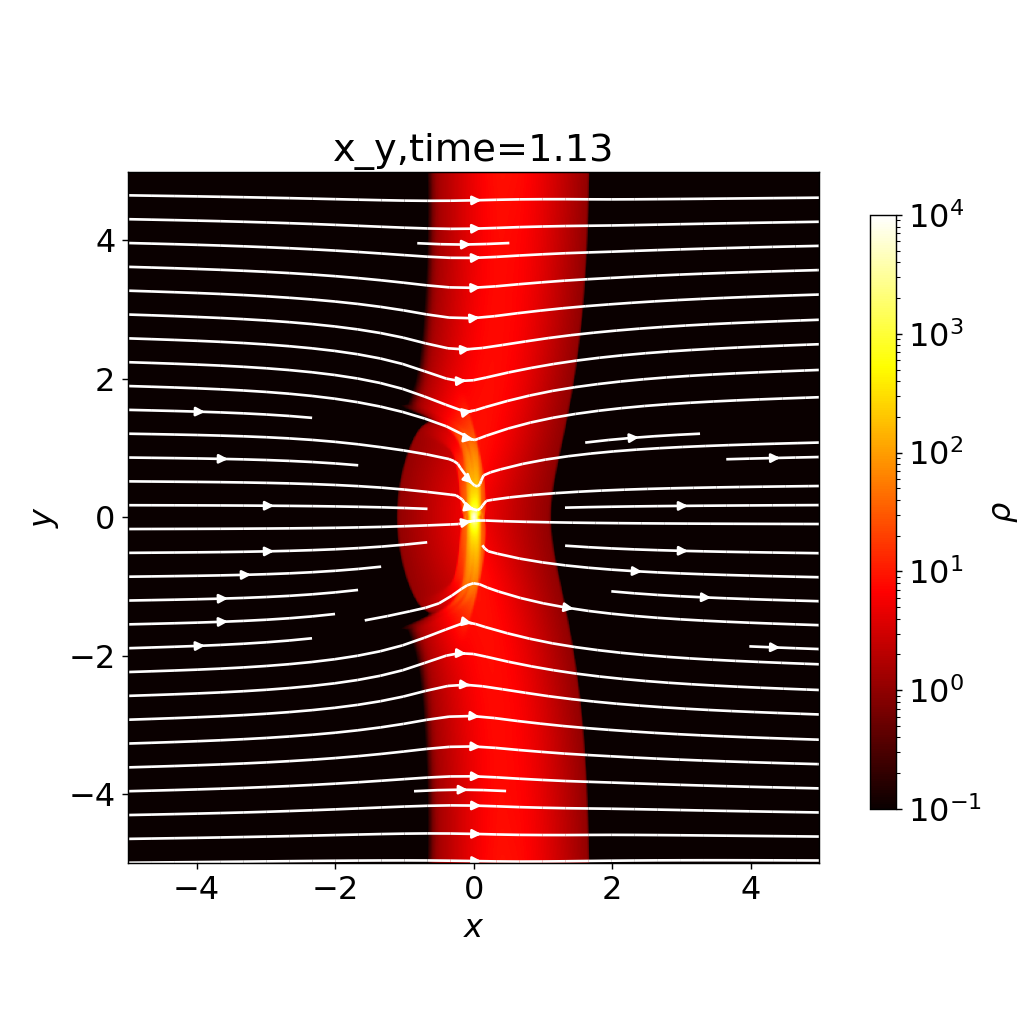} }& 
         \subfigure[]{
         \includegraphics[keepaspectratio,scale=0.28]{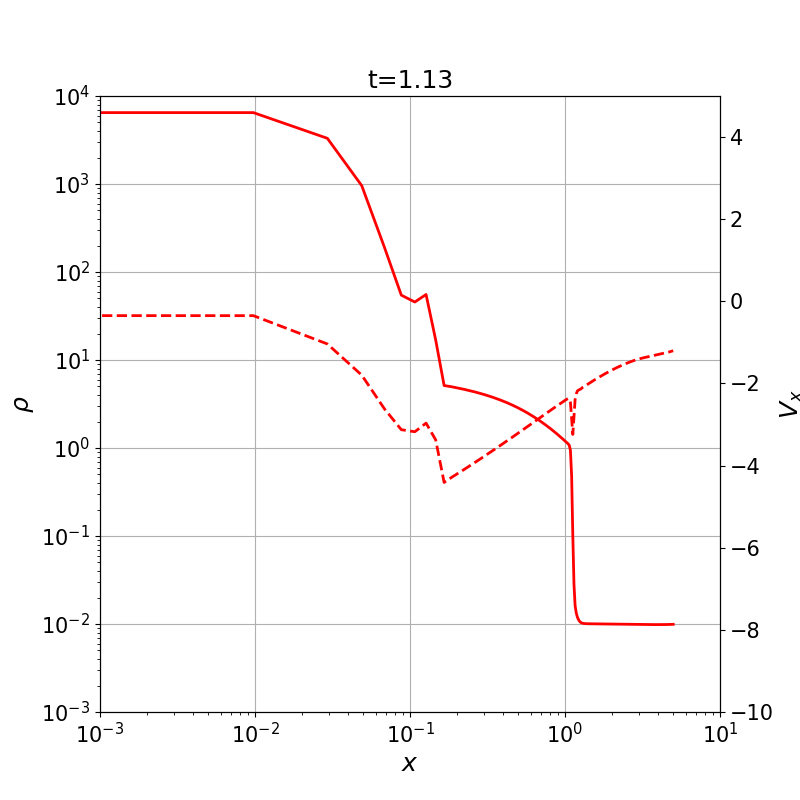}}\\
         \subfigure[]{
         \includegraphics[keepaspectratio,scale=0.3]{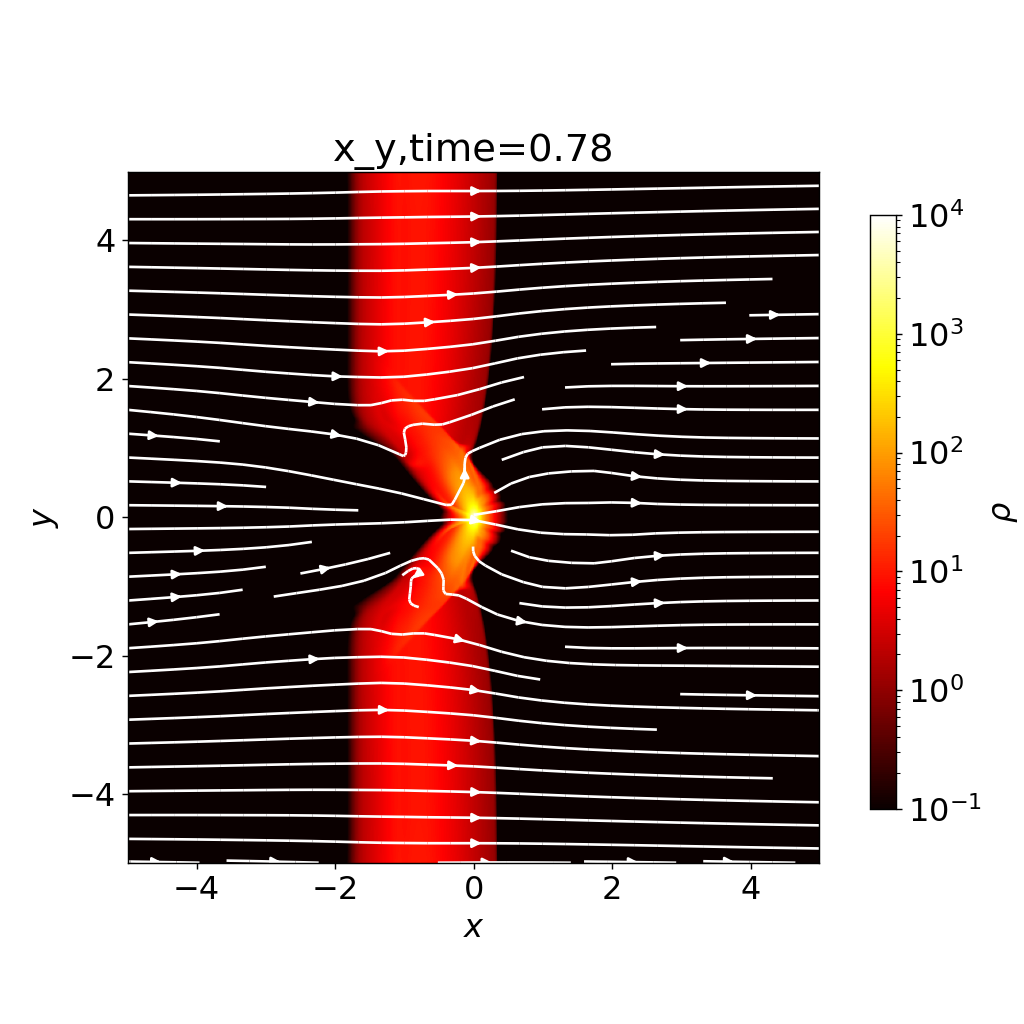}}& 
         \subfigure[]{
         \includegraphics[keepaspectratio,scale=0.28]{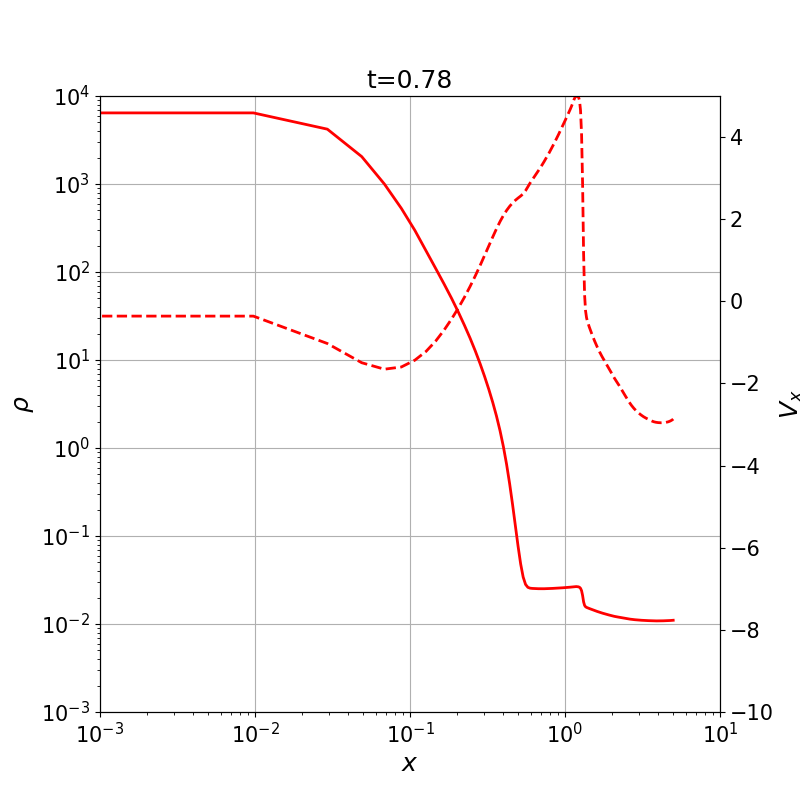} }        
     \end{tabular}
      \caption{
      Two-dimensional slice and one-dimensional profile of the models of B1L075V1 
      (rapid collapse mode, the upper panels) and B1L075V5 (moderate collapse mode, the lower panels)
      at which the maximum density reaches $\rho_\mathrm{lim}$.
      The left and right panels correspond to the slice at $z=0$ and the 1D profile along the $x$-axis, respectively.
      In the 1D profiles, the solid line represents the density distribution (the left vertical axis) and the dashed line shows the velocity distribution 
      (the right vertical axis).}
     \label{fig:rapid_and_moderate}
     \end{figure*}

In Figure \ref{fig:stability_for_orthogonal}, we find that with increasing $V_\mathrm{int}$ while fixing $\lambda_0$ and $\beta_0$,
the evolutionary mode changes from the rapid collapse (open star) to the moderate collapse (filled star) modes.
For example, we focus on B1L075V1 and B1L075V5 which are located in Figure \ref{fig:stability_for_orthogonal}(a).
Figure \ref{fig:rapid_and_moderate} shows the two-dimensional slice and one-dimensional profile of these models of B1L075V1 
(rapid collapse mode) and B1L075V5 (moderate collapse mode)
at the time of gravitational collapse $t_\mathrm{collapse}$ which we define as the time at which
the maximum density reaches the numerical threshold density $\rho_\mathrm{lim}$.

From Figure \ref{fig:rapid_and_moderate}(a), it is evident that B1L075V1 model shows the rapid collapse mode
because the filaments have not yet fully collided at $t=t_\mathrm{collapse}$. 
Additionally, the velocity profile along the $x$-axis in Figure \ref{fig:rapid_and_moderate}(b) shows only the inflow of gas toward the center.
At $t=t_\mathrm{collapse}$, the outward shock around $x \simeq 0.15$ is propagating in the unshocked filament in $x>0$.
In contrast, Figure \ref{fig:rapid_and_moderate}(c) shows that the filaments have fully collided by 
$t=t_\mathrm{collapse}$, and the velocity profile along the $x$-axis in Figure \ref{fig:rapid_and_moderate}(d) indicates that the outward shock wave ($x \simeq 1.2$) that has passed through the filament is propagating into the ambient medium outside the filaments as shown in Section \ref{sec:collapse_mode}.
Therefore, as shown in Figures \ref{fig:rapid_and_moderate}(c) and (d), 
B1L075V5 model shows the moderate collapse mode.

The moderate collapse is identified by examining the 1D velocity distribution along the $x$-axis at $t=t_\mathrm{collapse}$ and checking for the existence of a significant outward shock wave traveling in the ambient low-density gas.
This transition can be observed in relatively heavier filaments $(\lambda_0/\lambda_\mathrm{crit,B}\gtrsim 0.75)$ for each plasma beta.
Furthermore, this transition shifts upward with increasing the magnetic field strength or decreasing the plasma beta, as seen in Figure \ref{fig:stability_for_orthogonal}.

\subsubsection{Transition from Moderate Collapse Mode to Expansion Mode}\label{sec:collapse_to_expand}

In the preceding section, we showed the 
mode change from the rapid collapse to the moderate collapse modes 
with increasing the initial velocity while fixing $\lambda_0$ and $\beta_0$.
In this section, we examine the possibility of the transition from the moderate collapse (filled star) mode to 
the expansion (square) mode by increasing the initial velocity as seen in Figure \ref{fig:stability_for_orthogonal}.
A comparison is conducted using models in which $\beta_0=1$ and $\lambda_0=0.53 \lambda_\mathrm{crit}$ 
are fixed, but the initial velocity is varied as $V_\mathrm{int}=\pm 0.5$, $\pm2.5$, and $\pm 5$. 
This corresponds to the models of B1L045V1, B1L045V5, and B1L045V10 in Table \ref{tab:model_parameters} and middle part of Figure \ref{fig:stability_for_orthogonal}(a).

Figure \ref{fig:stability_for_orthogonal}(a) shows that the collision with $V_\mathrm{int}=\pm0.5$ results in the moderate collapse mode. 
However, at $V_\mathrm{int}=\pm2.5$ and $\pm5.0$, the evolution changes to the expansion mode. 
Therefore, the evolution mode of the shocked clouds changes depending on the initial velocity.
In addition to this, the initial velocity at which this transition occurs increases with the line mass ($\lambda_0/\lambda_\mathrm{crit,B}=0.4$ to 0.5) in Figure \ref{fig:stability_for_orthogonal}(a). 
Additionally, although this transition is also seen in Figure \ref{fig:stability_for_orthogonal}(b), it diminishes with further decreasing the plasma beta, as shown in Figure \ref{fig:stability_for_orthogonal}(c).

\subsubsection{Transition from Stable Mode to Expansion Mode}\label{sec:stable_to_expansion}
Here, we also comment on the transition between stable and expansion modes.
In Figure \ref{fig:stability_for_orthogonal}(a), the system exhibits the stable (circle) mode for the transonic collisions $V_\mathrm{int}=\pm 0.5$ and transitions to the expansion (square) mode for the supersonic collisions $V_\mathrm{int}=\pm 2.5$. 
This comes from the fact that the shocked cloud expands and eventually reaches the boundaries of the computational domain, thus classified as the expansion mode. 
However, in Figures \ref{fig:stability_for_orthogonal}(b) and (c), where the magnetic field gets strong (the line mass increases), such a transition from the stable to expansion modes is not observed even at supersonic initial velocities, and the evolutionary mode remains in the stable mode.

\subsubsection{Transition from Stable Mode to Moderate Collapse Mode}\label{sec:stable_to_collapse}
Figure \ref{fig:stability_for_orthogonal}(a) shows that the transition from the stable (circle) to the moderate collapse (filled star) modes occurs when increasing the line mass from $\lambda_0/\lambda_\mathrm{crit,B}\simeq0.34$ to 0.4 for $\beta_0 = 1$. 
For $\beta_0 = 0.1$ and 0.01, the necessary line mass for the collapse mode is between $\lambda_0/\lambda_\mathrm{crit,B}=0.4$  and 0.45, indicating an increase in the transition line mass compared to $\beta_0 = 1$. 
Additionally, as the initial velocity increases, the stable mode disappears for $\beta_0=1$, as shown in Section \ref{sec:stable_to_expansion}. 
However, for $\beta = 0.1$ and $\beta = 0.01$, the transition occurs within the same range of line mass regardless of the collision speed.

\section{Discussion}\label{sec:discussion}
\subsection{ A Criterion for Evolutional Modes of Shocked Clouds }

In this section, we investigate the conditions to specify the evolutional modes of the shocked clouds formed at the intersections.

\subsubsection{Condition for the Collapse and Expansion Modes}\label{sec:condition_for_collapse_and_expansion}

In this section, we discuss the condition that the collapse and expansion modes are realized.
In Section \ref{sec:effect_of_parameters}, we found that the models with higher initial velocities tend to exhibit the expansion mode, whereas they exhibit the moderate collapse mode for the transonic collisions.

In Figure \ref{fig:stability_for_orthogonal}, the models that result in the collapse (stars) and expansion (squares) modes are shown on the $(V_\mathrm{int},\lambda_0/\lambda_\mathrm{crit,B}$) plane
for each $\beta_0$.
In Figure \ref{fig:stability_for_orthogonal}(a), the transition can be observed from the moderate collapse mode to the expansion mode at approximately $0.5 - 2.5$ times the speed of sound in the small initial line mass.
In addition to this, as $\beta_0$ decreases (from panels (a) to (c)), the transition line moves rightward, and finally the expansion mode diminishes.

Here, we determine the conditions under which the transition to the expansion mode occurs from the standpoint of the energy balance of the colliding system.
In the early stage of filament collision, because the influence of 
the region outside the intersection must be small, we focus on the intersection region.
Especially, we pay attention to the state of the cloud immediately after the collision, since the subsequent evolution of the shocked cloud is potentially explained by the relative importance of the magnitude of gravitational energy over other energies.
The non-dimensional total energy of the shocked cloud immediately after it completes the collision is expressed as (for clarity, normalized variables are marked with prime)
\begin{equation}\label{eq:KE_after_PE_ME_GE}
    KE'_\mathrm{after}+PE'+ME'+GE'=2\lambda'_0 R'_0 \epsilon^2 V'^2_0 +6\lambda'_0 R'_0 +8\theta R'^3_0 B'^2_0 -\frac{6}{5}R'_0\lambda'^2_0.
\end{equation}
The first term represents the kinetic energy given as
\begin{equation}\label{eq:kin_E}
KE'_\mathrm{after}= 2\int dV' \frac{1}{2}\bar{\rho}' V'^2_\mathrm{out}= 2\lambda'_0 R'_0 \epsilon^2 V'^2_0,
\end{equation}
where the first coefficient of two comes from the fact that there are two colliding filaments, $\bar{\rho}'$ corresponds to the mean density of the filament, $\epsilon$ indicates the coefficient of restitution, and $V'_0$ denotes the initial velocity of each filament.
Here, we assumed that the expansion velocity after the collision ($V'_\mathrm{out}$) is expressed as $V'_\mathrm{out}=\epsilon V'_0$.
Given that the integration range corresponds to the filament volume that becomes the intersection as $\int\cdot\cdot\cdot dV'=\int^{\theta R_0'}_{-\theta R_0'}\int^{R_0'}_{-R_0'}\int^{R_0'}_{-R_0'}\cdot\cdot\cdot dx' dy' dz'$, where $\theta$ is a geometric factor that represents the minor-to-major axis ratio of the initial filament, then we use a relation $\lambda'_0\simeq 4\bar{\rho}'\theta R'^2_0$.
Similarly, the second term of Equation (\ref{eq:KE_after_PE_ME_GE}) 
represents the thermal energy and is given by
\begin{equation}\label{eq:th_E}
PE' = 2\int dV' \frac{3}{2}\bar{\rho}'=6\lambda'_0 R'_0,
\end{equation}
then the third term of Equation (\ref{eq:KE_after_PE_ME_GE}) 
represents the magnetic energy and is given by
\begin{equation}\label{eq:mag_E}
ME' = 2\int dV' \frac{1}{2}B'^2_0=8\theta R'^3_0B'^2_0,
\end{equation}
where $B'_0$ is the uniform magnetic field strength.
The fourth term of Equation (\ref{eq:KE_after_PE_ME_GE}) shows the gravitational energy of the shocked cloud that is approximated as the MacLaurin spheroid \citep[see][]{1969efe..book.....C,1987PThPh..78.1250N} since the shape of the shocked cloud formed by the collision looks oblate.
This is given in a dimensional form as 
\begin{equation}
    GE\simeq Mc_s^2\left(-\frac{\sin^{-1}e}{e}\frac{1}{\xi} \right),
\end{equation}
where using the semimajor, $a$, and semiminor, $b$, axes of the shocked cloud, $\xi \equiv (5c_s^2/3GM)a$ is the non-dimensional semimajor axis, $e\equiv (1-\eta^2)^{1/2}$, and $\eta=b/a$. 
This Maclaurin spheroid approximation is also adopted in \citet{1987PThPh..78.1250N}, who simulate cloud-cloud collisions that result in the formation of oblate shocked clouds.
Then, when we assume a sufficiently thin oblate spheroid is formed $\eta\simeq 0$ $(b\ll a)$, i.e, $e=1$ and $\sin^{-1}e=\pi/2$, $GE$ is written in a non-dimensional form as 
\begin{equation}
    GE'\simeq M'\left(-\frac{\pi}{2}\frac{1}{4\pi}\frac{3M'}{5 R'_0} \right)=-\frac{6}{5}R'_0\lambda'^2_0,
\end{equation}
where we use the relation $M'\simeq 2\times 2R'_0\times \lambda'_0$.

From this, we plotted the relationship between the line mass and initial velocity to satisfy the total energy equation of $E'_\mathrm{all}=KE'_\mathrm{after}+PE'+ME'+GE'=0$ as dash-dotted lines in Figure \ref{fig:stability_for_orthogonal}.
We expect that if $E'_\mathrm{all}<0$, the magnitude of gravitational energy is dominant, and the shocked cloud collapses; if $E'_\mathrm{all}>0$, the cloud does not collapse.
Here, we use the coefficient of restitution as $\epsilon=0.4$, which is a mean value estimated from calculations with the same settings of this study but without gravity\footnote{This calculation utilized six models—B1L05V5-10, B01L05V5-10, and B001L05V5-10—excluding gravitational effects and maintaining identical settings for all collisions. The coefficient of restitution ($\epsilon=0.397$) is estimated based on the square root of the ratio of the initial total kinetic energy $KE_\mathrm{int}$ to the post-collision kinetic energy $KE_\mathrm{post}$ as $\epsilon=\sqrt{KE_\mathrm{post}/KE_\mathrm{int}}$, where $KE_\mathrm{post}$ is estimated when the outward velocity component immediately after the collision.}.
For reference, in Figure \ref{fig:stability_for_orthogonal}, we also plot the relation for $\epsilon$, such as 0.5 and 0.3.

This dash-dotted line well separates the collapse and stable modes in the transonic speed regime.
This is because, in the case of a low-velocity collision, the stability of the shocked cloud is determined by the balance of thermal energy, magnetic energy, and gravitational energy in Equation (\ref{eq:KE_after_PE_ME_GE}), corresponding to a kind of equilibrium state.
For reference, the horizontal line in Figure \ref{fig:stability_for_orthogonal} shows the condition that the mass of the intersection coincides with the magnetically critical mass \citep[][details are shown in Appendix \ref{magnetized_mass}]{1976ApJ...210..326M,1988ApJ...335..239T}.
However, only in Figure \ref{fig:stability_for_orthogonal}(a), there is a bit of discrepancy between the dash-dotted line derived from the total energy argument and the actual boundary between the collapse and stable modes in the transonic speed regime. 
This is because, at the transonic collision with $\beta_0=1$, the shocked cloud has a thickness, although we assumed a thin oblate spheroid ($\eta\simeq 0$) to derive the gravitational energy.
For instance, changing the aspect ratio from $\eta=0$ to $\eta=1/2$ leads the y-intercept of the dash-dotted line upward from $\lambda_0/\lambda_\mathrm{crit,B}(\beta_0=1)\simeq 0.30$ to $\lambda_0/\lambda_\mathrm{crit,B}(\beta_0=1)\simeq 0.37$, making it consistent with the calculated results in Figure \ref{fig:stability_for_orthogonal}(a).
On the other hand, in models with small $\beta_0$, the initial filament shape is already flattened \citep{2014ApJ...785...24T}, and the resulting shocked cloud can be well approximated by the thin oblate spheroidal (disk-like) shape assumed in this study. 
Consequently, the condition of the total energy to be zero in Equation (\ref{eq:KE_after_PE_ME_GE}) explains the simulation results effectively.

In addition to this, the dash-dotted line is suitable for distinguishing the expansion mode from the collapse mode in the range of sufficiently supersonic collisions in all $\beta_0$ values. 
Then, from Equation (\ref{eq:KE_after_PE_ME_GE}), we estimate the critical initial velocity $V_\mathrm{crit}$ at which the kinetic energy becomes dominant over the thermal and magnetic energies for each $\beta_0$. 
For example, when we fix as $\lambda_0/\lambda_\mathrm{crit,B}=0.5,\, \epsilon=0.4$ and use $\theta$ for each $\beta_0$ as $\theta=0.65\,(\beta_0=1)$, $\theta=0.52\,(\beta_0=0.1)$, and $\theta=0.30\,(\beta_0=0.01)$,
the critical initial velocity is $V_\mathrm{crit}=5.3$ for $\beta_0=1$, $V_\mathrm{crit}=7.9$ for $\beta_0=0.1$, and
$V_\mathrm{crit}=11.4$ for $\beta_0=0.01$. 
This means that the initial velocity required for the kinetic energy to counterbalance the magnitude of gravitational energy (resulting in a steeper slope of the dash-dotted line) increases with the magnetic field strength. 
Therefore, in Figure \ref{fig:stability_for_orthogonal}, the slope of the dash-dotted line appears to become shallower as the magnetic field gets stronger, for the range of the initial speeds examined in this study.

\subsubsection{Condition for the Emergence of Rapid Collapse Mode}\label{sec:condition_for_rapid_collapse}

In Section \ref{sec:rapid_collapse}, we show there are two types of collapse modes: rapid and moderate.
In order for the rapid collapse mode to occur, it is necessary for the shocked cloud formed by the collision to be in collapse before the filaments collide completely. 
Therefore, to achieve the rapid collapse mode, it is necessary for the growth timescale of the gravitational instability of the shocked clouds to be shorter than the compression dynamical timescale.
In this section, we examine this by using the growth time scale of the gravitational instability appeared in the magnetized sheet with a magnetic field perpendicular to the sheet, $t_m$, and the dynamical time for the initial filament to pass the intersection,  $t_\mathrm{comp}$.

Comparing the time scales $t_\mathrm{comp}/t_m$ can be expressed as follows,
\begin{equation}
\begin{split}\label{eq:t_comp_vs_t_m}
  \frac{t_\mathrm{comp}}{t_m}&=\omega_\mathrm{max}t_\mathrm{comp}=\left(1-\alpha_0/2\mu_s^2\right)^{1.25}\left(0.139-0.022\mu_s+0.103\mu_s^2\right)^{1/2}\left(4\pi G\rho_c\right)^{1/2}\frac{\theta R_f}{c_s\mathcal{M}}, \\ 
\end{split}
\end{equation}
where, $\theta$ is a geometric factor that represents the axis ratio of the initial filament, $R_f$ is the half-width of the filament, $\mathcal{M}$ is the Mach number of the colliding filament, $\mu_s\simeq(2\pi G\theta\lambda_0)^{1/2}/c_s\mathcal{M}$ is a parameter indicating the influence of external pressure, originally defined as $\mu_s \equiv \left(1-p_\mathrm{ext}/c_s^2\rho_c\right)^{1/2}$, and $\alpha_0\equiv B^2_0/4\pi c_s^2\rho_c$ is a parameter for the strength of the magnetic field (details in deriving the growth rate of the gravitational instability in the magnetized sheet $\omega_\mathrm{max}$ are shown in Appendix \ref{sec:condition_for_rapid_collapse_Growth_timescale}).

In Equation (\ref{eq:t_comp_vs_t_m}), the first factor comes from the stabilization effect of the magnetic field of the maximum growth rate of the gravitational instability \citep{1983PASJ...35..187T,1988PASJ...40..593N}, the second factor is the maximum growth rate of a non-magnetized sheet \citep{1978ApJ...220.1051E}, and the last factor shows the time scale for the filament to complete its collision.  
When we use the assumption of the isothermal shock wave $\rho_c\simeq\bar{\rho}\mathcal{M}^2$, the relation of the mean density with the line mass, $\bar{\rho}=\lambda_0/4\theta R_f^2$, and the non-dimensional form of the line mass $\lambda'_0=\lambda_0(4\pi G)^{1/2}/c_s$, the non-dimensional form of Equation (\ref{eq:t_comp_vs_t_m}) is written as follows:
\begin{equation}
\begin{split}\label{eq:t_comp_vs_t_m_nondim}
  \frac{t_\mathrm{comp}}{t_m}=\left(1-\alpha_0/2\mu_s^2\right)^{1.25}\left(0.139-0.022\mu_s+0.103\mu_s^2\right)^{1/2}\frac{\lambda'^{1/2}_0\theta^{1/2}}{2}, 
\end{split}
\end{equation}
where $\mu_s=(\theta\lambda'_0/2)^{1/2}/\mathcal{M}$, and $\alpha_0= B'^2_0/\bar{\rho'}\mathcal{M}^2$.

Equation (\ref{eq:t_comp_vs_t_m_nondim}) predicts that when $t_\mathrm{comp}/t_m>1$, the gravitational growth is faster than compression, resulting in the rapid collapse mode.
Indeed, $t_\mathrm{comp}/t_m=1$, which corresponds to the solid lines in Figure \ref{fig:stability_for_orthogonal}, represents fairly well the critical state distinguishing between the rapid collapse and moderate collapse modes, especially models with $\beta_0=1$ and $0.1$. 
However, in Figure \ref{fig:stability_for_orthogonal}(c), when $\beta_0=0.01$, there is no rapid collapse mode even above the solid lines in the regime of the transonic collisions with strongly magnetized (heavier) filaments. 
This is reasonable if we take into account the point that the gravitational acceleration of each filament increases the collision speeds resulting in the shortening of the compression time scale. 
On the other hand, in the case of supersonic collisions, since the collision occurs before the gravity changes the collision speed sufficiently, the gravitational acceleration is expected to be unimportant. 
Therefore, using the comparison between the growth timescale of the gravitational instability in the magnetized sheet and the compression timescale, 
we can distinguish the rapid collapse mode from the moderate collapse mode.

\subsection{Self-similar Manner for Collapse Mode}
\begin{figure*}
    \centering
     \begin{tabular}{cc}
         \subfigure[]{
         \includegraphics[keepaspectratio,scale=0.5]{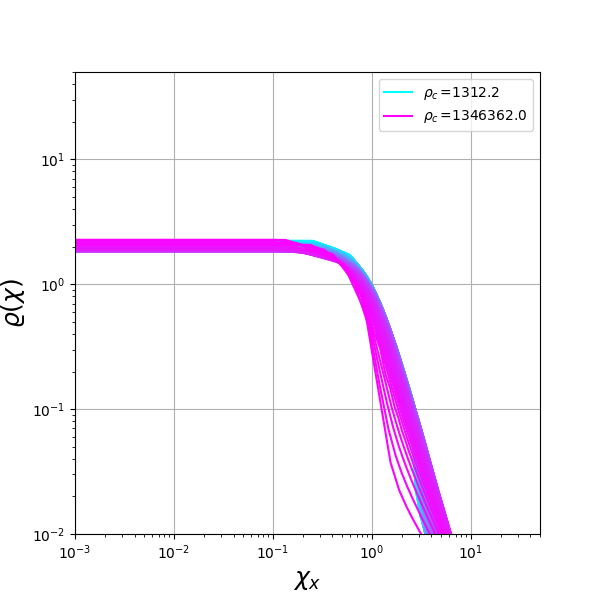}}&
         \subfigure[]{
         \includegraphics[keepaspectratio,scale=0.5]{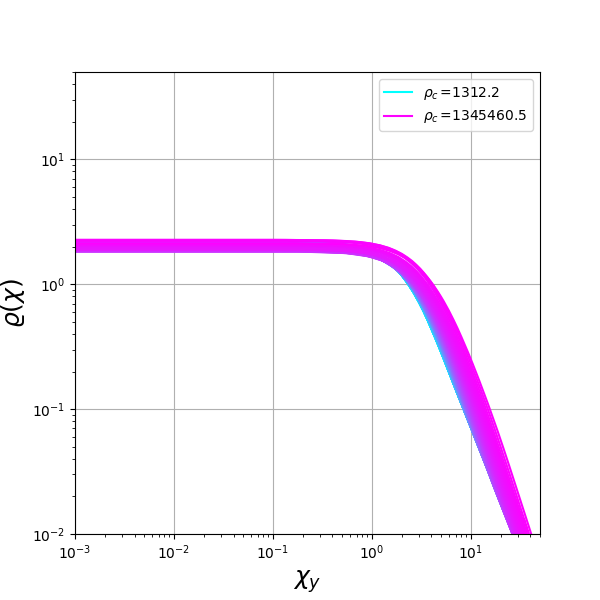}}
     \end{tabular}
      \caption{ 
      Density profiles of the collapse model resulting from the orthogonal collision, written in the similarity coordinate system using the model B1L05V1.
      Panels (a) and (b) correspond to the profiles on the $x$- and $y$-axes, respectively.
      We presented the densities ranging from $\rho_c\sim 1.3\times10^3$ to $\sim 1.3\times 10^6$, corresponding to the color gradation from cyan to magenta.
      }
     \label{fig:selfsimilar_solution}
\end{figure*}

In this section, we examine self-similar solutions for a collapsing shocked cloud.
Understanding the property of collapsing clouds is important because it relates to the characteristics of the stars that subsequently form.
This self-similar solution has been previously studied to describe the properties of the collapsing filament in \citet{1998PASJ...50..577K}, and collapsing magnetized clouds in \citet{1999ApJ...510..274N}.

In \citet{1998PASJ...50..577K}, the similarity coordinates are defined with the zooming coordinate $\bm{\chi}$ and the density in the zooming coordinate $\varrho$ as,
\begin{equation}\label{eq:zooming_coordinate}
   \bm{\chi}\equiv \frac{\bm{r}}{c_s|t-t_0|},
\end{equation}
\begin{equation}\label{eq:density_in_zooming_coordinate}
   \varrho(\chi)\equiv 2\pi G(t-t_0)^2\rho,
\end{equation}
where $\bm{r}$ represents the position vector and $t_0$ corresponds to the time when the central density becomes infinite.\footnote{
To determine $t_0$, we first examine the time evolution of $1/\sqrt{\rho_\mathrm{max}}$.
By identifying the part that follows a linear relation, $1/\sqrt{\rho_\mathrm{max}}=a_0t+b_0$, we find that the time at which $1/\sqrt{\rho_\mathrm{max}}=0$ corresponds to $t_0=-b_0/a_0$.
}

Here, we focus on Model B1L05V1, one of the collapse models. 
Additionally, we conducted calculations using adaptive mesh refinement to achieve a sufficient spatial resolution to cover enough density ratio.
Figure \ref{fig:selfsimilar_solution} indicates the time evolution of the density profiles plotted in the similarity coordinate for model B1L05V1, using Equations (\ref{eq:zooming_coordinate}) and (\ref{eq:density_in_zooming_coordinate}).
In Figure \ref{fig:selfsimilar_solution}, the density profiles on the $x$- and $y$-axes do not seem to depend on time significantly when the similarity coordinates are used, even though the central density varies by a factor of $\sim 1000$.
As a result, the high-density region of the shocked clouds collapses in a self-similar manner, keeping an ellipsoidal structure flattened along the $x$-axis, aligned with the direction of the magnetic field lines.

\citet{1999ApJ...510..274N} also reported that, in addition to this self-similar property, the collapsing clouds exhibited oscillations with a finite amplitude, due to the calculation with a wide density range 
(up to $18$ order of magnitude). 
Therefore, conducting higher-resolution calculations may enable us to capture these oscillations more accurately.

\subsection{Subsequent Accretion}\label{sec:filament_accretion}

In this section, we consider gas accretion along the major axis of the filament. 
In this study, we assumed infinitely long filaments, and although it is difficult to perform a detailed investigation of the accretion along the filament's long axis, we reported signs of this accretion in Section \ref{sec:collapse_mode}. 
Here, we provide a brief discussion on accretion, which will be useful for future detailed investigations.

In particular, focusing on the hub structure, the accretion rates attain $\sim220 M_\odot/\mathrm{Myr}$ in G22 \citep{2018ApJ...852...12Y} and $\sim2500 M_\odot/\mathrm{Myr}$ in SDC335 \citep{2013A&A...555A.112P}. 
These results indicate that, in the hub filament structure with massive stars, the accretion rate is around $10^{-3\sim-4} M_\odot/\mathrm{yr}$, which is necessary to form massive stars 
\citep[e.g.,][]{2003ApJ...585..850M}.
This implies that accretion along the filament axis is 
an important mass accumulation mechanism for the formation of massive stars.
Thus, it would be worthwhile to evaluate the gas accretion onto the intersection from the regions of the filaments that did not collide.

In orthogonal collisions, certain segments of the two filaments remain unaffected by the collision.
These segments are referred to as the non-colliding segments.
For simplicity, we assume that 
the non-colliding segments fall onto the intersection only due to gravity 
from the intersection and neglect the pressure gradient, Lorentz force, and the self-gravity of the non-colliding segments. 
In addition to this, we ignore the gas motion perpendicular to the vector from the non-colliding segments to the intersection.
The velocity of the gas flow $(V_\mathrm{inflow})$  along the non-colliding segments can be approximated as
\begin{equation}
V_\mathrm{inflow}\simeq \left[2G M_\mathrm{cross}\left( -\frac{1}{r_0}+\frac{1}{r_\mathrm{end}}\right)\right]^{1/2},
\end{equation}
where $r_0$ represents the initial position of a fluid element from the center of the intersection, and 
$r_\mathrm{end}$ corresponds to the position of the surface of the intersection.
For estimating $V_\mathrm{inflow}$, the initial filament width is 
fixed at $R_0\simeq 0.2~\mathrm{pc}$ and the relationship between the 
initial line mass ($\lambda_0$) and the mass of the intersection is 
derived from $M_\mathrm{cross} \simeq 2\times2R_0\times \lambda_0 $. 
We analyze the accretion from the surrounding filaments based on a model where gas at a distance of $2R_0$ from the center reaches the surface of the intersection, i.e., 
$r_0=2R_0$ and $r_\mathrm{end}=R_0$.
This is because, according to \citet{2019A&A...621A..42A}, filaments are defined as those with an aspect ratio, where the ratio of the filament length and width, of three or more.
Following this, we assume a filament with a minimum aspect ratio of three and derive the typical accretion rate by estimating the velocity of a gas element originally at $r_0=2R_0$, when it reaches the surface of the intersection ($r_\mathrm{end}=R_0$).
This position ($2R_0$) corresponds to the center of the non-colliding segment, which ranges from $R_0$ to $3R_0$, where the gravitational forces from the inner and outer sides of the non-colliding segment cancel each other out.
Therefore we expect that the fluid element is affected only by the gravity from $M_\mathrm{cross}$.
From the above assumptions, considering the accretion along the filament with $\lambda_0=\lambda_\mathrm{cri,B}$, the accretion velocity at the surface of the intersection is calculated as $V_\mathrm{inflow}(\beta_0=1)\simeq 3.1c_s$, $V_\mathrm{inflow}(\beta_0=0.1)\simeq 3.9c_s$, and $V_\mathrm{inflow}(\beta_0=0.01)\simeq 5.8c_s$.
The accretion rate $(\dot{M})$ can be obtained as $\dot{M}=4\times \lambda\times V_\mathrm{inflow}$, 
where the coefficient of four indicates the number of the non-colliding segments connected to the intersection. 
The accretion rates are approximately $\dot{M}(\beta_0=1)\simeq 4.8\times 10^{-5}M_\odot \mathrm{yr}^{-1}$, $\dot{M}(\beta_0=0.1)\simeq 9.6\times 10^{-5}M_\odot \mathrm{yr}^{-1}$, and $\dot{M}(\beta_0=0.01)\simeq 3.2\times 10^{-4}M_\odot \mathrm{yr}^{-1}$.
To summarize, the accretion rate for the range of filament line masses used in this calculation is up to $\sim 3\times10^{-4}M_\odot \mathrm{yr}^{-1}$ for filaments with $\beta_0=0.01$. 
This suggests that massive filaments with a line mass around the critical value at $\beta_0=0.01$ have the potential to exhibit observed high accretion rates.
However, to achieve the accretion rate around $10^{-3}M_\odot \mathrm{yr}^{-1}$ required for high-mass star formation, heavier filaments are necessary.

In addition, we estimate the timescale of accretion along the filament.
This is because if the accretion timescale is unrealistically longer than the star formation timescale, stars would form without the effect of the succeeding accretion.
If we assume that gas follows the free-fall motion, the time required to move from the initial position $r_0$ to $r$ is given as
\begin{equation}\label{eq:free_fall}
    t_\mathrm{infall}=\left(\theta+\frac{1}{2}\sin \theta \right) \left( \frac{r_0^3}{2GM_\mathrm{cross}}\right)^{1/2},
\end{equation}
where we use $r=r_0 \cos^2 \theta$.
Here, we consider $t_\mathrm{infall}$ necesary for moving from $r=2R_0$ to $r=R_0$, then substitute $\theta=\pi/4$ and $r_0=2R_0$ to Equation (\ref{eq:free_fall}).
The time scales, normalized by the free-fall time scale of the surface density of the filament $(t_\mathrm{ff}\equiv c_s/\sqrt{4\pi G \rho_s})$ for each plasma beta, are $t_\mathrm{infall}(\beta_0=1)=1.49t_\mathrm{ff}$, $t_\mathrm{infall}(\beta_0=0.1)=1.17t_\mathrm{ff}$, and $t_\mathrm{infall}(\beta_0=0.01)=0.78t_\mathrm{ff}$.
From this estimation, the timescale for accretion is roughly the same as $t_\mathrm{ff}$, suggesting that accretion along the filaments is feasible during star formation.
In addition to this, for accretion along the filaments to be significant, 
relatively massive filaments, or in other words, intersections containing a sufficient mass are required.

It is noted that the above discussion is done in which the gravity from the non-colliding segments is neglected. 
From our calculations, when the initial velocity is too high,
the non-colliding segments move away from the intersection (e.g., Figure \ref{fig:3d_L05B1_expand}),
making accretion onto the intersection less likely. 
Therefore, this discussion is expected to be valid only for collisions where the speed is around transonic.
However, it must be useful in understanding the characteristics of accretion along real filaments.

Additionally, gas accretion onto filaments from the surroundings may also be important. 
This phenomenon is supported by observations; for instance, \citet{2019A&A...623A..16S} found the accretion of surrounding gas onto filament B213/B213 in a low-mass star-forming region, and \citet{2019A&A...632A..83S} also suggest the possibility of gas accretion onto filaments in the high-mass star-forming region NGC6334.

Considering mass accretion after or together with the filament collision, gas supplied to high-density regions seems to encourage the formation of massive stars.

\subsection{Implication for Star Formation from Orthogonal and Parallel Collisions}\label{sec:implication_for_starformation}

In this study and \citepaperI{2023ApJ...954..129K}, we investigated the instability and subsequent evolution modes of the shocked clouds 
formed by orthogonal and parallel collisions of two identical filaments,
varying magnetic field strength, line mass, and initial velocity.
In parallel collisions, the instability of the shocked clouds is primarily influenced by whether the mass-to-magnetic flux ratio of the shocked region exceeds the critical one.
On the other hand, in Section \ref{sec:condition_for_collapse_and_expansion}, we revealed that the orthogonal collision model has a dependence on the initial velocity while the head-on collision models do not. 

Based on these findings, we proposed the required conditions for star formation triggered by 
filament-filament collisions in actual molecular clouds as follows:
When the filament long axes are almost parallel, \citepaperI{2023ApJ...954..129K} shows 
that star formation triggered by the collapse mode is expected only when the line mass of the shocked region exceeds the critical line mass.
In this scenario, it is anticipated that stars will form along the dense filamentary cloud formed by the collision.

When the filament long axes are nearly perpendicular to each other, it is shown that the outcome can be described by the total energy of the intersection ($E_\mathrm{all}$).
If $E_\mathrm{all}<0$, meaning that the magnitude of gravitational energy is dominant, the shocked cloud exhibits the collapse mode. 
In orthogonal collisions, stars are expected to form preferentially at the intersection of the two filaments. 

In Figure \ref{fig:stability_for_orthogonal}, the collapse symbols distribute even below the dashed horizontal line indicating the collapse conditions for parallel collisions, showing that filaments with the same line mass are more likely to collapse in orthogonal collisions than in parallel collisions. 
In the cases of supersonic orthogonal collisions, 
especially for weak magnetic fields, collisions between heavier filaments are required for collapse as the initial velocity increases (see Figure \ref{fig:stability_for_orthogonal}(a)).
Therefore, in the case of orthogonal collisions within molecular clouds with a large velocity dispersion (i.e., a large initial velocity), more massive filaments are required for the collapse mode at higher initial velocities.
However, in parallel collisions, the collapse is determined by the critical line mass, at least within the velocity range of this study.
In the case of a strong magnetic field, the velocity dependence of the outcomes of the orthogonal collisions becomes weak (see Figure \ref{fig:stability_for_orthogonal}(c)).

The above discussions can also be conducted using the speed ($V_\mathrm{eq}$) at which the dash-dotted line intersects the dashed horizontal line in Figure \ref{fig:stability_for_orthogonal}, where the dash-dotted line corresponds to $\lambda_0/\lambda_\mathrm{crit,B}=0.5$.
According to the total energy equation, this speed is $V_\mathrm{eq}\simeq5.3$ for $\beta=1$, $V_\mathrm{eq}\simeq5.3$ for $\beta=0.1$, and $V_\mathrm{eq}\simeq8.3$ for $\beta=0.01$.
From these, we expect that the easiness of collapse can be mainly divided into two cases: (1) When $V_\mathrm{int}\lesssim V_\mathrm{eq}$, orthogonal collisions are more likely to collapse than parallel collisions.
(2) When $V_\mathrm{int}\gtrsim V_\mathrm{eq}$, parallel collisions are more likely to collapse than orthogonal collisions.

\section{Summary}\label{sec:summary}

In this study, we perform three-dimensional ideal MHD simulations to explore orthogonal 
collisions between two identical isothermal filaments in a lateral magnetic field. 
The long axes of the two filaments are tilted 90 degrees.
They share the same magnetic flux tube and 
the collision direction is parallel to the global magnetic field.
As parameters, we consider the magnetic field strength, the initial line mass, and 
the initial velocity.
We investigate 48 models listed in Table \ref{tab:model_parameters}.
Our findings are summarized as follows:

\begin{enumerate}
\item  Orthogonal collisions lead to the formation of sheet-like shocked clouds in the intersection of two filaments. 
Subsequently, these shocked clouds exhibit rapid collapse, moderate collapse, stable, or expansion modes as shown in Figure \ref{fig:stability_for_orthogonal}.
In the rapid collapse, moderate collapse, and stable modes, the shocked cloud evolves into an oblate shape flattened along the magnetic field lines, whereas, in the expansion mode, it expands in the collision direction significantly.

\item For the collapse, stable, and expansion modes, we must carefully consider the energy balance of the shocked clouds. 
This is because a model that exhibited collapse mode at transonic collisions may show the expansion mode at supersonic velocities. 
If the magnitude of gravitational energy exceeds the total of kinetic, thermal, and magnetic energies immediately after the collision is completed, the shocked clouds undergo collapse. 
Conversely, if the magnitude of gravitational energy is decreased, the clouds transition from collapse to expansion modes (see Figure \ref{fig:stability_for_orthogonal}). 
This transition is primarily governed by the kinetic energy of the shocked cloud, which includes the outward expansion component immediately after the complete collision and can be predicted from the initial velocity.
Additionally, in cases of relatively low initial velocity, when the total of thermal and magnetic energies surpasses the gravitational energy, a stable mode is observed.

\item The criteria separating the rapid collapse and moderate collapse modes are determined by the time scale of the gravitational instability of the magnetized sheet ($t_m$) which is formed at the intersection, and the dynamical time scale for the filaments to complete the collision $(t_\mathrm{comp}$). 
If $t_\mathrm{comp}/t_m>1$, the shocked cloud evolves in the rapid collapse mode; if not, it evolves in the moderate collapse mode. The collapsing clouds exhibit the self-similar manner.

\item For the accretion along the filament's long axis, assuming gas elements undergo free-fall due to the gravity of the intersection, we estimate that the maximum accretion rate achievable with our model is $\dot{M}\simeq 3\times 10^{-4}M_\odot \rm{yr}^{-1}$.
However, a higher accretion rate would require heavier filaments. Additionally, our discussion assumes the initial velocity of the filaments is around transonic; supersonic collisions are unlikely to result in significant accretion.

\item By comparing orthogonal and parallel collisions, we investigated the differences in star formation and the resulting distribution of stars. 
If $V_\mathrm{int} \lesssim V_\mathrm{eq}$, orthogonal collisions are more likely to collapse than parallel collisions, whereas if $V_\mathrm{int} \gtrsim V_\mathrm{eq}$, parallel collisions are more likely to collapse than orthogonal collisions.
In orthogonal collisions, stars form near the intersection where oblate shocked clouds develop. 
In contrast, parallel collisions likely lead to star formation along the filamentary shocked clouds after fragmentation.
\end{enumerate}

In addition, even in models exhibiting the stable mode in this study, when considering the ambipolar diffusion, it is anticipated that during oscillation, the magnetic field may escape from the shocked cloud. 
This escape leads to an increase in the mass-to-flux ratio, and if this ratio exceeds the critical value, then the shocked cloud begins to collapse.
In practice, it is reasonable to consider collisions between filaments for an origin of massive star formation, taking into account of accretion, turbulence, and other realistic factors.
In addition to these considerations, simulations that take into account of feedback from the formed stars are necessary for a comprehensive understanding of hub structures and massive star formation.

\begin{acknowledgments}
We would like to express our sincere gratitude to the anonymous referee for the valuable comments and suggestions to improve this paper.
We are grateful to the following scholars for their valuable discussions regarding this research: Tomida, K., Inoue, T., Hacar, A., Minamidani, T., Ishii, S., Nakamura, F., Arzoumanian, D., Hanawa, T., and Peretto, N.
This study was supported by JST, the establishment of university fellowships towards the creation of science technology innovation, Grant Number JPMJFS2136, and by JSPS KAKENHI (Grant Numbers: 22J11106(RK), 19K03929, 19H01938 (KI), and 19K03919(KT)).
Numerical computations were performed in Cray XC50 at the Center for Computational Astrophysics, National Astronomical Observatory of Japan.
\end{acknowledgments}

%




\appendix

\section{Condition of the Collapse mode in Transonic Collisions Using Magnetized Critical Mass}\label{magnetized_mass}

In this section, we estimate the condition of the collapse mode in transonic collisions where the effect of the initial velocity is relatively low, using the magnetized critical mass, therefore we focus on the models with $V_\mathrm{int}=\pm0.5$ in Figure \ref{fig:stability_for_orthogonal}.
\citepaperI{2023ApJ...954..129K} conclude that in the case of head-on collisions, the instability of the shocked region is determined only by the condition of whether the total line mass exceeds the magnetized critical line mass.
Nonetheless, even in cases where the total line mass is smaller than the magnetized critical line mass ($\lambda_\mathrm{tot}(=2\lambda_0)<\lambda_\mathrm{crit,B}$), there are some parameter ranges where the shocked clouds exhibit the collapse mode.

A reason why the condition for gravitational collapse is not determined by $\lambda_\mathrm{tot}/\lambda_\mathrm{crit,B}$ for the orthogonal-collision 
models is that the shocked region is no more filamentary, but sheet-like with a finite lateral size.
It is thus reasonable to compare the total mass of the shocked region, estimated by $M_\mathrm{cross}=4R_f\lambda_0$ from the initial condition 
with the maximum mass of an oblate-shaped cloud $M_\mathrm{crit,B}$ that can be supported 
by magnetic fields, known as the magnetized critical mass \citep{1976ApJ...210..326M,1988ApJ...335..239T}.
\citet{1988ApJ...335..239T} found that $M_\mathrm{crit,B}$ is well approximated in terms of the mass-to-flux ratio at the cloud center $(dm/d\phi_\mathrm{B})_\mathrm{center}$ and the pressure of the cloud surface $p_\mathrm{surf}$ as follows:
\begin{equation}\label{eq:TIN1988_dimention}
  M_\mathrm{crit,B}= 62\left\{ 1-\left[\frac{0.17}{\left. dm/d(\phi_\mathrm{B}/G^{1/2})\right|_\mathrm{center}}\right]^2\right\}^{-3/2}\frac{c_s^4}{p^{1/2}_\mathrm{surf}(4\pi G)^{3/2}},
\end{equation}
and this equation represents the dimensional form of the magnetized critical mass.

The relationship between the differential mass-to-magnetic flux ratio and the average mass-to-magnetic flux ratio of the filament is shown
as 
\begin{equation}\label{eq:dlambda_dphi}
    \frac{d\lambda}{d\Phi}=\frac{2\lambda_0}{\pi \Phi_\mathrm{cl}}\left[1-\left(\frac{\Phi}{\Phi_\mathrm{cl}}\right)^2\right]^{1/2},
\end{equation}
where the mass-to-magnetic flux distribution is derived from Equation (21) in \citet{2014ApJ...785...24T}.  
This Equation (\ref{eq:dlambda_dphi}) assumes a cylindrical parental cloud with a uniform density $\bar{\rho}$ and the radius $R_0$, which is threaded by a uniform magnetic field $B_0$. 
Thus the differential mass-to-magnetic flux ratio at the center of the intersection is shown as
\begin{equation}\label{eq:dm_dphi_r_0}
    \left. \frac{dM}{d\phi}\right|_{r=0}=\left.  2\frac{d\lambda}{d\Phi}\right|_{\Phi=0}=\frac{4\lambda_0}{\pi \Phi_\mathrm{cl}},
\end{equation}
where the coefficient of two comes from the fact that there are two colliding filaments.
By substituting Equation (\ref{eq:dm_dphi_r_0}) into Equation (\ref{eq:TIN1988_dimention}), the magnetized critical mass is expressed in terms of the line mass and the magnetic flux. 

Then,the relationship between the mass of the intersection ($M_\mathrm{cross}$) and the line mass is provided in the non-dimensional form as follows:
\begin{equation}\label{eq:TIN1988_nondimention}
\frac{M_\mathrm{cross}}{M_\mathrm{crit,B}}\simeq\frac{4R_f\lambda_0}{62}\left\{ 1-\left[\frac{0.17}{\lambda_0/(\sqrt{2}\pi^2R_0\beta_0^{-1/2})}\right]^2\right\}^{3/2},
\end{equation}
where $R_0$ is the radius of the parental filament (here we assume $R_0\simeq R_f$). 
Then, in Equation (\ref{eq:TIN1988_nondimention}), we use the approximation $p_\mathrm{surf}\simeq c^2_s{\rho_s}$, assuming a quiescent coalescence. 
Equation (\ref{eq:TIN1988_nondimention}) shows whether the total mass of the intersection exceeds the magnetically critical mass, which is determined by the magnetic flux contained in the intersection.
Therefore, if $M_\mathrm{cross}/M_\mathrm{crit,B}>1$, it is expected that the shocked cloud at the intersection cannot support itself and will collapse.
The horizontal solid line of Figure \ref{fig:stability_for_orthogonal} corresponds to the line mass in which the mass of the intersection is equal to the magnetized critical mass, $M_\mathrm{cross}=M_\mathrm{crit,B}$ in Equation (\ref{eq:TIN1988_nondimention}).

In Figure \ref{fig:stability_for_orthogonal}, the solid line separates the collapse mode and the stable mode when the initial velocity is around the sound speed, except for the $\beta_0=0.01$ models.
In the $\beta_0=0.01$ models of Figure \ref{fig:stability_for_orthogonal} (c), gravitational collapse happens even below the solid line. 
Specifically, the model of $\lambda_0/\lambda_\mathrm{crit,B}(\beta_0=0.01)=0.45$ indicates the collapse mode, however the mass of the intersection smaller than the magnetized critical mass.
This fact comes from that the empirical formula for the magnetized critical mass is based on results in the range of $\beta_0=1-0.02$, reproducing numerical results within an error of around $10\%$ \citep{1988ApJ...335..239T}. 
Therefore, the larger error in Equation (\ref{eq:TIN1988_dimention}) for low $\beta_0$ may contribute to the discrepancy observed in the results in Figure \ref{fig:stability_for_orthogonal} (c).
However, in the case of $\beta_0=0.01$, the solid line of $M_\mathrm{cross}/M_\mathrm{crit,B}=1$ can barely distinguish the collapse mode within an error of approximately $15\%$.
Thus, in the case of collisions at $V_\mathrm{int}=\pm0.5$ with ranging from $\beta_0=1$ to 0.01, the instability of the shocked clouds can be distinguished using the ratio of the mass of the intersection to the magnetized critical mass $\left(M_\mathrm{cross}/M_\mathrm{crit,B}\right)$, as follows:
if $M_\mathrm{cross}/M_\mathrm{crit,B}\lesssim 1$, the shocked clouds will remain in the stable mode, whereas if $M_\mathrm{cross}/M_\mathrm{crit,B}\gtrsim 1$, the shocked clouds will evolve into the rapid or moderate collapse mode.

\section{Derivation of the Growth Timescale of Gravitational Instability in a Magnetized Sheet and the Compression Timescale}\label{sec:condition_for_rapid_collapse_Growth_timescale}

To determine the growth timescale for the gravitational instability of the magnetized sheet with the lateral 
magnetic field, we first derive the maximum growth rate. 
The density distribution in the thickness direction $(x)$
of the sheet is given by 
\begin{equation}
\rho(x)= \rho_c \mathrm{sech}^2\left[x/\left(\sqrt{2}H_0\right)\right]=\rho_c\left(1-\mu^2\right),
\end{equation}
where the scale height is defined as $H_0\equiv c_s/\left(4\pi G\rho_c\right)^{1/2}$, and $\mu\equiv \mathrm{tanh}\left(x/\sqrt{2}H_0\right)$.
The sheet is assumed to be confined by the external pressure $p_\mathrm{ext}$, and 
$\mu=\mu_\mathrm{s}$ at the surface is given by 
\begin{equation}
\mu_s=\left(1-\frac{p_\mathrm{ext}}{c_s^2 \rho_c}\right)^{1/2}.
\end{equation}
The relationship between the central density ($\rho_c$), the surface density ($\sigma_0$), and the external pressure ($p_\mathrm{ext}$) is given as follows:
\begin{equation}\label{eq:rho_c_and_surface_den}
    \rho_c=\frac{p_\mathrm{ext}}{c_s^2}+\frac{\pi G}{2 c_s^2}\sigma_0^2.
\end{equation}

For the magnetized sheet without the external pressure ($\mu_s=1$), 
\citet{1988PASJ...40..593N} obtained the following approximate expression for the maximum growth rate,
\begin{equation}
\omega^2_\mathrm{max}=-0.225\left(1-\alpha_0/2\right)^{2.5} \times 4\pi G\rho_c,
\label{omega2woPext}
\end{equation}
where $\alpha_0\equiv B^2_0/4\pi c_s^2\rho_c$ is a parameter for 
the strength of the magnetic field.
The factor $(1-\alpha_0/2)^{2.5}$ shows the stabilization effect due to the magnetic field.
The critical value $\alpha_\mathrm{0,crit}$ at which the instability vanishes is $\alpha_\mathrm{0,crit}\equiv \alpha_0=2$.\footnote{Although $\alpha_0$ is defined as $\alpha_N\equiv B^2_0/8\pi c_s^2\rho_c$ in \citet{1988PASJ...40..593N}, for convenience, we use the $\alpha_0$ from \citet{1983PASJ...35..187T}.}

When considering external pressure ($\mu_s<1$), from Equation (\ref{eq:rho_c_and_surface_den}), the surface density is written by using $\mu_s$ as
\begin{equation}\label{eq:surf_den_and_mu_s}
\sigma_0=
2\sqrt{2}\rho_cH_0\mu_s.
\end{equation}
Similar to Equation (\ref{omega2woPext}), even when the external pressure is not zero, 
we assume that the maximum growth rate $\omega_\mathrm{max}^2$ can be divided into 
a stabilization factor due to the magnetic field $f_\mathrm{mag}$ and the maximum 
growth rate without magnetic fields $\omega_\mathrm{max,hydro}^2$, or $\omega_\mathrm{max}^2 = f_\mathrm{mag}\times 
\omega_\mathrm{max,hydro}^2$. 
\citet{1983PASJ...35..187T} indicated the value of $\alpha_\mathrm{0,crit}$ where the instability disappears 
decreases with decreasing $\mu_s$, and can be approximated as 
\begin{equation}\label{eq:tomisaka_1983}
\alpha_\mathrm{0,crit}\simeq 2\mu_s^2.
\end{equation}
Since $f_\mathrm{mag}$ should be $(1-\alpha_0/2)^{2.5}$ for $\mu_\mathrm{s}=1$ (see 
Equation (\ref{omega2woPext})), 
we simply adopt $f_\mathrm{mag} = [1-\alpha_0/(2\mu_\mathrm{s}^2)]^{2.5}$ that satisfies Equation (\ref{eq:tomisaka_1983}).
An analytic expression of maximum growth rate of
a non-magnetized sheet derived by \citet{1978ApJ...220.1051E} is
\begin{equation}
\omega^2_\mathrm{max,hydro}
\simeq -\left(0.139-0.022\mu_s+0.103\mu_s^2\right)\times 4\pi G \rho_c.
\end{equation}
Finally, when the external pressure is taken into account, 
the maximum growth rate of the gravitational instability of magnetized sheets becomes
\begin{equation}\label{eq:omega_for_magnetized_sheet}
\omega^2_\mathrm{max}\simeq - \left(1-\alpha_0/2\mu_s^2\right)^{2.5}\left(0.139-0.022\mu_s+0.103\mu_s^2\right)\times 4\pi G \rho_c .
\end{equation}
Consequently, the growth timescale of the gravitational instability in the magnetized sheet is given by $t_m\equiv(-i\omega_\mathrm{max})^{-1}$.

Then we rewrite Equation (\ref{eq:omega_for_magnetized_sheet}) using the parameters used in our calculations.
The cross-section of the intersection is 
\begin{equation}
  S\simeq\left(2R_f\right)^2.
\end{equation}
The average surface density ($\sigma_0$), the average density before collision ($\bar{\rho}$), 
and the average magnetic flux density ($B_0$) corresponds to 
\begin{equation}\label{eq:surface_density}
   \sigma_0\simeq \frac{M_\mathrm{cross}}{S}=\frac{\lambda_0}{R_f},
\end{equation}
\begin{equation}\label{eq:mean_density}
    \bar{\rho}\simeq \frac{M_\mathrm{cross}/2}{2\theta R_f S}=\frac{\lambda_0}{4\theta R^2_f},
\end{equation}
and 
\begin{equation}
    B_0\simeq \frac{\Phi}{S}=\frac{\phi_\mathrm{cl}}{R_f},
\end{equation}
respectively.
Utilizing these parameters, $\mu_s$ is written as
\begin{equation}\label{eq:mu_s}
\begin{split}
    \mu_s&=\frac{\sigma_0}{2\sqrt{2}\rho_cH_0}\simeq\frac{(2\pi G\theta \lambda_0)^{1/2}}{c_s\mathcal{M}}=\frac{(\theta \lambda'_0/2)^{1/2}}{\mathcal{M}},
\end{split}
\end{equation}
where we use Equations (\ref{eq:surface_density}), (\ref{eq:mean_density}), $\lambda_0=\lambda'_0c_s/(4\pi G)^{1/2}$, and the approximation $\rho_c\simeq \bar{\rho}\mathcal{M}^2$ under the assumption of the isothermal shock wave (i.e. Rankine-Hugoniot relation for the isothermal gas).
Similarly, $\alpha_0$ is
\begin{equation}\label{eq:alpha_0}
    \alpha_0=\frac{B^2_0}{4\pi c_s^2\rho_c}\simeq \frac{B_0^2}{4\pi c_s^2\bar{\rho}\mathcal{M}^2}.
\end{equation}
From Equations (\ref{eq:mu_s}) and (\ref{eq:alpha_0}),
\begin{equation}\label{eq:alpha_0_mu_s}
  \frac{\alpha_0}{2\mu_s^2}\simeq 
  \frac{8R'^2_f}{\beta_0\lambda'^2_0}.
\end{equation}

Next, the time for the filament to complete the collision can be given by
\begin{equation}\label{eq:comp_time}
t_\mathrm{comp}=\frac{\theta R_f}{c_s\mathcal{M} },
\end{equation}
where $\theta$ is a geometric factor (a thickness) of the initial filament. 

From Equations (\ref{eq:omega_for_magnetized_sheet}) and (\ref{eq:comp_time}), the ratio of the timescales $t_\mathrm{comp}/t_m$ can be expressed as 
\begin{equation}\label{eq:t_comp_t_m_}
\begin{split}
\frac{t_\mathrm{comp}}{t_m}=\omega_\mathrm{max}t_\mathrm{comp}
=\left(1-\alpha_0/2\mu_s^2\right)^{1.25}\left(0.139-0.022\mu_s+0.103\mu_s^2\right)^{1/2}\frac{\lambda'^{1/2}_0\theta^{1/2}}{2}.\\   
\end{split}
\end{equation}
Equation (\ref{eq:t_comp_t_m_}) is the same as Equation (\ref{eq:t_comp_vs_t_m_nondim}).
By substituting Equations (\ref{eq:mu_s}) and (\ref{eq:alpha_0_mu_s}) into Equation (\ref{eq:t_comp_vs_t_m_nondim}), the initial velocity which satisfies $t_\mathrm{comp}/t_m=1$ is obtained as a function of the line mass.


\bibliography{sample631}{}

\begin{thebibliography}{}
\expandafter\ifx\csname natexlab\endcsname\relax\def\natexlab#1{#1}\fi
\providecommand{\url}[1]{\href{#1}{#1}}
\providecommand{\dodoi}[1]{doi:~\href{http://doi.org/#1}{\nolinkurl{#1}}}
\providecommand{\doeprint}[1]{\href{http://ascl.net/#1}{\nolinkurl{http://ascl.net/#1}}}
\providecommand{\doarXiv}[1]{\href{https://arxiv.org/abs/#1}{\nolinkurl{https://arxiv.org/abs/#1}}}

\bibitem[{{Andr{\'e}} {et~al.}(2014){Andr{\'e}}, {Di Francesco}, {Ward-Thompson}, {Inutsuka}, {Pudritz}, \& {Pineda}}]{2014prpl.conf...27A}
{Andr{\'e}}, P., {Di Francesco}, J., {Ward-Thompson}, D., {et~al.} 2014, in Protostars and Planets VI, ed. H.~{Beuther}, R.~S. {Klessen}, C.~P. {Dullemond}, \& T.~{Henning}, 27, \dodoi{10.2458/azu_uapress_9780816531240-ch002}

\bibitem[{{Andr{\'e}} {et~al.}(2010){Andr{\'e}}, {Men'shchikov}, {Bontemps}, {K{\"o}nyves}, {Motte}, {Schneider}, {Didelon}, {Minier}, {Saraceno}, {Ward-Thompson}, {di Francesco}, {White}, {Molinari}, {Testi}, {Abergel}, {Griffin}, {Henning}, {Royer}, {Mer{\'\i}n}, {Vavrek}, {Attard}, {Arzoumanian}, {Wilson}, {Ade}, {Aussel}, {Baluteau}, {Benedettini}, {Bernard}, {Blommaert}, {Cambr{\'e}sy}, {Cox}, {di Giorgio}, {Hargrave}, {Hennemann}, {Huang}, {Kirk}, {Krause}, {Launhardt}, {Leeks}, {Le Pennec}, {Li}, {Martin}, {Maury}, {Olofsson}, {Omont}, {Peretto}, {Pezzuto}, {Prusti}, {Roussel}, {Russeil}, {Sauvage}, {Sibthorpe}, {Sicilia-Aguilar}, {Spinoglio}, {Waelkens}, {Woodcraft}, \& {Zavagno}}]{2010A&A...518L.102A}
{Andr{\'e}}, P., {Men'shchikov}, A., {Bontemps}, S., {et~al.} 2010, \aap, 518, L102, \dodoi{10.1051/0004-6361/201014666}

\bibitem[{{Arzoumanian} {et~al.}(2019){Arzoumanian}, {Andr{\'e}}, {K{\"o}nyves}, {Palmeirim}, {Roy}, {Schneider}, {Benedettini}, {Didelon}, {Di Francesco}, {Kirk}, \& {Ladjelate}}]{2019A&A...621A..42A}
{Arzoumanian}, D., {Andr{\'e}}, P., {K{\"o}nyves}, V., {et~al.} 2019, \aap, 621, A42, \dodoi{10.1051/0004-6361/201832725}

\bibitem[{{Arzoumanian} {et~al.}(2023){Arzoumanian}, {Arakawa}, {Kobayashi}, {Iwasaki}, {Fukuda}, {Mori}, {Hirai}, {Kunitomo}, {Kumar}, \& {Kokubo}}]{2023ApJ...947L..29A}
{Arzoumanian}, D., {Arakawa}, S., {Kobayashi}, M. I.~N., {et~al.} 2023, \apjl, 947, L29, \dodoi{10.3847/2041-8213/acc849}

\bibitem[{{Berger} \& {Colella}(1989)}]{1989JCoPh..82...64B}
{Berger}, M.~J., \& {Colella}, P. 1989, Journal of Computational Physics, 82, 64, \dodoi{10.1016/0021-9991(89)90035-1}

\bibitem[{{Bontemps} {et~al.}(2010){Bontemps}, {Motte}, {Csengeri}, \& {Schneider}}]{2010A&A...524A..18B}
{Bontemps}, S., {Motte}, F., {Csengeri}, T., \& {Schneider}, N. 2010, \aap, 524, A18, \dodoi{10.1051/0004-6361/200913286}

\bibitem[{{Chandrasekhar}(1969)}]{1969efe..book.....C}
{Chandrasekhar}, S. 1969, {Ellipsoidal Figures of Equilibrium} (New Haven: Yale University Press)

\bibitem[{{Dewangan}(2019)}]{2019ApJ...884...84D}
{Dewangan}, L.~K. 2019, \apj, 884, 84, \dodoi{10.3847/1538-4357/ab4189}

\bibitem[{{Dewangan} {et~al.}(2024){Dewangan}, {Bhadari}, {Maity}, {Eswaraiah}, {Sharma}, \& {Jadhav}}]{2024MNRAS.527.5895D}
{Dewangan}, L.~K., {Bhadari}, N.~K., {Maity}, A.~K., {et~al.} 2024, \mnras, 527, 5895, \dodoi{10.1093/mnras/stad3384}

\bibitem[{{Duarte-Cabral} {et~al.}(2011){Duarte-Cabral}, {Dobbs}, {Peretto}, \& {Fuller}}]{2011A&A...528A..50D}
{Duarte-Cabral}, A., {Dobbs}, C.~L., {Peretto}, N., \& {Fuller}, G.~A. 2011, \aap, 528, A50, \dodoi{10.1051/0004-6361/201015477}

\bibitem[{{Elmegreen} \& {Elmegreen}(1978)}]{1978ApJ...220.1051E}
{Elmegreen}, B.~G., \& {Elmegreen}, D.~M. 1978, \apj, 220, 1051, \dodoi{10.1086/155991}

\bibitem[{{Evans} \& {Hawley}(1988)}]{1988ApJ...332..659E}
{Evans}, C.~R., \& {Hawley}, J.~F. 1988, \apj, 332, 659, \dodoi{10.1086/166684}

\bibitem[{{Gardiner} \& {Stone}(2008)}]{2008JCoPh.227.4123G}
{Gardiner}, T.~A., \& {Stone}, J.~M. 2008, Journal of Computational Physics, 227, 4123, \dodoi{10.1016/j.jcp.2007.12.017}

\bibitem[{{Gottlieb} {et~al.}(2009){Gottlieb}, {Ketcheson}, \& {Shu}}]{2009JSCom.38..251G}
{Gottlieb}, S., {Ketcheson}, D.~I., \& {Shu}, C.-W. 2009, Journal of Scientific Computing, 38, 251, \dodoi{10.1007/s10915-008-9239-z}

\bibitem[{{Hacar} {et~al.}(2023){Hacar}, {Clark}, {Heitsch}, {Kainulainen}, {Panopoulou}, {Seifried}, \& {Smith}}]{2023ASPC..534..153H}
{Hacar}, A., {Clark}, S.~E., {Heitsch}, F., {et~al.} 2023, in Astronomical Society of the Pacific Conference Series, Vol. 534, Protostars and Planets VII, ed. S.~{Inutsuka}, Y.~{Aikawa}, T.~{Muto}, K.~{Tomida}, \& M.~{Tamura}, 153, \dodoi{10.48550/arXiv.2203.09562}

\bibitem[{{Hoemann} {et~al.}(2021){Hoemann}, {Heigl}, \& {Burkert}}]{2021MNRAS.507.3486H}
{Hoemann}, E., {Heigl}, S., \& {Burkert}, A. 2021, \mnras, 507, 3486, \dodoi{10.1093/mnras/stab1698}

\bibitem[{{Hoemann} {et~al.}(2024){Hoemann}, {Socci}, {Heigl}, {Burkert}, \& {Hacar}}]{2024MNRAS.532L..42H}
{Hoemann}, E., {Socci}, A., {Heigl}, S., {Burkert}, A., \& {Hacar}, A. 2024, \mnras, 532, L42, \dodoi{10.1093/mnrasl/slae045}

\bibitem[{{Inutsuka} \& {Miyama}(1992)}]{1992ApJ...388..392I}
{Inutsuka}, S.-I., \& {Miyama}, S.~M. 1992, \apj, 388, 392, \dodoi{10.1086/171162}

\bibitem[{{Kashiwagi} {et~al.}(2023){Kashiwagi}, {Iwasaki}, \& {Tomisaka}}]{2023ApJ...954..129K}
{Kashiwagi}, R., {Iwasaki}, K., \& {Tomisaka}, K. 2023, \apj, 954, 129, \dodoi{10.3847/1538-4357/ace7bd}

\bibitem[{{Kashiwagi} \& {Tomisaka}(2021)}]{2021ApJ...911..106K}
{Kashiwagi}, R., \& {Tomisaka}, K. 2021, \apj, 911, 106, \dodoi{10.3847/1538-4357/abea7a}

\bibitem[{{Kawachi} \& {Hanawa}(1998)}]{1998PASJ...50..577K}
{Kawachi}, T., \& {Hanawa}, T. 1998, \pasj, 50, 577, \dodoi{10.1093/pasj/50.6.577}

\bibitem[{{K{\"o}nyves} {et~al.}(2015){K{\"o}nyves}, {Andr{\'e}}, {Men'shchikov}, {Palmeirim}, {Arzoumanian}, {Schneider}, {Roy}, {Didelon}, {Maury}, {Shimajiri}, {Di Francesco}, {Bontemps}, {Peretto}, {Benedettini}, {Bernard}, {Elia}, {Griffin}, {Hill}, {Kirk}, {Ladjelate}, {Marsh}, {Martin}, {Motte}, {Nguy{\^e}n Luong}, {Pezzuto}, {Roussel}, {Rygl}, {Sadavoy}, {Schisano}, {Spinoglio}, {Ward-Thompson}, \& {White}}]{2015A&A...584A..91K}
{K{\"o}nyves}, V., {Andr{\'e}}, P., {Men'shchikov}, A., {et~al.} 2015, \aap, 584, A91, \dodoi{10.1051/0004-6361/201525861}

\bibitem[{{Kumar} {et~al.}(2020){Kumar}, {Palmeirim}, {Arzoumanian}, \& {Inutsuka}}]{2020A&A...642A..87K}
{Kumar}, M.~S.~N., {Palmeirim}, P., {Arzoumanian}, D., \& {Inutsuka}, S.~I. 2020, \aap, 642, A87, \dodoi{10.1051/0004-6361/202038232}

\bibitem[{{Machida} {et~al.}(2004){Machida}, {Tomisaka}, \& {Matsumoto}}]{2004MNRAS.348L...1M}
{Machida}, M.~N., {Tomisaka}, K., \& {Matsumoto}, T. 2004, \mnras, 348, L1, \dodoi{10.1111/j.1365-2966.2004.07402.x}

\bibitem[{{McKee} \& {Tan}(2003)}]{2003ApJ...585..850M}
{McKee}, C.~F., \& {Tan}, J.~C. 2003, \apj, 585, 850, \dodoi{10.1086/346149}

\bibitem[{{Miyoshi} \& {Kusano}(2005)}]{2005JCoPh.208..315M}
{Miyoshi}, T., \& {Kusano}, K. 2005, Journal of Computational Physics, 208, 315, \dodoi{10.1016/j.jcp.2005.02.017}

\bibitem[{{Mouschovias} \& {Spitzer}(1976)}]{1976ApJ...210..326M}
{Mouschovias}, T.~C., \& {Spitzer}, L., J. 1976, \apj, 210, 326, \dodoi{10.1086/154835}

\bibitem[{{Nagasawa}(1987)}]{1987PThPh..77..635N}
{Nagasawa}, M. 1987, Progress of Theoretical Physics, 77, 635, \dodoi{10.1143/PTP.77.635}

\bibitem[{{Nagasawa} \& {Miyama}(1987)}]{1987PThPh..78.1250N}
{Nagasawa}, M., \& {Miyama}, S.~M. 1987, Progress of Theoretical Physics, 78, 1250, \dodoi{10.1143/PTP.78.1250}

\bibitem[{{Nakamura} {et~al.}(1999){Nakamura}, {Matsumoto}, {Hanawa}, \& {Tomisaka}}]{1999ApJ...510..274N}
{Nakamura}, F., {Matsumoto}, T., {Hanawa}, T., \& {Tomisaka}, K. 1999, \apj, 510, 274, \dodoi{10.1086/306553}

\bibitem[{{Nakamura} {et~al.}(2014){Nakamura}, {Sugitani}, {Tanaka}, {Nishitani}, {Dobashi}, {Shimoikura}, {Shimajiri}, {Kawabe}, {Yonekura}, {Mizuno}, {Kimura}, {Tokuda}, {Kozu}, {Okada}, {Hasegawa}, {Ogawa}, {Kameno}, {Shinnaga}, {Momose}, {Nakajima}, {Onishi}, {Maezawa}, {Hirota}, {Takano}, {Iono}, {Kuno}, \& {Yamamoto}}]{2014ApJ...791L..23N}
{Nakamura}, F., {Sugitani}, K., {Tanaka}, T., {et~al.} 2014, \apjl, 791, L23, \dodoi{10.1088/2041-8205/791/2/L23}

\bibitem[{{Nakano}(1988)}]{1988PASJ...40..593N}
{Nakano}, T. 1988, \pasj, 40, 593

\bibitem[{{Ostriker}(1964)}]{1964ApJ...140.1056O}
{Ostriker}, J. 1964, \apj, 140, 1056, \dodoi{10.1086/148005}

\bibitem[{{Palmeirim} {et~al.}(2013){Palmeirim}, {Andr{\'e}}, {Kirk}, {Ward-Thompson}, {Arzoumanian}, {K{\"o}nyves}, {Didelon}, {Schneider}, {Benedettini}, {Bontemps}, {Di Francesco}, {Elia}, {Griffin}, {Hennemann}, {Hill}, {Martin}, {Men'shchikov}, {Molinari}, {Motte}, {Nguyen Luong}, {Nutter}, {Peretto}, {Pezzuto}, {Roy}, {Rygl}, {Spinoglio}, \& {White}}]{2013A&A...550A..38P}
{Palmeirim}, P., {Andr{\'e}}, P., {Kirk}, J., {et~al.} 2013, \aap, 550, A38, \dodoi{10.1051/0004-6361/201220500}

\bibitem[{{Pattle} {et~al.}(2023){Pattle}, {Fissel}, {Tahani}, {Liu}, \& {Ntormousi}}]{2023ASPC..534..193P}
{Pattle}, K., {Fissel}, L., {Tahani}, M., {Liu}, T., \& {Ntormousi}, E. 2023, in Astronomical Society of the Pacific Conference Series, Vol. 534, Protostars and Planets VII, ed. S.~{Inutsuka}, Y.~{Aikawa}, T.~{Muto}, K.~{Tomida}, \& M.~{Tamura}, 193, \dodoi{10.48550/arXiv.2203.11179}

\bibitem[{{Peretto} {et~al.}(2013){Peretto}, {Fuller}, {Duarte-Cabral}, {Avison}, {Hennebelle}, {Pineda}, {Andr{\'e}}, {Bontemps}, {Motte}, {Schneider}, \& {Molinari}}]{2013A&A...555A.112P}
{Peretto}, N., {Fuller}, G.~A., {Duarte-Cabral}, A., {et~al.} 2013, \aap, 555, A112, \dodoi{10.1051/0004-6361/201321318}

\bibitem[{{Pilbratt} {et~al.}(2010){Pilbratt}, {Riedinger}, {Passvogel}, {Crone}, {Doyle}, {Gageur}, {Heras}, {Jewell}, {Metcalfe}, {Ott}, \& {Schmidt}}]{2010A&A...518L...1P}
{Pilbratt}, G.~L., {Riedinger}, J.~R., {Passvogel}, T., {et~al.} 2010, \aap, 518, L1, \dodoi{10.1051/0004-6361/201014759}

\bibitem[{{Shimajiri} {et~al.}(2019{\natexlab{a}}){Shimajiri}, {Andr{\'e}}, {Ntormousi}, {Men'shchikov}, {Arzoumanian}, \& {Palmeirim}}]{2019A&A...632A..83S}
{Shimajiri}, Y., {Andr{\'e}}, P., {Ntormousi}, E., {et~al.} 2019{\natexlab{a}}, \aap, 632, A83, \dodoi{10.1051/0004-6361/201935689}

\bibitem[{{Shimajiri} {et~al.}(2019{\natexlab{b}}){Shimajiri}, {Andr{\'e}}, {Palmeirim}, {Arzoumanian}, {Bracco}, {K{\"o}nyves}, {Ntormousi}, \& {Ladjelate}}]{2019A&A...623A..16S}
{Shimajiri}, Y., {Andr{\'e}}, P., {Palmeirim}, P., {et~al.} 2019{\natexlab{b}}, \aap, 623, A16, \dodoi{10.1051/0004-6361/201834399}

\bibitem[{{Stod{\'o}{\l}kiewicz}(1963)}]{1963AcA....13...30S}
{Stod{\'o}{\l}kiewicz}, J.~S. 1963, \actaa, 13, 30

\bibitem[{{Stone} {et~al.}(2020){Stone}, {Tomida}, {White}, \& {Felker}}]{2020ApJS..249....4S}
{Stone}, J.~M., {Tomida}, K., {White}, C.~J., \& {Felker}, K.~G. 2020, \apjs, 249, 4, \dodoi{10.3847/1538-4365/ab929b}

\bibitem[{{Sugitani} {et~al.}(2011){Sugitani}, {Nakamura}, {Watanabe}, {Tamura}, {Nishiyama}, {Nagayama}, {Kandori}, {Nagata}, {Sato}, {Gutermuth}, {Wilson}, \& {Kawabe}}]{2011ApJ...734...63S}
{Sugitani}, K., {Nakamura}, F., {Watanabe}, M., {et~al.} 2011, \apj, 734, 63, \dodoi{10.1088/0004-637X/734/1/63}

\bibitem[{{Sugitani} {et~al.}(2019){Sugitani}, {Nakamura}, {Shimoikura}, {Dobashi}, {Nguyen-Luong}, {Kusune}, {Nagayama}, {Watanabe}, {Nishiyama}, \& {Tamura}}]{2019PASJ...71S...7S}
{Sugitani}, K., {Nakamura}, F., {Shimoikura}, T., {et~al.} 2019, \pasj, 71, S7, \dodoi{10.1093/pasj/psz072}

\bibitem[{{Tokuda} {et~al.}(2019){Tokuda}, {Fukui}, {Harada}, {Saigo}, {Tachihara}, {Tsuge}, {Inoue}, {Torii}, {Nishimura}, {Zahorecz}, {Nayak}, {Meixner}, {Minamidani}, {Kawamura}, {Mizuno}, {Indebetouw}, {Sewi{\l}o}, {Madden}, {Galametz}, {Lebouteiller}, {Chen}, \& {Onishi}}]{2019ApJ...886...15T}
{Tokuda}, K., {Fukui}, Y., {Harada}, R., {et~al.} 2019, \apj, 886, 15, \dodoi{10.3847/1538-4357/ab48ff}

\bibitem[{{Tokuda} {et~al.}(2023){Tokuda}, {Harada}, {Tanaka}, {Inoue}, {Shimonishi}, {Zhang}, {Sewi{\l}o}, {Kunitoshi}, {Konishi}, {Fukui}, {Kawamura}, {Onishi}, \& {Machida}}]{2023ApJ...955...52T}
{Tokuda}, K., {Harada}, N., {Tanaka}, K. E.~I., {et~al.} 2023, \apj, 955, 52, \dodoi{10.3847/1538-4357/acefb7}

\bibitem[{{Tomida} \& {Stone}(2023)}]{2023ApJS..266....7T}
{Tomida}, K., \& {Stone}, J.~M. 2023, \apjs, 266, 7, \dodoi{10.3847/1538-4365/acc2c0}

\bibitem[{{Tomisaka}(1998)}]{1998ApJ...502L.163T}
{Tomisaka}, K. 1998, \apjl, 502, L163, \dodoi{10.1086/311504}

\bibitem[{{Tomisaka}(2014)}]{2014ApJ...785...24T}
---. 2014, \apj, 785, 24, \dodoi{10.1088/0004-637X/785/1/24}

\bibitem[{{Tomisaka} \& {Ikeuchi}(1983)}]{1983PASJ...35..187T}
{Tomisaka}, K., \& {Ikeuchi}, S. 1983, \pasj, 35, 187

\bibitem[{{Tomisaka} {et~al.}(1988){Tomisaka}, {Ikeuchi}, \& {Nakamura}}]{1988ApJ...335..239T}
{Tomisaka}, K., {Ikeuchi}, S., \& {Nakamura}, T. 1988, \apj, 335, 239, \dodoi{10.1086/166923}

\bibitem[{{Truelove} {et~al.}(1997){Truelove}, {Klein}, {McKee}, {Holliman}, {Howell}, \& {Greenough}}]{1997ApJ...489L.179T}
{Truelove}, J.~K., {Klein}, R.~I., {McKee}, C.~F., {et~al.} 1997, \apjl, 489, L179, \dodoi{10.1086/310975}

\bibitem[{{Yuan} {et~al.}(2018){Yuan}, {Li}, {Wu}, {Ellingsen}, {Henkel}, {Wang}, {Liu}, {Liu}, {Zavagno}, {Ren}, \& {Huang}}]{2018ApJ...852...12Y}
{Yuan}, J., {Li}, J.-Z., {Wu}, Y., {et~al.} 2018, \apj, 852, 12, \dodoi{10.3847/1538-4357/aa9d40}

\end{thebibliography}
\bibliographystyle{aasjournal}



\end{document}